\documentclass{emulateapj}
\usepackage{amsmath}
\usepackage{natbib}
\def\galfit{{\sc Galfit}}
\def\sersic{{S\'ersic}}

\begin{document}

\slugcomment{Submitted to {\it The Astronomical Journal}}

\title{DETAILED DECOMPOSITION OF GALAXY IMAGES. II. Beyond Axisymmetric Models}

\author {Chien Y. Peng\altaffilmark{1}, Luis C. Ho\altaffilmark{2}, Chris D.
Impey\altaffilmark{3}, and Hans-Walter Rix\altaffilmark{4}}

\altaffiltext{1}{Herzberg Institute of Astrophysics, National Research Council
of Canada, 5071 West Saanich Road, Victoria, British Columbia, Canada V9E 2E7;
cyp@nrc-cnrc.gc.ca}

\altaffiltext{2} {The Observatories of the Carnegie Institution for Science,
813 Santa Barbara St., Pasadena, CA 91101;  lho@obs.carnegiescience.edu}

\altaffiltext{3}{Steward Observatory, University of Arizona, 933 N. Cherry
Av., Tucson, AZ 85721;  cimpey@as.arizona.edu}

\altaffiltext{4}{Max-Planck-Institut f\"{u}r Astronomie, Koenigstuhl 17,
        Heidelberg, D-69117, Germany;  rix@mpia-hd.mpg.de}

\begin {abstract}

We present a two-dimensional (2-D) fitting algorithm (\galfit, Version 3) with
new capabilities to study the structural components of galaxies and other
astronomical objects in digital images.  Our technique improves on previous
2-D fitting algorithms by allowing for irregular, curved, logarithmic and
power-law spirals, ring and truncated shapes in otherwise traditional
parametric functions like the \sersic, Moffat, King, Ferrer, etc., profiles.
One can mix and match these new shape features freely, with or without
constraints, apply them to an arbitrary number of model components and of
numerous profile types, so as to produce realistic-looking galaxy model
images.  Yet, despite the potential for extreme complexity, the meaning of the
key parameters like the \sersic\ index, effective radius or luminosity remain
intuitive and essentially unchanged.  The new features have an interesting
potential for use to quantify the degree of asymmetry of galaxies, to quantify
low surface brightness tidal features beneath and beyond luminous galaxies, to
allow more realistic decompositions of galaxy subcomponents in the presence of
strong rings and spiral arms, and to enable ways to gauge the uncertainties
when decomposing galaxy subcomponents.  We illustrate these new features by
way of several case studies that display various levels of complexity.

\end {abstract}

\keywords{galaxies: bulges --- galaxies: fundamental parameters --- galaxies: 
structure --- techniques: image processing --- techniques: photometric}

\section {INTRODUCTION}

Images of astronomical objects store a wealth of information that encodes the
physical conditions and fossil records of their evolution.  Over the past
decade, the ability of optical/near-infrared telescopes to resolve objects
improved by a factor of 10, and to detect faint surface brightnesses by at
least 2 orders of magnitude.  These advances now enable the study of highly
intricate details on subarcsecond scales (e.g., nuclear star cluster, spiral
structure, bars, inner ring, profile cusps, etc.), and extremely faint outer
regions of galaxies.  Moreover, new integral-field imaging capabilities blur
the traditional boundary of obtaining, analyzing, and interpreting imaging and
spectroscopic data.  Faced with the convergence in volume, quality, and
multiwavelength datasets like never before, one of the main challenges toward
making full use of the investments is developing sophisticated ways to extract
information from the data to facilitate new science.

\subsection {Parametric and Non-Parametric Analysis}

Analyzing astronomical images is challenging because of the diversity in
object sizes and shapes, and nowhere is it more difficult than for galaxies.
Since the early era of photographic plates, one of the key methods for
studying the light distribution of galaxies is to model it by using analytic
functions---a technique known as parametric fitting.  This technique was first
applied to galaxies by \citet{devauc48} who showed that the light distribution
of elliptical galaxies tended to follow a power-law form of ${\rm exp}
\left(-r^{1/4}\right)$.  Subsequently, one of the breakthroughs in our
understanding of galaxy structure and evolution came when \citet{freeman70}
showed that dynamically ``hot'' stars in galaxies make up spheroidal bulges
having a de~Vaucouleurs light profile, whereas ``cold'' stellar components
make up the more flattened, rotationally supported, exponential disk region.

From that simple beginning, parametric fitting has been the mainstay for
galaxy imaging studies, and expanding into many applications whenever the
science calls for detailed and rigorous analysis.  Among some of the examples,
past investigations delved into the structural parameters of disk galaxies
\citep[e.g.,][]{dejong96}, the Tully-Fisher relation
\citep[e.g.,][]{tully77, hinz03, bedregal06}, the evolution of disky galaxies
\citep{simard02, ravindranath04, barden05}, the cosmic evolution of galaxy
morphology \citep[e.g.,][]{lilly98,marleau98,hathi09} in both groundbased
surveys and Hubble (Ultra-)Deep Fields \citep{williams96,beckwith06}, the
morphological transformation of galaxies in cluster environments
\citep[e.g.,][]{dressler80}, the fundamental plane of spheroids
\citep{djorgovski87, dressler87, bender92}, the red sequence of galaxies
\citep{bell04b, bell04c, faber07}, morphological dissimilarities between
spheroidal galaxies and ellipticals \citep{kormendy85, kormendy87}, the
central structure of early-type galaxies \citep{kormendy85, lauer95, faber97,
ferrarese06b, ferrarese06c, lauer07c} and implications for the formation of
massive black holes \citep{ravindranath02, kormendy09}, black hole vs. bulge
relations \citep{kormendy95} and their evolution \citep{rix01, peng06b,
peng06a}, the ``extra light'' due to gas dissipation in galaxy centers
\citep{kormendy99, kormendy09b, hopkins08a, hopkins08b}, quasar host galaxies
\citep[e.g.,][]{hutchings84, mcleod94a, mclure00, jahnke04, sanchez04,
kim08b}, gravitational lensing of quasar host galaxies \citep{rix01, peng06a},
and the clustering of dark matter through weak lensing
\citep[e.g.,][]{heymans06, heymans08}.

Since the original development of galaxy fitting nearly 70 years ago, where
the analysis was performed on one-dimensional (1-D) surface brightness
profiles \citep[see also][]{kormendy77, burstein79, boroson81, kent85,
andredakis94, macarthur03}, newer techniques have emerged to directly analyze
two-dimensional (2-D) images \citep[e.g.,][]{shaw89, mcleod94a, byun95,
dejong96, moriondo98, simard98, ratnatunga99, wadadekar99, simard02,
desouza04, laurikainen04, gadotti08}.  The benefit of performing 2-D image
analysis is to potentially make full use of all spatial information and to
properly account for image smearing by the point-spread function (PSF).

Even though 2-D analysis can be quite sophisticated, there are legitimate
questions about whether it is more beneficial than 1-D for profile analysis.
Proponents of the 1-D technique are skeptical that perfect ellipsoid models
are suitable to use for galaxies that show isophotal twists, or that are
non-elliptical in shape.  They note that not only is 1-D analysis more
appealing because it is more straightforward to implement, the surface
brightness profiles serve as visual confirmation about the reality of fitting
multiple component models.  

However, beneath the apparent simplicity there are a number of important
subtleties to weigh.  For instance, the decision about how to extract 1-D
profiles is often not unique, nor are there strong reasons to prefer major or
minor axis profiles, or a profile along some arc traced by spiral arms or
isophote twists that result from the superposition of multiple components
oriented at different angles.  When symmetry is broken it is also unclear that
there is an optimal or unique way to extract a 1-D profile, such as in
irregular galaxies, overlapping galaxies, and galaxies with double nuclei.
Another factor to consider is that the process of extracting 1-D profiles
reduces spatial information content:  in many situations, a bulge, disk, and
bar can all have different axis ratios, position angles (PAs), and profiles
that help to break model degeneracies, but this information is lost when the
data are collapsed into 1-D.  Lastly, for compact galaxies, 1-D profile
fitting cannot properly correct for image smearing by the PSF because 1-D
profile convolution is not mathematically equivalent to convolution in a 2-D
image.  While some of the above concerns also affect 2-D analysis (i.e.
irregular galaxies), most others benefit from treatment using 2-D techniques.
When it comes to judging which models are more plausible, there are few
diagnostics more discerning than a moment's glance at 2-D models and residual
images; a good fit in 2-D always means that 1-D profiles are necessarily a
good fit.  Proponents of 2-D analysis therefore believe that the benefits
outweigh the drawbacks.  Moreover, many drawbacks can be mitigated by breaking
free from axisymmetry in 2-D analysis, which is the purpose of this study to
show.

In the box of tools for morphology analysis, a complimentary approach is
non-parametric analysis.  While we do not use non-parametric methods in this
study, it is useful to understand the conceptual differences between the two
approaches.  We thus provide a brief overview.  In contrast to function
fitting, the non-parametric approach does not involve deciding what functional
form to use or how many.  One method is to decompose an image into
``shapelets'' or "wavelets" \citep[e.g.,][]{refregier03, massey04}, which is
analogous to taking a 2-D Fourier transform of an image using mathematically
orthogonal basis functions.  The main conceptual difference with parametric
fitting is that the shapelet basis functions do not represent physical
subcomponents of a galaxy.  Moreover, the power spectrum of the basis
functions is quite useful for diagnosing the degree of galaxy distortions.
There are also other non-parametric techniques \citep[e.g.,][]{abraham94,
rudnick98, conselice00, lotz04}.  To measure concentration non-parametrically,
one way is to compare fluxes within apertures of different radii; whereas to
measure asymmetry one can rotate an image by $180^\circ$ and subtract it from
the original image and measure the residuals \citep[e.g.,][]{abraham94,
conselice00}.  Toward the same goals, two studies, \citet{abraham03} and
\citet{lotz04}, introduce the Gini index to measure the concentration of a
galaxy by comparing the relative distribution of pixel flux values within a
certain area.  \citet{lotz04} also introduce a method for measuring asymmetry
through the $M_{20}$ parameter, which is the second-order moment of the
brightest 20\% of the a galaxy's flux.

The application of non-parametric analysis has mostly been to quantify galaxy
mergers \citep[e.g.,][]{conselice03b, lotz08}.  These techniques are generally
much simpler to implement than parametric fitting and have a strong virtue
that no assumptions are made about the galaxy profiles and shapes.  The
tradeoff is that the techniques often do not deal with image smearing by the
PSF and different sensitivity thresholds between different surveys.
Consequently, one has to take particular care to compare compact with extended
objects, measured in different apertures, or measured from images of different
surface brightness depths \citep{lisker08}.  One also should guard against
contamination by intervening galaxies or stars because the techniques do not
have a rigorous way to separate overlapping objects.  For separating objects,
extracting structural components of a galaxy, and extrapolating galaxy wings
well into the background noise, there are few, if any, alternatives to
parametric analysis that are more rigorous.

When comparing the merits of non-parametric and parametric analysis, the idea
of ellipsoid models in parametric analysis is sometimes considered to be a
weakness, because galaxies, after all, are not perfect ellipses in projection.
However, it is worth pointing out that the notion of there being a {\it global
average} size inherently implies comparison against some kind of approximate
shape.  Indeed, even in non-parametric techniques, to measure a size in an 2-D
image, one assumes a basic shape either explicitly (through using aperture
photometry) or implicitly (through calculating flux moments, which requires a
center to be defined).  An ellipsoid is one of the simplest and most natural
low-order shapes against which all galaxies can be compared, especially for
measuring an average size.  This notion is useful: deviations from the basic
ellipsoid shape can then be considered as higher order modifications, even for
highly irregular galaxies.

Nevertheless, there are many situations where it is desirable to use models
that deviate from ellipsoid shapes.  Contrary to the common practice, there
are numerous ways to break from axisymmetry.  However, the harder challenge is
to devise a scheme that is intuitive to grasp and well motivated.  Breaking
free from axisymmetry allows for other interesting science applications,
including a promising new way to quantify asymmetry.

\subsection {The Next Generation of Parametric Imaging Fitting}

In this study we present, as a proof of concept, new capabilities in 2-D image
fitting that progress beyond the limitations of traditional parametric fitting
models.  One key aspect of our approach is to first identify a minimum basis
set of features that spans the range of galaxy morphologies and shapes.  From
experience, we determine those four ``basic elements'' to be bending, Fourier,
coordinate rotation, and truncation modes.  Secondly, one of the main reasons
why parametric fitting is useful is that the profile parameters are intuitive
to grasp (e.g., concentration index, effective radius, total luminosity, etc.).
Therefore, another key requirement is that the traditional profile parameters
must retain their original, intuitive, meaning even under detailed shape
refinements, and even under such extreme cases as irregular galaxies.  This
can be accomplished if the basic premise starts with the traditional ellipsoid
function, on top of which one can add perturbations, rotations,
irregularities, and curvature.  This is possible because of the fact that
simple ellipsoid fits are a reasonable way to quantify global average
properties, and other details can be considered to be higher order
perturbations that may be of other practical interest.

As we attempt to demonstrate, combining just the four basic mophology elements
can quickly yield a dizzying array of possibilities for fitting galaxies.  The
end result can look highly ``realistic.''  Indeed, it is now possible to fit
many spiral galaxies, asymmetric tidal features, irregular galaxies, ring
galaxies, dust lanes, truncated galaxies, arcs, among others (though,
certainly, there are limitations).  However, it is important to realize that
``being possible'' often does not mean ``being necessary'' or ``being
practical.''  Necessity ought to be judged in the scientific context of
whether it is worth the extra effort to obtain diminishing returns.  For
instance, to measure total galaxy luminosity, it is often unnecessary to fit
high order Fourier modes or spiral rotations.  For many science studies
interested in global parameterization, often a single ellipsoid component
would suffice.  It is therefore important to always let the science determine
what kind of analysis is required, rather than to use the new capabilities in
the absence of a clearly defined goal.  Having provided some foregoing
disclaimers, some of the key scientific reasons motivating the new
capabilities are to:

\begin{itemize} 

    \item Quantify global asymmetry or substructure asymmetry.

    \item Quantify bending modes for weak-lensing applications, or fit
	  arcs in the image plane for strong gravitational lenses.

    \item Obtain more accurate substructure decomposition in the presence of
	  bars, spirals, rings, etc.

    \item Obtain more accurate global photometry.

    \item Quantify profile deviation in inner or outer regions of a galaxy, 
          such as disk truncations, deviations from a \sersic\ function, etc.

    \item Extract parametric information to the limits imposed by resolution,
	  signal-to-noise, and other small scale fluctuations.

    \item Quantify model-dependent errors in the decomposition.

\end{itemize}

We thus begin by giving an overview of the \galfit\ software
(\S~\ref{sect:overview}).  Then, we introduce the radial profile functions
that one can use (\S~\ref{sect:radial}), and illustrate how symmetric and
asymmetric shapes can be generated by modifying the coordinate system in
(\S~\ref{sect:azimuthal}).  Next, we introduce a new capability that allows
for radial profile truncation (\S~\ref{sect:truncation}).  Enabling all the
capabilities may result in extremely complex galaxy shapes, the interpretation
of which may give concerns to those new to the analysis.  Therefore, we
discuss the interpretations and model degeneracies of the parameters in
\S~\ref{interpretation}.  We then apply these new features to real galaxies in
\S~\ref{sect:examples}, followed by concluding remarks
(\S~\ref{sect:conclusion}).

\section {The 2-D FITTING PROGRAM \galfit}

\label{sect:overview}

This study builds on an existing algorithm named \galfit \footnote
{http://users.obs.carnegiescience.edu/peng/work/galfit/galfit.html}
\citep{peng02}, which is a 2-D parametric galaxy fitting algorithm, in the
same spirit as other widely used 2-D image-fitting algorithms (e.g.,
\citep[GIM2D:][]{simard98, simard02}; \citep[BUDDA:][]{desouza04}).  \galfit\
is a stand-alone program written in the C language, and can be run on most
modern operating systems.  To read and produce FITS images, \galfit\ calls
upon the CFITSIO package \citep{pence99}.  \galfit\ is designed to allow for
complex image decomposition tasks:  by allowing for an arbitrary number and
mix of parametric functions (\sersic, Moffat, Gaussian, exponential, Nuker,
etc.), it can simultaneously fit any number of galaxies and their
substructures.  It is possible to use \galfit\ for both interactive analysis
and galaxy surveys where complete automation is required.  However, automation
requires the use of an external ``wrapping'' algorithm written by the user
that takes care of both the pre-processing (object identification, initial
parameter estimation) and post-processing (extracting and tabulating fitting
parameters) of the fitting results.

\subsection {$\chi^2_\nu$ and Error Analysis}

\galfit\ is a non-linear least-squares fitting algorithm that uses the
Levenberg-Marquardt technique to find the optimum solution to a fit.  The
Levenberg-Marquardt algorithm is currently the most efficient one for
searching large parameter spaces, allowing for the possibility to fit complex
images with multiple components and a large number of parameters.  \galfit\
determines the goodness of fit by calculating $\chi^2$ and computing how to
adjust the parameters for the next step.  It continues to iterate until the
$\chi^2$ no longer decreases appreciably.  The indicator of goodness of fit is
the {\it normalized} or reduced $\chi^2$, $\chi^2_\nu$:

\begin{equation}
\label{chi2}
\chi^2_\nu = \frac{1} {N_{\rm dof}}\sum_{x=1}^{nx}\sum_{y=1}^{ny} \frac {\left(f_{\rm data}(x,y) - f_{\rm model}(x,y)\right)^2} {{\sigma(x,y)}^2},
\end{equation}

\noindent where

\begin{equation}
f_{\rm model} (x,y) = \sum_{\nu=1}^{m} f_{\nu}(x, y; \alpha_1...\alpha_n).
\end{equation}

\noindent $N_{\rm dof}$ is the number of degrees of freedom in the fit; $nx$
and $ny$ are the $x$ and $y$ image dimensions; and $f_{\rm data}(x,y)$ is the
image flux at pixel $(x,y)$.  The $f_{\rm model}(x,y)$ is the sum of $m$
functions of $f_{\nu}(x, y; \alpha_1...\alpha_n)$, with $n$ free parameters
$(\alpha_1...\alpha_n)$ in the 2-D model.  The uncertainty as a function of
pixel position, $\sigma(x,y)$, is the Poisson error at each pixel, which can
be provided as an input image.  If no $\sigma$-image is given, one is
generated based on the gain and read-noise parameters contained in the image
header.  Pixels in the image marked as being bad do not enter into the
calculation of $\chi^2$.

In the Levenberg-Marquardt algorithm, the minimization process involves
computing a Hessian matrix, which is closely related to the covariance matrix
of the parameters \citep[e.g., see][]{press92}.  The covariance matrix is then
directly related to the formal uncertainty in the fitting parameters that
\galfit\ reports.  However, the usefulness of the formal uncertainty is
limited to ideal situations where the fluctuations in the residual image are
only due to Poisson noise after removing the model.  This situation is mostly
realized in idealized situations, such as image simulations.  In real images,
the residuals are due to structures like stars and galaxies that are not
fitted, flat-fielding errors, and imperfect functional match to the data.
These factors cause formal uncertainties reported in numerical fits to be only
lower estimates.  In image fitting, more realistic uncertainties are
necessarily obtained by other processes, such as comparing fit results based
on different assumptions about the model rather than through a formal
covariance matrix.

In summary, the three images \galfit\ takes as input to calculate least
squares are the data, a $\sigma$-image, and an optional bad pixel mask.  To
account for image smearing by the PSF, \galfit\ will also require a PSF image.

\subsection {Accounting for Telescope Optics and Atmospheric Seeing}

The wavefront of light from distant sources is always perturbed by the act of
producing an image, distortions due to imperfect optics, and sometimes by the
Earth's atmosphere, resulting in some blurring.  To accurately compare the
intrinsic shape of an object with a model, image blurring must be taken into
account.  In image fitting this is often done by convolving a model image with
the input PSF before comparing with the data.  The process of performing
convolution is mathematically rigorous, but the actual implementation has
several subtleties.

\begin{figure*}
\footnotesize
\begin{verbatim}
# IMAGE and GALFIT CONTROL PARAMETERS
A) gal.fits            # Input data image (FITS file)
B) imgblock.fits       # Output data image block
C) none                # Sigma image name (made from data if blank or "none") 
D) psf.fits   #        # Input PSF image and (optional) diffusion kernel
E) 1                   # PSF fine sampling factor relative to data 
F) none                # Bad pixel mask (FITS image or ASCII coord list)
G) none                # File with parameter constraints (ASCII file) 
H) 1    200  1    100  # Image region to fit (xmin xmax ymin ymax)
I) 100    100          # Size of the convolution box (x y)
J) 20.000              # Magnitude photometric zeropoint 
K) 1.000  1.000        # Plate scale (dx dy)   [arcsec per pixel]
O) both                # Display type (regular, curses, both)
P) 0                   # Choose: 0=optimize, 1=model, 2=imgblock, 3=subcomps

# INITIAL FITTING PARAMETERS
#
#   For component type, the allowed functions are: 
#       sersic, expdisk, edgedisk, devauc, king, nuker, psf, 
#       gaussian, moffat, ferrer, and sky. 
#  
#   Hidden parameters will only appear when they're specified:
#       Bn (n=integer, Bending Modes).
#       C0 (diskiness/boxiness), 
#       Fn (n=integer, Azimuthal Fourier Modes).
#       R0-R10 (coordinate rotation, for creating spiral structures).
#       To, Ti, T0-T10 (truncation function).
# 
# ------------------------------------------------------------------------------
#   par)    par value(s)    fit toggle(s)    # parameter description 
# ------------------------------------------------------------------------------

# Component number: 1
 0) sersic3 /              #  Component type
 1) 50.0000  50.0000  1 1  #  Position x, y
 3) 15.0000     1          #  Surface brghtnss @ outer R_break [mag/arcsec^2]
 4) 30.0000     1          #  R_e (effective radius)   [pix]
 5) 4.0000      1          #  Sersic index n (de Vaucouleurs n=4) 
 9) 0.7000      1          #  Axis ratio (b/a)  
10) -30.0000    1          #  Position angle (PA) [deg: Up=0, Left=90]
Ti)  2                     #  Inner truncation by component number(s)
F5) 0.1500   20.0000  1 1  #  Azim. Fourier mode 5, amplitude, & phase angle

# Component number: 2
T0) radial                 #  Truncation type (radial, length, height)
T1) 45.0000  45.0000  1 1  #  Position x, y 
T4) 8.0000      1          #  Break radius (99% normal flux)    [pixels]
T5) 5.0000      1          #  Softening length (1% normal flux) [pixels]
T9) 0.7000      1          #  Axis ratio (optional)
T10) 45.0000    1          #  Position angle (optional) [deg: Up=0, Left=90]
F1) 0.6000   20.0000  1 1  #  Azim. Fourier mode 1, amplitude, & phase angle
B2) -5.000e+00  1          #  Bending mode 2 amplitude

# Component number: 3
 0) sersic                 #  Component type
 1) 150.0000 50.0000  1 1  #  Position x, y
 3) 7.0000      1          #  Integrated magnitude 
 4) 15.0000     1          #  R_e (effective radius)   [pix]
 5) 2.0000      1          #  Sersic index n (de Vaucouleurs n=4) 
 9) 0.5000      1          #  Axis ratio (b/a)  
10) 0.0000      1          #  Position angle (PA) [deg: Up=0, Left=90]

R0) power                  #  PA rotation func. (power, log, none)
R1) 0.0000      1          #  Spiral inner radius [pixels]
R2) 15.0000     1          #  Spiral outer radius [pixels]
R3) 180.0000    1          #  Cumul. rotation out to outer radius [degrees]
R4) 0.3000      1          #  Asymptotic spiral powerlaw 
R9) 10.0000     1          #  Inclination to L.o.S. [degrees]
R10) 45.0000    1          #  Sky position angle
F1) 0.3000   45.0000  1 1  #  Azim. Fourier mode 1, amplitude, & phase angle
F5) 0.1000   90.0000  1 1  #  Azim. Fourier mode 5, amplitude, & phase angle

\end{verbatim}

\figcaption[] {Example of an input file.  The object list is dynamic and can
be extended as needed.  Each model is modified by a mix of higher order
Fourier modes, bending modes, truncation, or spiral structure.  These
parameters produce the models shown in Figure~\ref{fig:example}.
\label{fig:menu}}
\normalsize
\end{figure*}

One consideration is the computation speed, as the process of convolving a
model is frequently the most time consuming part of parametric fitting.  The
trade-off is that the smaller the convolution region the faster the
computation time, but also the less accurate.  To achieve a compromise,
\galfit\ allows the user to decide on the size of the convolution region.  This
gives the flexibility for one to hone in on a solution quickly before trying
to obtain higher accuracy in the final step.

Another important issue to consider is whether to convolve each component
separately or all of them together just once in the final image.  This is an
important consideration because even though the model functions are analytic,
they are resampled by discrete pixel grids, resulting in a ``pixellated''
profile instead of one that is infinitely smooth.  If an intrinsic model is
sufficiently sharp, the curvature may not be critically subsampled by the
pixels prior to convolution, regardless of whether the recorded data are
Nyquist sampled.  The resulting profile after image convolution therefore can
depend very sensitively on how the model is centered on a pixel.  If such a
model is created off-center, pixellation effectively broadens out the model
ever so slightly more than normal once convolution is applied, but the effect
is noticeable in high-contrast imaging studies.  Therefore, the better way to
deal with ``pixellation broadening'' is to convolve each model component
individually rather than the entire image at once.  To do so, \galfit\ creates
every model on a pixel center; the pixel fluxes near the center of the models
are integrated over the pixel area adaptively.  Then to effect an off-centered
model, \galfit\ makes use of the convolution theorem by shifting the PSF by
the required amount before convolving it with the model.  This process
circumvents the problem of artificial pixellation broadening because whereas
the model core region may not be sufficiently resolved, the PSF ought to
be\footnote{If the PSF is not resolved then the convolution process will not
be accurate regardless of the technique.}.

\begin{figure}
    \epsscale {1.2}
    \plotone {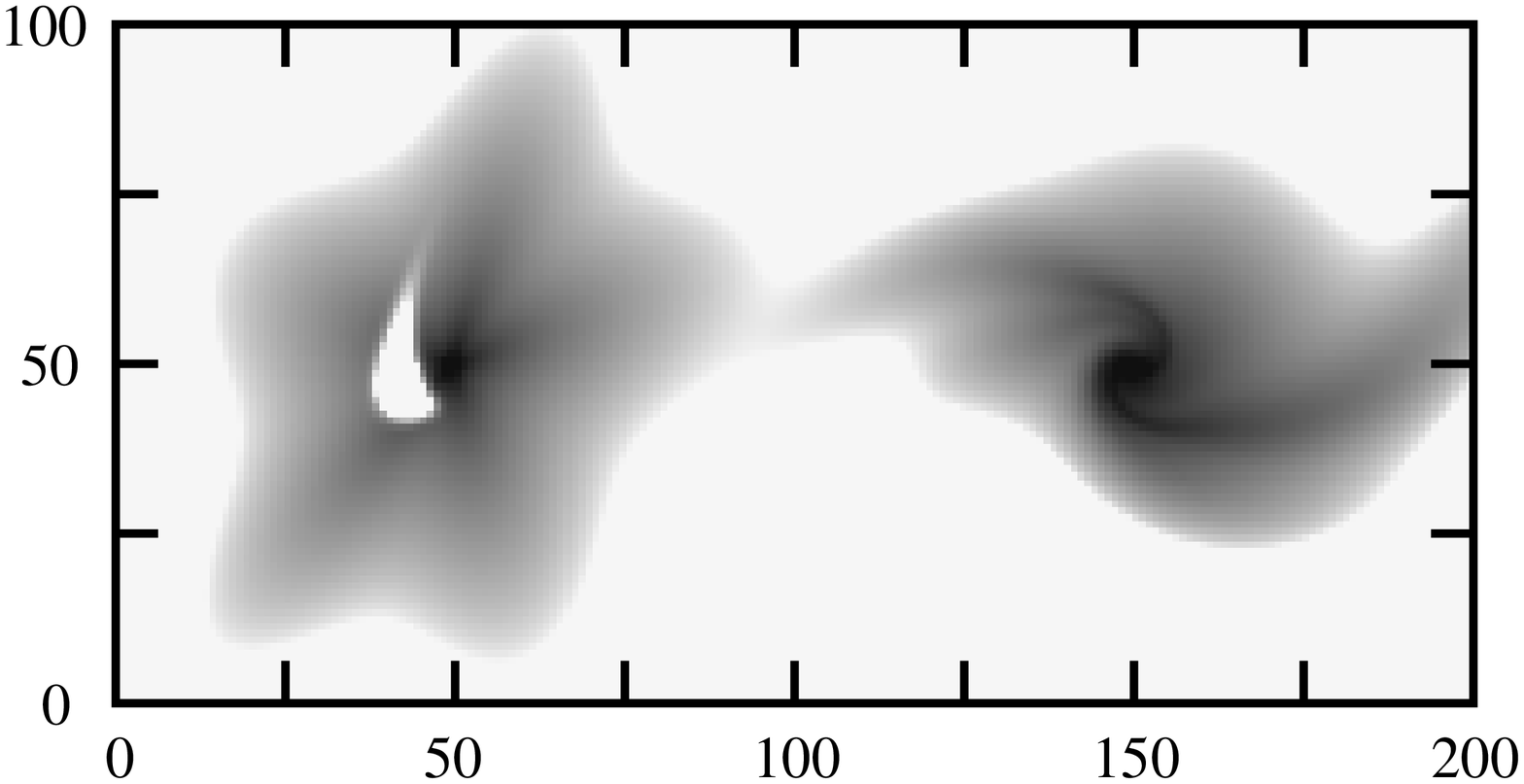}

    \figcaption{Shapes produced by parameters in the \galfit\ input file of
    Figure~\ref{fig:menu}. {\it Left}: a \sersic\ light profile modified by a
    single Fourier mode $m=5$, creating the star shape.  It is truncated in
    the inner region by a truncation function, which is modified by a bending
    mode $m=2$, with a lopsided Fourier mode of $m=1$.  {\it Right}: a
    \sersic\ light profile with Fourier modes $m=1$ and $m=5$ is modified by a
    coordinate rotation function to create a lopsided, multi-armed, spiral
    structure.  \label{fig:example}}

\end{figure}

Shifting the PSF, however, can be quite problematic when it is marginally
Nyquist sampled, or if the diffraction patterns are not critically sampled.
Accurate shifting of the PSF is of basic importance in high contrast imaging
studies.  For instance, in the case of studying active galaxies with a strong
central point source, issues of contrast, resolution, and sampling all
conspire to make the PSF fitting crucial to deriving a reliable host model.
In such situations, the standard interpolation techniques (e.g., linear or
spline) tend to broaden out the PSF core, so they are only accurate in the
extended outskirts where the gradient is shallow.  One alternative is to
interpolate using the sinc kernel, which is theoretically the perfect
interpolation kernel for critically sampled images, and preserves the
intrinsic width of the data.  However, significant ``ringing'' appears around
sharp features (i.e. PSF or galaxy core).  This effect can be nearly as bad on
a fit as pixellation broadening.  An improved solution is to taper the wing of
the sinc kernel using a windowing function (e.g., Lanczos), but the ringing
often may still be quite large beyond the PSF core, which must be further
suppressed.

\galfit\ seeks a compromise by using a hybrid scheme where the interpolation
in the PSF core is done by using a sinc kernel with a Kaiser window function
so as to faithfully preserve the width, but a bicubic spline interpolation is
used in the wings.  The result of this scheme is that for a Gaussian having a
full width at half-maximum (FWHM) of 2 pixels, the interpolation is accurate
to 0.1\% in the center, and 0.03\% at the distance of the FWHM relative to the
peak (or 1\% relative to the local flux).  For oversampled PSFs, the
interpolation is even more accurate.  Compared to bicubic spline
interpolation, our scheme is about 20 times more accurate.  From a more
practical standpoint, the mismatch in the PSFs between data taken using the
{\it Hubble Space Telescope (HST)} imaging cameras and synchronously observed
PSFs is rarely better than 3\% in the core.  For all practical purposes, our
interpolation scheme therefore will more than suffice for the most demanding
high contrast studies of quasar host galaxies at high redshift.

When the data are undersampled, convolution of the model can still be done
correctly if the convolution PSF provided to \galfit\ is either critically
sampled or oversampled.  In this situation, \galfit\ will generate a model on
a finer grid, convolve it with the PSF, then bin the result down to the
resolution of the data for comparison.  One way for users to obtain an
oversampled PSF compared to the data is to dither the PSF observations by
fractional pixels.  Another way is to numerically reconstruct a more
oversampled PSF star by extracting multiple stars from the data image itself
\citep[e.g., via DAOphot,][]{stetson87}.

However, lastly, we note that when the data and the convolution PSF are {\it
both} undersampled (i.e.  with PSF FWHM $<2$ pixels), convolution cannot be
done accurately. In such a situation, for the purpose of image fitting, it is
often better to broaden out the data and the PSF to critical sampling than to
perform the analysis in the original resolution \citep{kim08a}.  

\subsection {The Concept of a Model Component}

\label{subsect:concept}

Using the new features, each model can take on a shape that is completely
unrecognizable from a traditional ellipsoid shape.  It is therefore necessary
to clarify what constitutes a single model component.  In \galfit, {\it each
model component} is referred to by the name of the {\it surface brightness
profile}, just as it is standard practice to call something a
\sersic, Gaussian or exponential component in traditional models.  As implied
by this notion, {\it no matter how complex the shape}, the flux declines
monotonically (unless modified by a truncation function,
Section~\ref{sect:truncation}) from a peak in every radial direction in a
non-rotating frame, or along an arc in a rotating frame, strictly following
the functional form specified by the user.  The radial profile parameters are
mathematically decoupled from the azimuthal shape because the radial profile
functions are self-similar in the expression of the radius parameter, i.e.
with powers of ($r/r_{e}$), whereas the complex azimuthal shapes are obtained
by simply stretching the coordinate metric into more exotic grids than the
standard Cartesian grid.  This idea is in fact implicit in all 2-D
image-fitting algorithms, where the axis ratio parameter, $q$, turns a
circular profile into an ellipse by compressing the coordinate axis along one
direction, even though the functional form of the profile remains the same in
every direction.  In the same manner, the definition of a scale or effective
radius in a component, no matter how complex the shape, corresponds closely to
that of the best-fitting ellipse in the direction of the semi-major axis.

Figure~\ref{fig:menu} demonstrates how \sersic\ profiles can be modified by
bending modes, Fourier modes, and a spiral rotation function in
\galfit---the results of which are shown in Figure~\ref{fig:example}.  In the
example, there are only two \sersic\ model components, despite the appearance
of numerous parameters: both the ``radial'' and ``power'' functions are
modifications to the \sersic\ profiles.  Furthermore, the Fourier and bending
modes can modify the \sersic\ profiles, or modify the modifiers to the light
profiles.  Each radial surface brightness profile has a single peak and the
flux decline is monotonic radially (unless truncated by a truncation function
called ``radial'' in Figure~\ref{fig:menu}) or in a rotating coordinate system
(called ``power'' in Figure~\ref{fig:menu}).  Therefore, for each component,
it is still meaningful to talk about, for example, an {\it ``average'' light
profile}\ (e.g., S\'ersic), with {\it an average S\'ersic concentration index
$n$}---no matter what the galaxy may look like azimuthally.  In this manner
even irregular galaxies can be parameterized in terms of their average light
profile.  When the average peak of an irregular galaxy is not located at the
geometric center, it has a high-amplitude $m=1$ Fourier mode (i.e.
lopsidedness).

In such a way, no matter how complex the azimuthal shape, interpreting the
surface brightness profile parameters is just as straightforward as the
traditional ellipsoid.

\section {THE RADIAL PROFILE FUNCTIONS}

\label{sect:radial}

The radial profile functions describe the intensity fall-off of a model away
from the peak, such as the S\'ersic, Nuker, or exponential models, among others.
For example, early-type galaxies typically have steep radial profiles whereas
late-type galaxies have shallower intensity slope near the center.  The rate of
decline is governed by a scale-length parameter.  The radial profile is often
of primary interest in galaxy studies from the standpoint of classification,
and because the exact functional form may have some bearing on the path of
galaxy evolution.  In \galfit\ the radial profile can have the following
functional forms, which are some of the most frequently seen in literature.

\begin{figure}
    \epsscale {1.2}
    \plotone {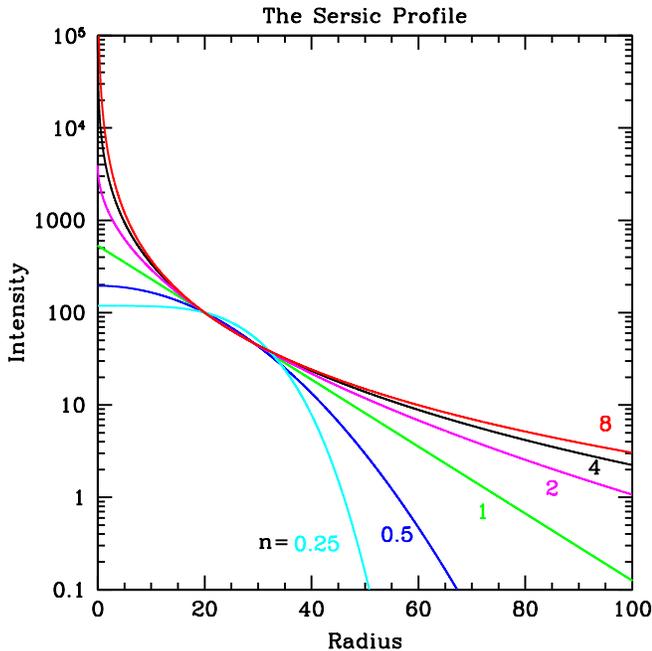}

    \figcaption{The \sersic\ profile, where $r_e$ and $\Sigma_e$ are held
    fixed.  Notice that the larger the \sersic\ index value $n$, the steeper
    the central core, and more extended the outer wing.  A low $n$ has a
    flatter core and a more sharply truncated wing.  Large \sersic\ index
    components are very sensitive to uncertainties in the sky background level
    determination because of the extended wings.  \label{fig:sersic}}

\end{figure}

\bigskip

\noindent {\bf The S\'ersic Profile}\ \ \ \ \ The S\'ersic power law is one of
the most frequently used to study galaxy morphology, and has the following
functional form:

\begin {equation}
\label{eqn:sersic}
\Sigma(r)=\Sigma_e \exp{\left[-\kappa \left(\left({\frac{r}{r_e}}\right)^{1/n} - 1\right)\right]}.
\end {equation}

\noindent $\Sigma_e$ is the pixel surface brightness at the effective radius
$r_e$.  The parameter $n$ is often referred to as the concentration parameter.
When $n$ is large, it has a steep inner profile and a highly extended outer
wing.  Inversely, when $n$ is small, it has a shallow inner profile and a
steep truncation at large radius.  The parameter $r_e$ is known as the
effective radius such that half of the total flux is within $r_e$.  To make
this definition true, the dependent variable $\kappa$ is coupled to $n$; thus,
it is not a free parameter.  The classic de~Vaucouleurs profile that describes
a number of galaxy bulges is a special case of the S\'ersic profile when $n$ =
4 (corresponding to $\kappa = 7.67$).  As explained below, both the
exponential and Gaussian functions are also special cases of the S\'ersic
function when $n=1$ and $n=0.5$, respectively.  As such the S\'ersic profile
is a common favorite when fitting a single component.

The flux integrated out to $r=\infty$ for a S\'ersic profile is:

\begin{equation} 
F_{\rm tot} = 2\pi r_e^2 \Sigma_e {\mbox e}^\kappa n \kappa^{-2n}\Gamma(2n) q/R(C_0; m).
\end{equation}

\noindent The term $R(C_0; m_i)$ is a geometric correction factor when the
azimuthal shape deviates from a perfect ellipse.  As the concept of azimuthal
shapes will be discussed in Section~\ref{sect:azimuthal}, we will only comment
here that $R(C_0; m_i)$ is simply the ratio of the {\it area} between a
perfect ellipse with the area of the more general shape, having the same axis
ratio $q$ and unit radius.  The shape can be modified by Fourier modes ($m$
being the mode number) or diskiness/boxiness.  For instance, when the shape is
modified by diskiness/boxiness, $R(C_0)$ has an analytic solution given by:

\begin{equation}
R(C_0) = {\frac{\pi (C_0+2)} {4 \beta (1/(C_0+2), 1+1/(C_0+2))}},
\end{equation}

\noindent where $\beta$ is the Beta function.  In general, when the Fourier
modes are used to modify an ellipsoid shape, there is no analytic solution for
$R(m_i)$, and so the area ratio must be integrated numerically.

In \galfit, the flux parameter that one can use for the \sersic\ function is
either the integrated magnitude $m_{\rm tot}$ or some kind of surface
brightness magnitude, for example at the center ($\mu_0$), at the effective
radius ($\mu_e$), or at the break radius ($\mu_{\rm break}$) for truncated
profiles (see Section~\ref{sect:truncation}).  The integrated magnitude
follows the standard definition:

\begin {equation}
m_{\rm tot} = -2.5  {\mbox {log}_{10}} \left( \frac{F_{\rm tot}}{t_{\rm exp}}\right) + {\mbox {mag zpt}},
\end {equation}

\noindent where $t_{\rm exp}$ is the exposure time from the image header.
Each \sersic\ function can thus potentially have 7 classical free parameters
in the fit:  $x_0$, $y_0$, $m_{\rm tot}$, $r_e$, $n$, $q$, and $\theta_{\rm
PA}$.  The non-classical parameters, $C_0$, Fourier modes, bending modes, and
coordinate rotation may be added as needed.  There is no restriction on the
number of Fourier modes, and bending modes, but each S\'ersic component can
only have a single set of $C_0$ and coordinate rotation parameters (see
Section~\ref{sect:azimuthal} for details).

\bigskip

\noindent{\bf The Exponential Disk Profile}\ \ \ \ \ The exponential profile
has some historical significance, so \galfit\ is explicit about calling this
profile an {\it exponential disk}, even though an object that has an
exponential profile need not be a classical disk.  Historically, an
exponential disk has a scale length $r_s$, which is not to be confused with
the effective radius $r_e$ used in the S\'ersic profile.  For situations where
one is not trying to fit a classical disk it would be less confusing
nomenclature-wise to use the S\'ersic function with $n=1$, and quote the
effective radius $r_e$.  But because the exponential disk profile is a special
case of the S\'ersic function for $n=1$ (see Figure~\ref{fig:sersic}),
there is a relationship between $r_e$ and $r_s$, given by

\begin {equation}
r_e = 1.678 r_s  \mbox{\ \ \ \ \ \ \ \ \ \ \ \ (For $n=1$ only).}
\end {equation}

\noindent The functional form of the exponential profile is

\begin {equation}
\label{eqn:expdisk}
\Sigma(r)=\Sigma_0 \exp{\left(-\frac{r}{r_s}\right)}, 
\end{equation}

\noindent
and the total flux is given by

\begin{equation}
F_{\rm tot} = 2\pi r_s^2 \Sigma_0 q/R(C_0; m).
\end {equation}

\noindent The 6 free parameters of the profile are:  $x_0$, $y_0$, $m_{\rm
tot}$, $r_s$, $\theta_{\rm PA}$, and $q$.

\bigskip

\noindent{\bf The Gaussian Profile}\ \ \ \ \ The Gaussian profile is another
special case of the S\'ersic function with $n=0.5$ (see
Figure~\ref{fig:sersic}), but here the size parameter is the FWHM instead of
$r_e$.  The functional form is

\begin{equation}
\Sigma(r) = \Sigma_0 \exp{\left(\frac{-r^2} {2 \sigma^2}\right)},
\end{equation}

\noindent
and the total flux is given by

\begin{equation}
F_{\rm tot} = 2\pi\sigma^2\Sigma_0 q/R(C_0; m),
\end {equation}

\noindent where FWHM = 2.354$\sigma$.  The 6 free parameters of the profile
are:  $x_0$, $y_0$, $m_{\rm tot}$, FWHM, $q$, and $\theta_{\rm PA}$.

\bigskip

\begin{figure}
    \epsscale {1.2}
    \plotone {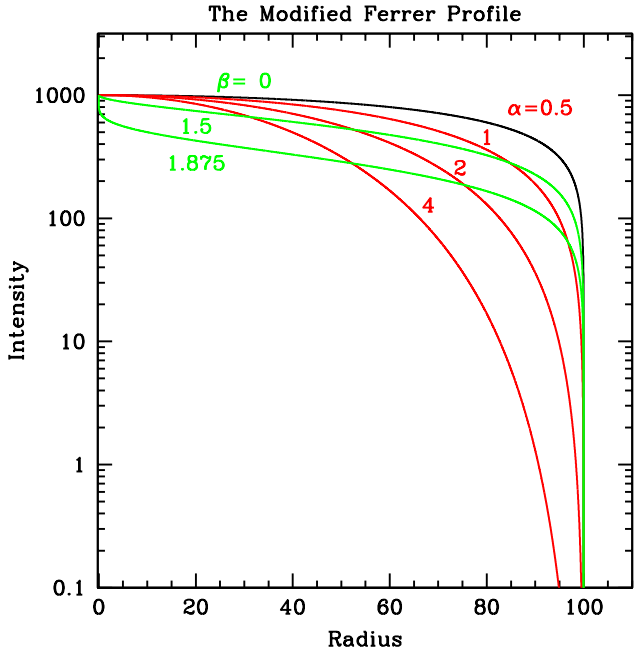}

    \figcaption{The modified Ferrer profile.  The black reference curve has
    parameters $r_{\rm out} = 100$, $\alpha=0.5$, $\beta=2$, and $\Sigma_0 =
    1000$.  The red curves differ from the reference only in the $\alpha$
    parameter, as indicated by the red numbers.  Likewise, the green curves
    differ from the reference only in the $\beta$ parameter, as indicated by
    the green numbers. \label{fig:ferrer}}

\end{figure}

\begin{figure}
    \epsscale {1.2}
    \plotone {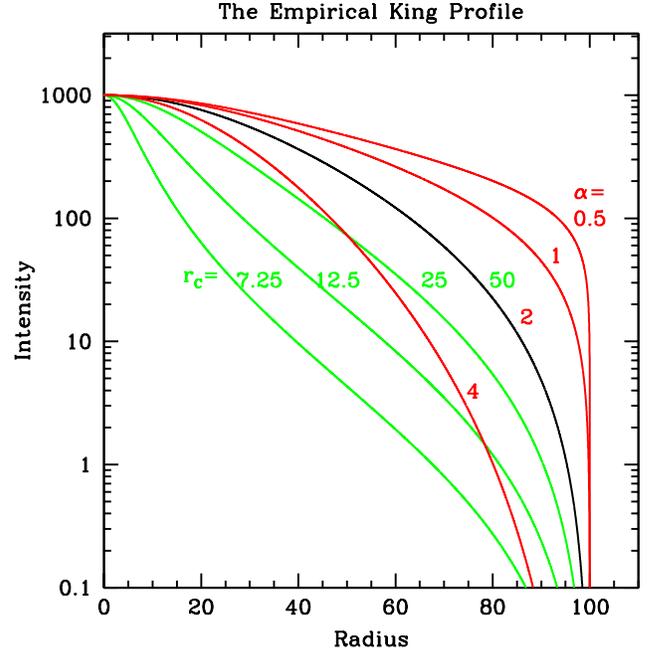}

    \figcaption{The empirical King profile.  The black reference curve has
    parameters $r_c = 50$, $r_t = 100$, $\alpha=2$, and $\Sigma_0 = 1000$.  The
    red curves differ from the reference curve only in the $\alpha$ parameter,
    as indicated by the red numbers.  Likewise, the green curves differ from
    the reference only in the $r_c$ parameter, as indicated by the green
    numbers. \label{fig:king}}

\end{figure}

\begin{figure}
    \epsscale {1.2}
    \plotone {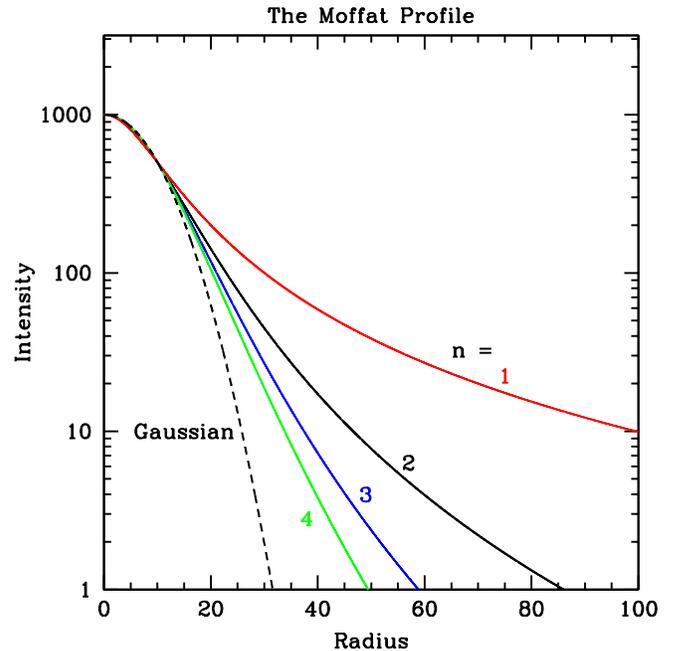}

    \figcaption{The Moffat profile.  The black reference curve has parameters
    $n = 2$, FWHM = 20, and $\Sigma_0 = 1000$.  The other colored lines differ
    only in the concentration index $n$, as shown by the numbers.  The dashed
    line shows a Gaussian profile of the same FWHM. \label{fig:moffat}}

\end{figure}

\begin{figure*}
    \epsscale {1.2}
    \plotone {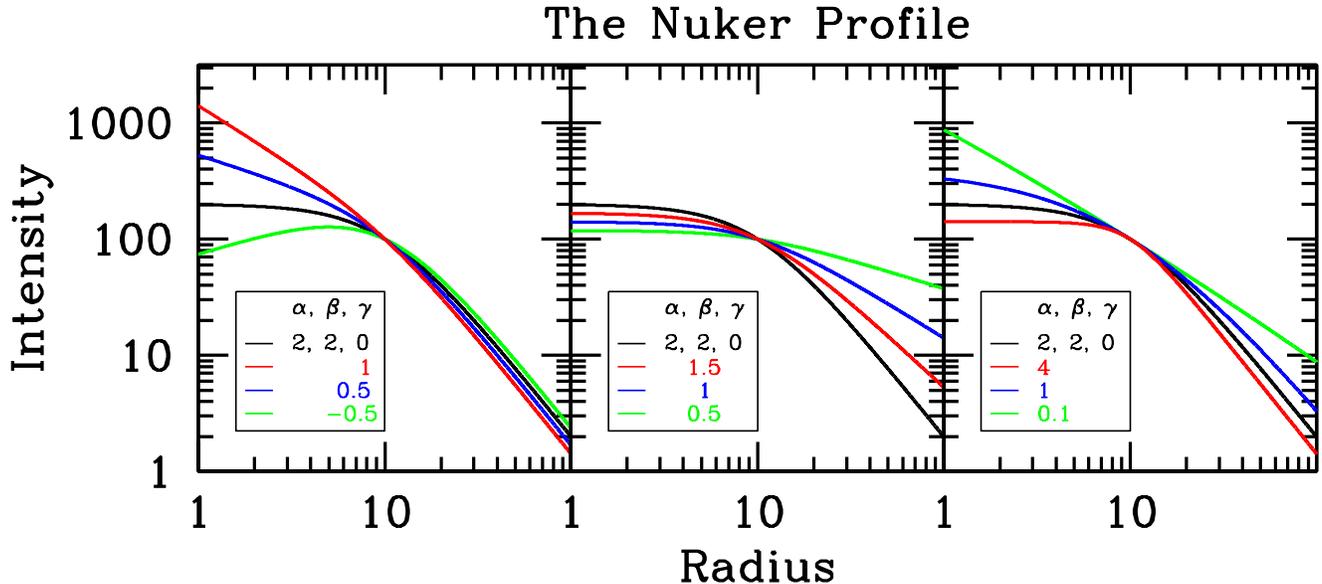}

    \figcaption{The Nuker profile.  The black reference curve has parameters
    $r_b = 10$, $\alpha=2$, $\beta=2$, $\gamma = 0$, and $I_b=100$.  For
    the other colored lines, only one value differs from the reference, as
    shown in the legend. \label{fig:nuker}}

\end{figure*}

\noindent{\bf The Modified Ferrer Profile}\ \ \ \ \ The Ferrer profile
\citep[Figure~\ref{fig:ferrer};][]{binney87} has a nearly flat core and an
outer truncation.  The sharpness of the truncation is governed by the
parameter $\alpha$, whereas the central slope is controlled by the parameter
$\beta$.  Because of the flat core and sharp truncation behavior, historically
it is often used to fit galaxy bars and ``lenses.''  The profile

\begin{equation}
\Sigma(r) = \Sigma_0 \left(1 - \left(r/r_{\rm out}\right)^{2-\beta} \right)^{\alpha}
\end{equation}

\noindent is only defined within $r \le r_{\rm out}$, beyond which the
function has a value of 0.  The 8 free parameters of the Ferrer profile are:
$x_0$, $y_0$, central surface brightness, $r_{\rm out}$, $\alpha$, $\beta$,
$q$, and $\theta_{\rm PA}$.

It is worth mentioning that a \sersic\ profile with low index $n<0.5$ has
similar profile shapes, thus it is often used instead of the Ferrer function.

\bigskip

\noindent{\bf The Empirical (Modified) King Profile}\ \ \ \ \ The empirical
King profile (Figure~\ref{fig:king}) is often used to fit the light profile of
globular clusters.  It has the following form \citep{elson99}:

\begin{eqnarray}
\Sigma(r) & = & \Sigma_0 \left[1 - \frac{1}{(1+(r_t/r_c)^2)^{1/\alpha}}\right]^{-\alpha} \times \notag \\
     & & \left[\frac{1}{(1+(r/r_c)^2)^{1/\alpha}} - \frac{1}{(1+(r_t/r_c)^2)^{1/\alpha}}\right]^\alpha.
\end{eqnarray}

\noindent The standard empirical King profile has a power law with index
$\alpha= 2$.  In \galfit, $\alpha$ can be a free parameter.  In this model,
the flux parameter to fit is the central surface brightness, $\mu_0$,
expressed in mag~arcsec$^{-2}$ (see Equation~\ref{eqn:mu0}).  The other free
parameters are the core radius ($r_c$) and the truncation radius ($r_t$), in
addition to the geometrical parameters.  Outside the truncation radius, the
function is set to 0.  Thus, the total number of classical free parameters is
8:  $x_0$, $y_0$, $\mu_0$, $r_c$, $r_t$, $\alpha$, $q$, and $\theta_{\rm
PA}$.

\bigskip

\noindent{\bf The Moffat Profile}\ \ \ \ \ The profile of the {\it HST}\ WFPC2
PSF is well described by the Moffat function (Figure~\ref{fig:moffat}).  Other
than that, the Moffat function \citep{moffat69} is less frequently used than
the above functions for galaxy fitting.  The functional profile is

\begin{equation}
\Sigma(r) = {\frac{\Sigma_0} {\left[1+(r/r_d)^2\right]^n}},
\end{equation}

\noindent
and the total flux is given by

\begin {equation}
F_{\rm tot} = {\frac {\Sigma_0 \pi r_d^2 q} {(n-1) R(C_0; m)}}.
\end {equation}

\noindent In \galfit\ the size parameter to fit is the FWHM, where the
relation between $r_d$ and FWHM is

\begin{equation}
r_d = \frac {{\rm FWHM}}{ 2 \sqrt{2^{1/n} - 1} }.
\end{equation}

\noindent The 7 free parameters are:  $x_0$, $y_0$, $m_{\rm tot}$ (i.e. total
magnitude, instead of $\mu_0$) FWHM (instead of $r_d$), the concentration
index $n$, $q$, and $\theta_{\rm PA}$.

\bigskip

\noindent {\bf The Nuker Profile}\ \ \ \ \ The Nuker profile
(Figure~\ref{fig:nuker}) was introduced by \citet{lauer95} to fit the central
light distribution of nearby galaxies, and it has the following form:

\begin{equation}
\label{eqn:nuker}
I(r) = I_b \ 2^{\frac{\beta - \gamma} {\alpha}}
\left({\frac{r}{r_b}}\right)^{-\gamma}\left[{1+\left(\frac{r}
{r_b}\right)^{\alpha}} \right] ^{\frac{\gamma-\beta}{\alpha}}.
\end{equation}

\noindent The flux parameter to fit is $\mu_b$, the surface brightness of the
profile at $r_b$, which is defined as

\begin{equation}
	\mu_b = -2.5\ {\rm{log}_{10}} \left(\frac {I_b} {t_{\rm exp} \Delta x \Delta y}\right) + \rm{mag\ zpt},
\end{equation}

\noindent where $t_{\rm exp}$ is the exposure time from the image header, and
$\Delta x$ and $\Delta y$ are the platescale in arcsec.  The Nuker profile is
a double power law, where (in Equation~\ref{eqn:nuker}) $\beta$ is the outer
power law slope, $\gamma$ is the inner slope, and $\alpha$ controls the
sharpness of the transition.  The motivation for using this profile is that
the nuclei of many galaxies appear to be fit well in 1-D
\citep[see][]{lauer95} by a double power law.  However, caution should be
exercised when using this function because, for example, a low value of
$\alpha$ ($\alpha \lesssim 2$) can be mimicked by a combination of high
$\gamma$ and low $\beta$ (compare Figure~\ref{fig:nuker}{\it c} with the other
two panels), which presents a serious potential for degeneracy.  In all there
are there are 9 free parameters:  $x_0$, $y_0$, $\mu_b$, $r_b$, $\alpha$,
$\beta$, $\gamma$, $q$, and $\theta_{\rm PA}$.

\bigskip

\noindent{\bf The Edge-On Disk Profile}\ \ \ \ \ Both the \sersic\
(Equation~\ref{eqn:sersic}) and exponential disk profile
(Equation~\ref{eqn:expdisk}) are merely empirical descriptors of a galaxy
light profile.  However, for edge-on disk galaxies, there is a more physically
motivated light profile: under the assumption that the disk component is
locally isothermal and self-gravitating, the light profile distribution is
given by \citep[][]{vanderkruit81}:

\begin {equation}
    \Sigma(r,h)=\Sigma_0 \left(\frac{r}{r_s}\right) 
        K_1\left(\frac{r}{r_s}\right) \mbox{sech}^2\left(\frac{h}{h_s}\right),
\end{equation}

\noindent where $\Sigma_0$ is the pixel central surface brightness, $r_s$ is
the major-axis disk scale length, $h_s$ is the perpendicular disk scale
height, and $K_1$ is a Bessel function.  The flux parameter being fitted in
\galfit\ is the central surface brightness:

\begin{equation} 
\label{eqn:mu0}
\mu_0 = -2.5\ {\rm{log}_{10}} \left(\frac{\Sigma_0}{t_{\rm exp} \Delta x \Delta y}\right) + \rm{mag\ zpt}.
\end{equation}

Note that if the disk is oriented horizontally the coordinate $r$ is the
$x$-distance (as opposed to the radius) of a pixel from the origin.  There are
6 free parameters in the profile model:  $x_0$, $y_0$, $\mu_0$, $r_s$,
$h_s$, and $\theta_{\rm PA}$.

\bigskip

\noindent{\bf The PSF Profile} \ \ \ \ \ For unresolved sources, one can fit
pure stellar PSFs to an image (as opposed to functions with narrow FWHM
convolved with the PSF).  The PSF function is simply the convolution PSF image
that the user provides, so there is no prescribed analytical functional form.
This is also the only profile that is not convolved in \galfit.  The PSF has
only 3 free parameters: $x_0$, $y_0$, and $m_{\rm tot}$.  Because there is no
analytical form, the total magnitude is determined by integrating over the PSF
image and assuming that it contains 100\% of the light.  If the PSF wing is
vignetted, there will be a systematic offset between the flux \galfit\ reports
and the actual value.

If one wants to fit this ``function,'' it is important to make sure that the
input PSF is close to, or super-, Nyquist sampled.  The PSF interpolation used
in shifting is done by a sinc function with a Kaiser window, which can
preserve the widths of the PSF even under subpixel shifting.  This is in
principle better then spline interpolation or other high-order interpolants.
However, if the PSF is undersampled, aliasing will occur, and the PSF
interpolation will be poor.  In this situation, it is better to provide an
oversampled PSF to \galfit\ (and to specify the amount of oversampling), even
if the data are undersampled.  With {\it HST}\ data this can be done using
TinyTim \citep{krist97} or by combining stars. \galfit\ will take care of
rebinning during the fitting.

Note that the alternative to fitting a PSF is to fit a Gaussian with a small
width (e.g., 0.4--0.5 pixels), which \galfit\  will convolve with the PSF.  This is
generally not advisable if a source is a pure point source because convolving
a narrow function with the PSF will broaden out the overall profile, even if
slightly.  The convergence can also be poor if the FWHM parameter starts
becoming smaller than 0.5 pixels.  However, this technique can still be useful
to see if a source is truly resolved.

\bigskip

\noindent{\bf The Background Sky}\ \ \ \ \ The background sky is a flat plane
with flux gradient along $x$ and $y$ directions.  Thus it has a total of 3
free parameters.  The pivot point for the sky is {\it fixed} to the geometric
center $(x_c, y_c)$ of the image, calculated by $(n_{\rm pix} + 1)/2$, where
$n_{\rm pix}$ is the number of pixels along one dimension.  The tip and tilt
are calculated relative to that center.  Because the galaxy centroid located
at $(x,y)$ is in general not at the geometric center $(x_c, y_c)$ of the
image, the sky value directly beneath the galaxy centroid is calculated by:

\begin {equation} 
{\rm sky}(x,y) = {\rm sky} (x_c, y_c) + (x - x_c){{d\rm sky}\over{dx}} + (y -
    y_c) {{d\rm sky}\over{dy}}.
\end {equation}

\bigskip

\section {THE AZIMUTHAL SHAPE FUNCTIONS}

\label{sect:azimuthal}

Whereas the radial profile governs the decline of galaxy flux radially from a
central peak, the azimuthal functions generate the projected shape in the
$x-y$ plane of the image.  For instance, ellipsoidal, irregular, spiral,
disky, and boxy shapes are all created by azimuthal functions.  All
traditional 2-D image-fitting techniques use an ellipse as the fundamental
shape, which is obtained by stretching the coordinate grid along one dimension
compared to the orthogonal direction.  Indeed, all azimuthal functions are
coordinate transformations.  Therefore, to change a shape from an ellipse into
more exotic shapes, the coordinate system [$r(x,y)$] can be further stretched
or shrunk radially from the peak, as a function of azimuth angle.  This
coordinate transformation preserves the functional form of the surface
brightness profile in every direction because the profiles are self-similar---
that is, they are functions of $(r/r_{\rm scale})$.  Thus defined, the radial
profile parameters (e.g., $r_e$, $q$, central concentration, etc.) retain
their original meaning no matter the complexity of the azimuthal shape.

We introduce four new ways to modify the azimuthal shape of a model, beginning
with the traditional ellipsoidal model.  On top of an ellipsoid, this section
describes how one can add Fourier modes, bending modes, and coordinate
rotation functions (power law and logarithmic).  Each component can be
modified by any one or all of the azimuthal functions simultaneously,
depending on the complexity of the galaxy one is trying to analyze.  The next
section will cover truncation functions.

\bigskip

\begin{figure*}
    \centerline{\includegraphics[angle=0,height=8.cm,width=8.cm]{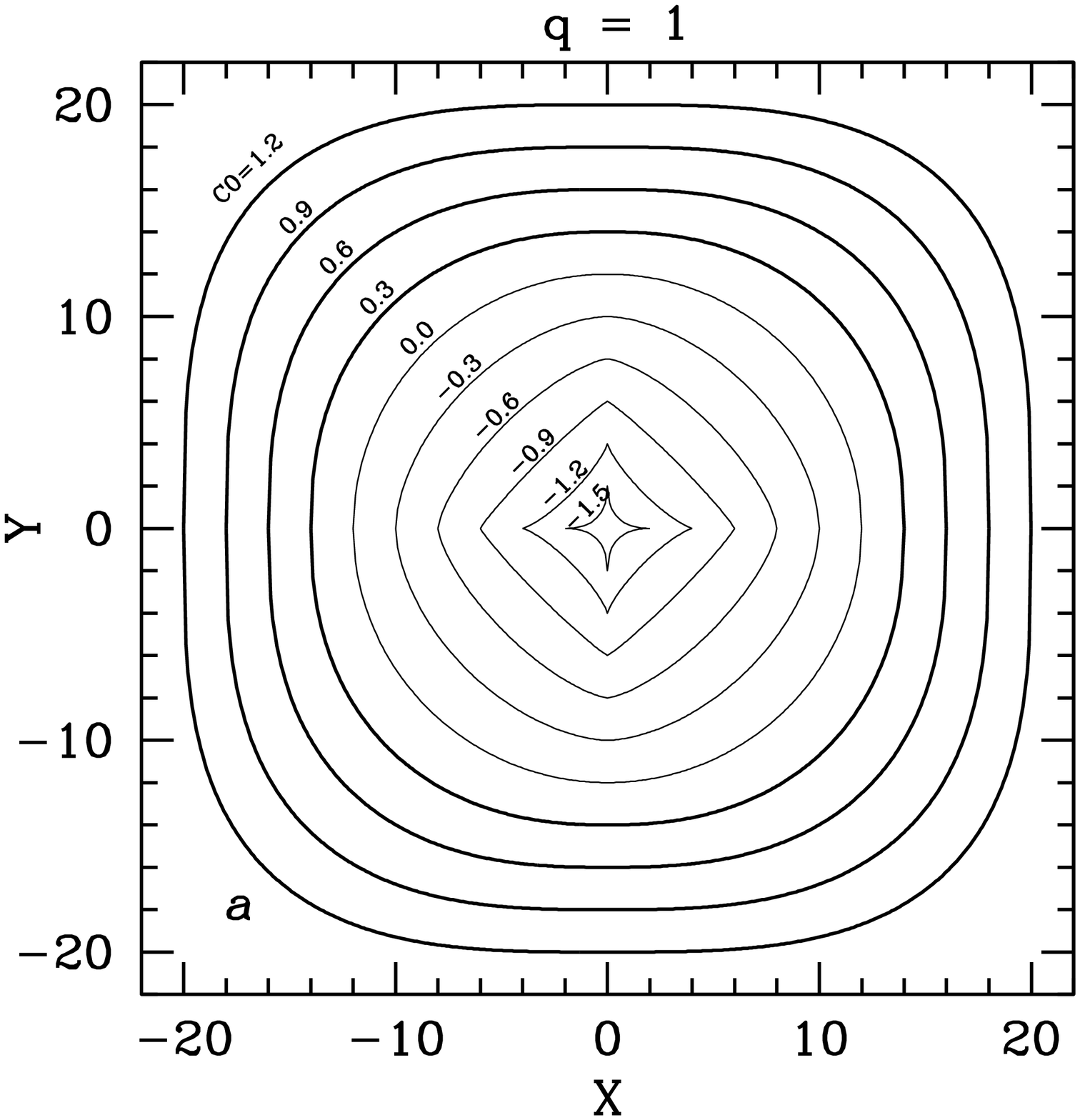}\hfill
                \includegraphics[angle=0,height=8.cm,width=8.cm]{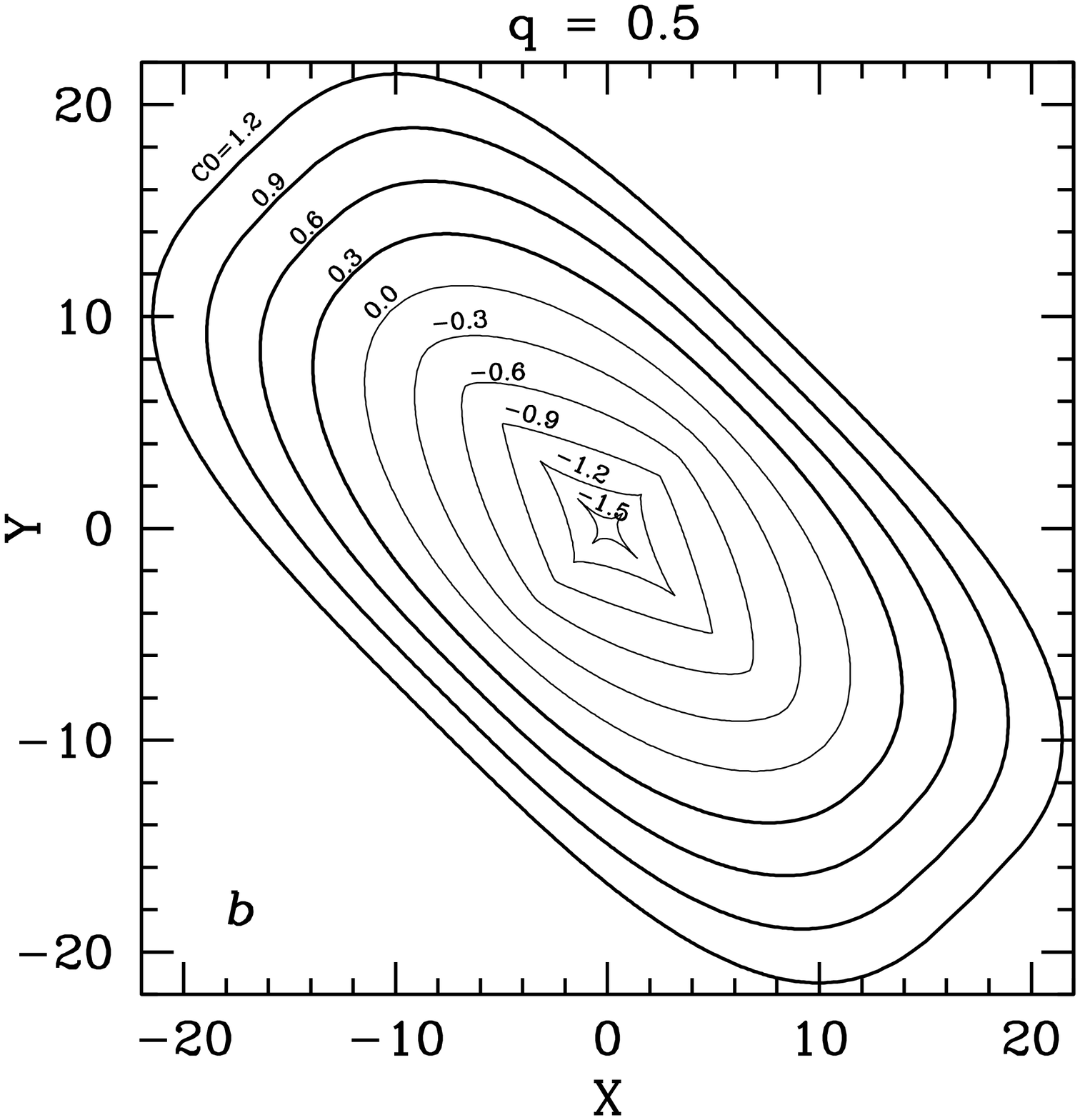}}

    \figcaption{Generalized ellipses with ({\it a}) axis ratio $q=1$ and ({\it
    b}) axis ratio $q=0.5$. Various values of the 
    diskiness/boxiness parameter $C_0$ are labeled. \label{fig:ellipse}}
\end{figure*}

\noindent {\bf Generalized Ellipses} \ \ \ \ \ The simplest azimuthal shape in
\galfit\ is the traditional generalized ellipse.  This is the starting point
for all \galfit\ analysis, no matter how complex is the final outcome.  The
radial coordinate of the generalized ellipse is defined by:

\begin {equation}
\label{eqn:ellipse}
r(x,y) = \left(\left|x-x_0\right|^{C_0+2}+\left| \frac{y-y_0}{q}\right|^{C_0+2}\right)^{\frac{1}{C_0+2}}.
\end{equation}

\noindent Here, the ellipse axes are aligned with the coordinate axes, and
($x_0, y_0$) is the centroid of the ellipse.  Defined by
\citet{athanassoula90}, the ellipse is called ``general'' in the sense that
$C_0$ is a free parameter, which controls the diskiness/boxiness of the
isophote.  When $C_0=0$ the isophotes are pure ellipses.  With decreasing
$C_0$ ($C_0 < 0$), the shape becomes more disky (diamond-like), and
conversely, more boxy (rectangular) as $C_0$ increases ($C_0>0$) (see
Figure~\ref{fig:ellipse}).  The major axis of the ellipse can be oriented to
any PA.  Thus, there are a total of 4 free parameters ($x_0,
y_0, q, \theta_{\rm{PA}}$) in the standard ellipse and an additional one,
$C_0$, for the generalized ellipse.

\bigskip

\noindent {\bf Fourier Modes} \ \ \ \ \ Few galaxies look like perfect
ellipsoids, and one can better refine the azimuthal shape by adding
perturbations in the form of Fourier modes.  The Fourier perturbation on a
perfect ellipsoid shape is defined in the following way:

\begin {equation} 
\label{eqn:fourier}
r(x,y) = r_0(x,y) \left (1 + \sum_{m=1}^{N} a_m \mbox { cos} \left(m(\theta+\phi_m) \right)\right).
\end {equation}

\begin{figure*}
    \plotone{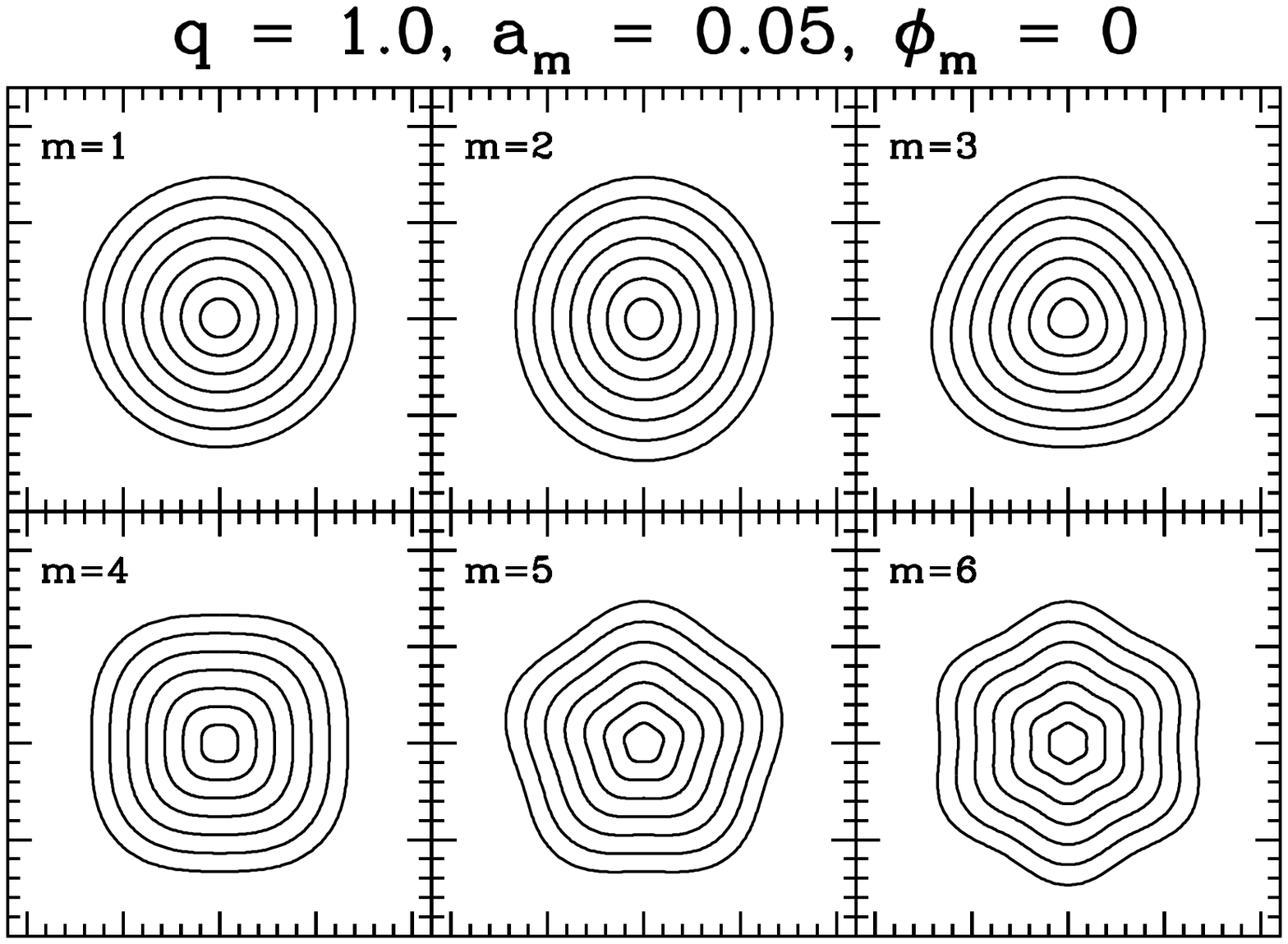}
    \plotone{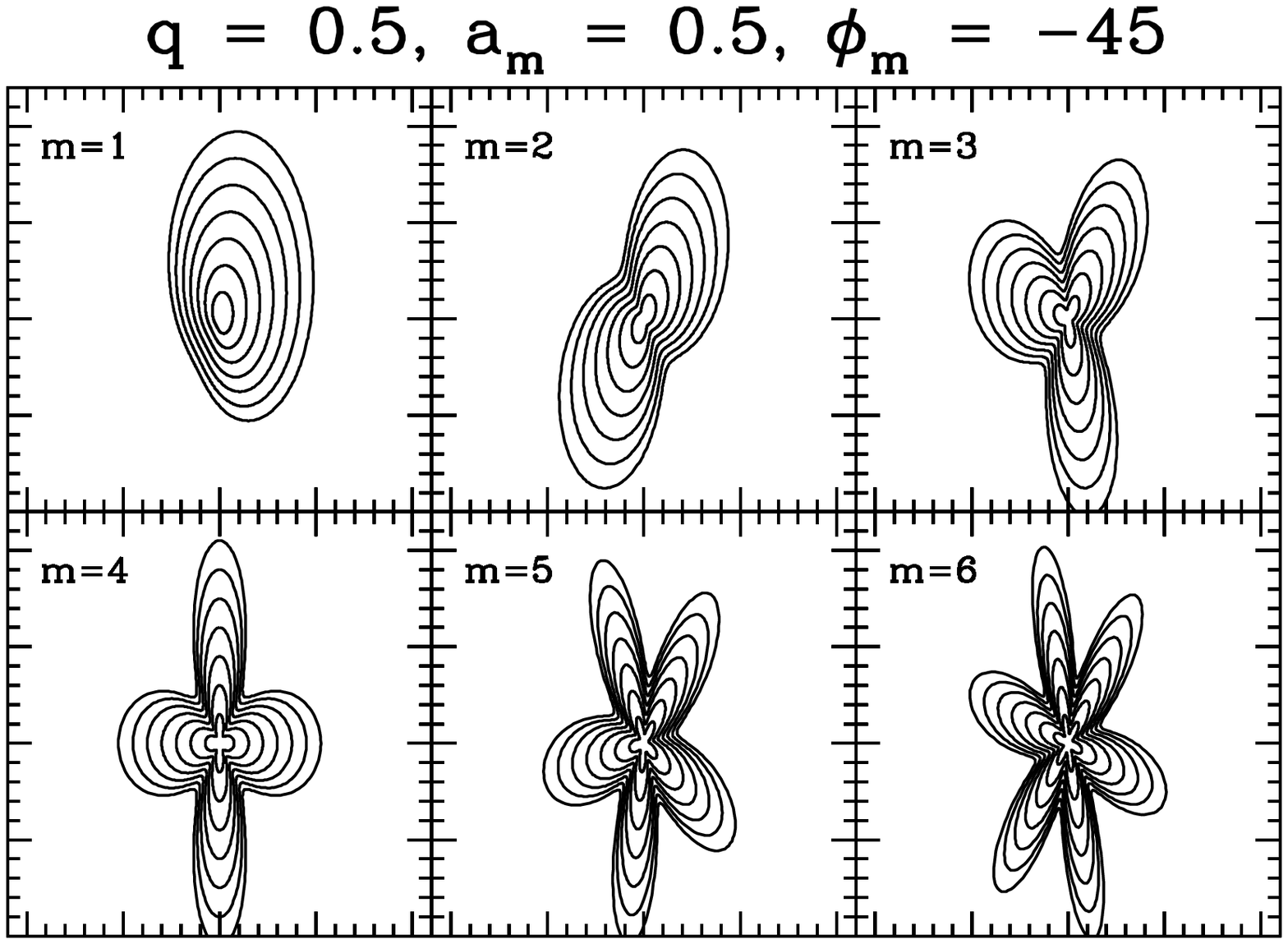}

    \figcaption{Examples of Fourier modes.  ({\it Top})  Low-amplitude ($a_m=0.05$)
    Fourier modes modifying a circular profile ($q=1.0$) with phase angle
    $\phi_m=0^\circ$.  ({\it Bottom})  High-amplitude ($a_m=0.5$)
    Fourier modes modifying an elliptical profile ($q=0.5$) with phase angle
    $\phi_m=-45^\circ$. \label{fig:fourier}}

\end{figure*}

\noindent In the absence of Fourier modes in the parenthesis, the $r_0(x,y)$
term is the radial coordinate for a traditional ellipse, and $\theta =$ arctan
$((y-y_0)/((x-x_0) q))$ defined in Equation~\ref{eqn:ellipse}.  The Fourier 
amplitude for mode $m$ is $a_m$.  Defined as such, $a_m$ is the fractional
deviation in radius from a generalized ellipse of Equation~\ref{eqn:ellipse}.  
The
number of modes $N$ is unrestricted, and any mode can be left out.  The ``phase
angle,'' $\phi_m$, is the relative alignment of mode $m$ relative to the
PA of the generalized ellipse; the phase angle is $0^\circ$ in the
direction of the semi-major axis of the generalized ellipse (rather than up),
increasing counter-clockwise.  Figure~\ref{fig:fourier} shows some examples of
how Fourier modes modify a circle and an ellipse into other shapes.

Each Fourier mode has 2 free parameters, $a_m$ and $\phi_m$, and the number of
modes the user can add is unrestricted.  However, the most useful modes are
low-order ones ($m=1, 3...6$).  We note that the $m=2$ mode is partially
degenerate with the classical axis ratio parameter, $q$, for an ellipse.
Therefore the use of $m=2$ and $q$, together, should be largely avoided except
in some situations (e.g., peanut-looking bulges).

The phase angles of the Fourier modes are also useful information to keep in
mind. Modes with the following phase angles have the following symmetry
properties:

\begin{itemize}

\item Symmetry about a central point:  $a_1=0$, regardless of other mode phase
     and amplitude.

\item For all modes $m$, there is reflection symmetry at:  $\phi_m = 0^\circ$,
    $\pm\frac {180^\circ}{m}$.  For $m=$ even, this symmetry is about
    both the major and minor axes, whereas for $m=$odd, the reflection
    symmetry is only about the major axis.

\item For odd modes of $m$, there is additional reflection symmetry
    about the minor axis at:  $\phi_m = \pm\frac{90^\circ}{m}$.

\end{itemize}

\noindent An irregular galaxy has angles that are ``out of phase'' whereas
regular galaxies have angles that are more ``in phase'' (i.e.  reflectionally
symmetric around either minor or major axis).  Therefore, it is possible to
quantify various forms and degree of asymmetry by constructing indices based
on the amplitude and phase angles of the Fourier modes.  The most intuitively
obvious asymmetry index is the $m=1$ mode, which captures the lopsidedness
($A_L$) of a galaxy, i.e. the positioning of the brightest central region
relative to the fainter outer region of a galaxy:

\begin{equation}
    A_L = \left| a_1 \right|.
\end{equation}

\noindent Asymmetric galaxies are also characterized by overall deviation from
an ellipse; thus, another intuitively useful quantity to measure is the sum of
the Fourier amplitudes:

\begin{equation}
    A_E =  \sum_{m}^{N} \left|a_m\right|.
\end{equation}

\noindent Asymmetric galaxies by definition have high $A_E$.  However, it is
possible for galaxies with both high $A_E$ and $A_L$ to be reflectionally
symmetric; the degree of reflectional symmetry may be an indicator for how
well the galaxies is relaxed.  Reflection {\it asymmetry} is given by the
index $A_R$:

\begin{eqnarray}
    A_R & = \sum_{m=\rm{even}} \left|a_m\right|\ {\rm sin }^2\left(\pi m \frac{\phi_m}{180^\circ}\right) + \notag \\
        &   \sum_{m=\rm{odd}} \left|a_m\right|\ {\rm sin }^2\left(\pi m \frac{\phi_m}{90^\circ} \right),
\end{eqnarray}

\noindent where $\phi_m$ is in degrees.  In this formulation, the higher the
reflectional asymmetry, the higherthe index $A_R$.  Used together, these three
descriptors provide highly useful ways to quantify the degree galaxies are
irregular.  For instance, high values of $A_R$ and $A_L$ most likely imply
high global asymmetry in the intuitive sense.  Whereas a high value of $A_E$
with low $A_R$ implies high regularity, but large deviation from an ellipse,
such as edge-on disky galaxies or a disky/boxy ellipticals.

\begin{figure*}
    \plotone{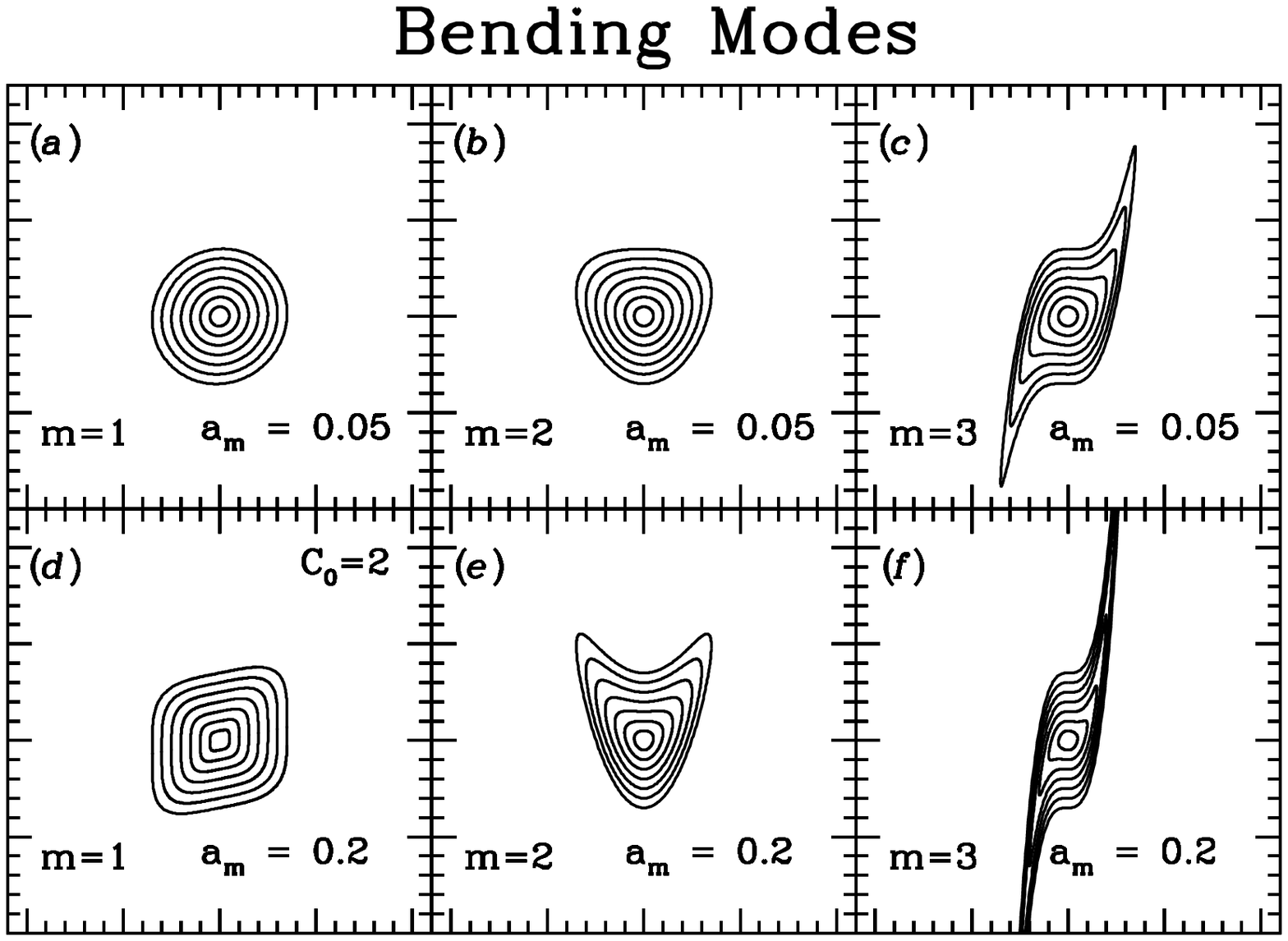}

    \figcaption{Examples of bending modes modifying a circular profile
    ($q=1.0$) with $C_0=0$ (unless indicated otherwise).  ({\it Top row})  
    Low-amplitude ($a_m=0.05r^m_{\rm scale}$) bending modes.  ({\it Bottom 
    row})  High-amplitude ($a_m=0.2r^m_{\rm scale}$) bending modes.  Bending 
    modes can be combined with Fourier modes to change the higher order shape.
    \label{fig:bending}}

\end{figure*}

\begin{figure*}
    \centerline{\includegraphics[angle=0,height=8.cm,width=8.cm]{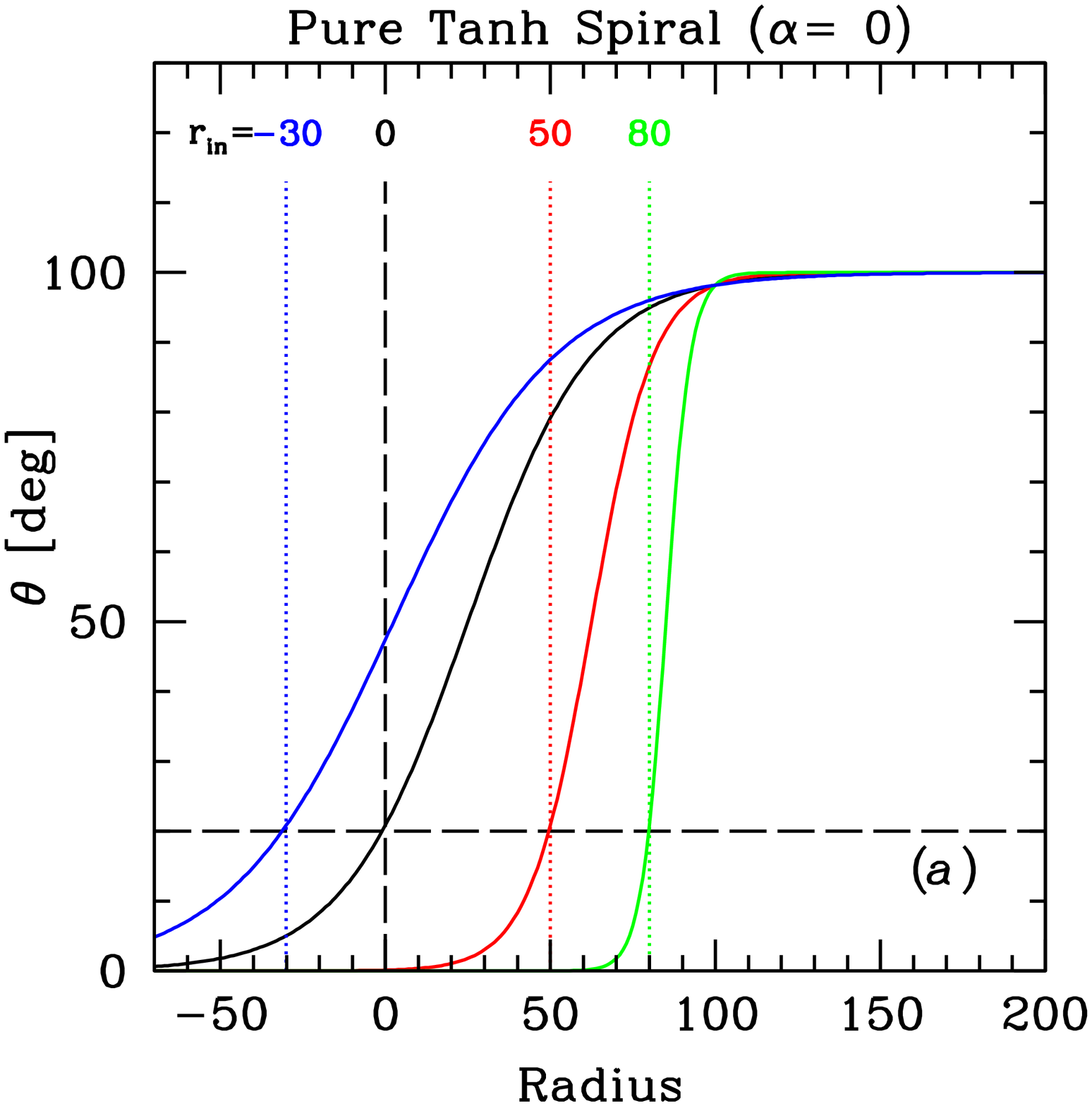}\hfill
                \includegraphics[angle=0,height=8.cm,width=8.cm]{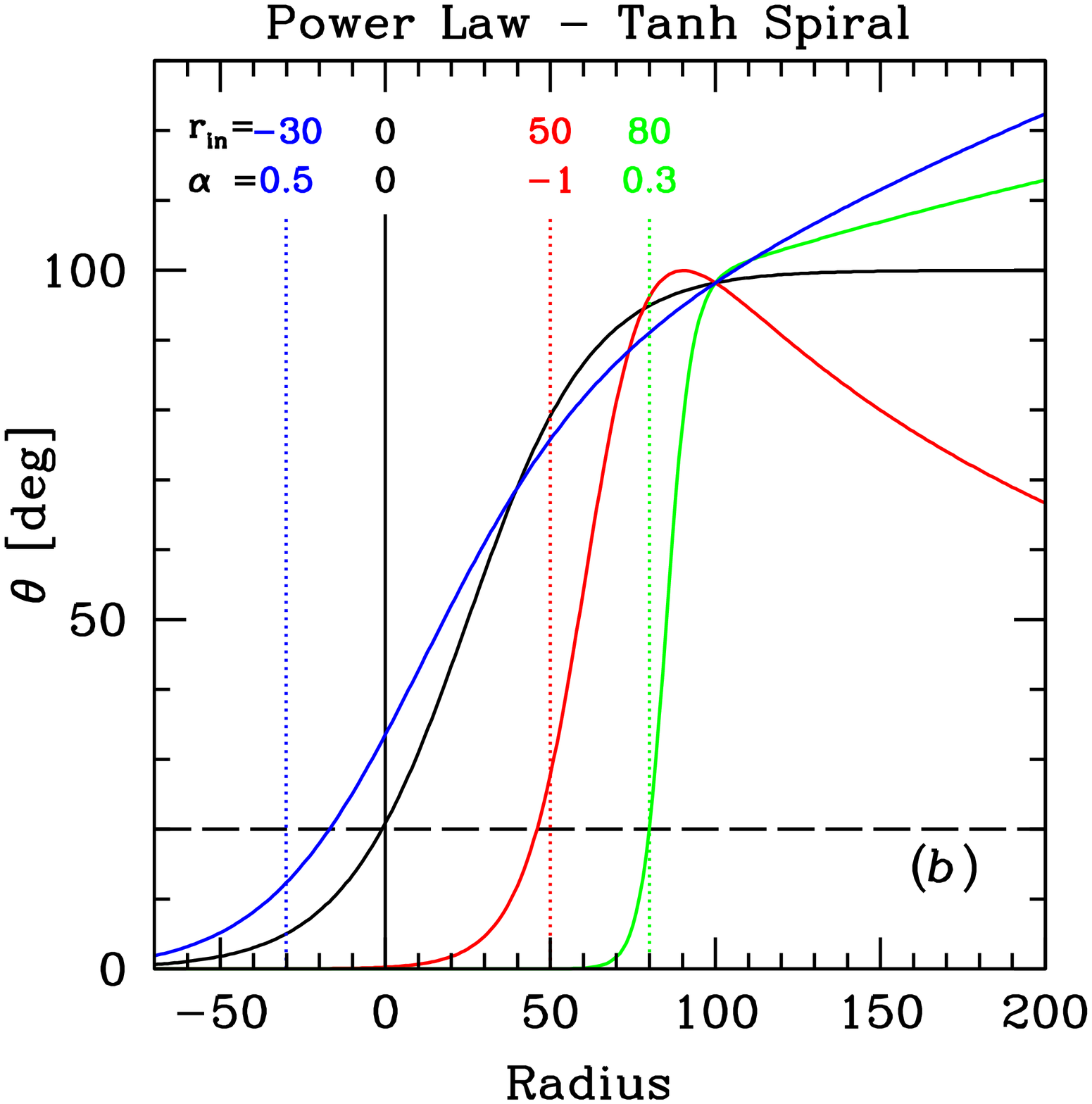}}

    \figcaption{Hyperbolic tangent-power-law spiral angular rotation functions
    with outer spiral radius of $r_{\rm out}=100$. ({\it a}) Examples
    of pure hyperbolic tangent spirals ($\alpha=0$) with different bar radii
    ($r_{\rm in}$).  ({\it b}) Examples with different bar radii and
    asymptotic power law $\alpha$, as indicated. See Figures~\ref{fig:spiral1}
    and \ref{fig:spiral2} for examples of how these parameters translate into
    2-D images. \label{fig:spiralpow}}

\end{figure*}

\bigskip

\noindent {\bf Bending Modes} \ \ \ \ \ Bending modes allow for
power-law-shape curvatures in the model, as opposed to spiral windings.  The
coordinate transformation ($x,y$) $\Longrightarrow$ ($x', y'$) is obtained by
only perturbing the $y$-axis (in a rotated frame) in the following way:

\begin{equation}
\label{eqn:bending} 
    y' = y + \sum_{m=1}^{N} a_m \left(\frac{x}{r_{\rm scale}}\right)^m,
\end{equation}

\noindent where $x'=x$, $r_{\rm scale}$ is the scale radius of the model
(i.e. $r_{\rm eff}$ for \sersic, $r_s$ for exponential, etc.).  Some examples
of this perturbation are shown in Figure~\ref{fig:bending}.  Note that $m=1$
resembles quite closely to the axis ratio parameter, $q$.  However, the $m=1$
bending mode is actually a shear term, the effect of which is most easily seen
when it operates on a purely boxy profile with $C_0 \approx 2$
(Figure~\ref{fig:ellipse}{\it a}), shearing it into a more disky shape (see
Figure~\ref{fig:bending}{\it d}).  The bending modes can be modified by
Fourier modes or diskiness/boxiness to change the higher order shape of the
overall model.  This kind of coordinate transformation again preserves the
original meaning of the radial profiles.  Here, the object size parameter
refers to the unstretched size, i.e.  projected onto the original ($x,y$)
Cartesian frame, as opposed to a length along the curvature.

\bigskip

\begin{figure*}
    \plotone{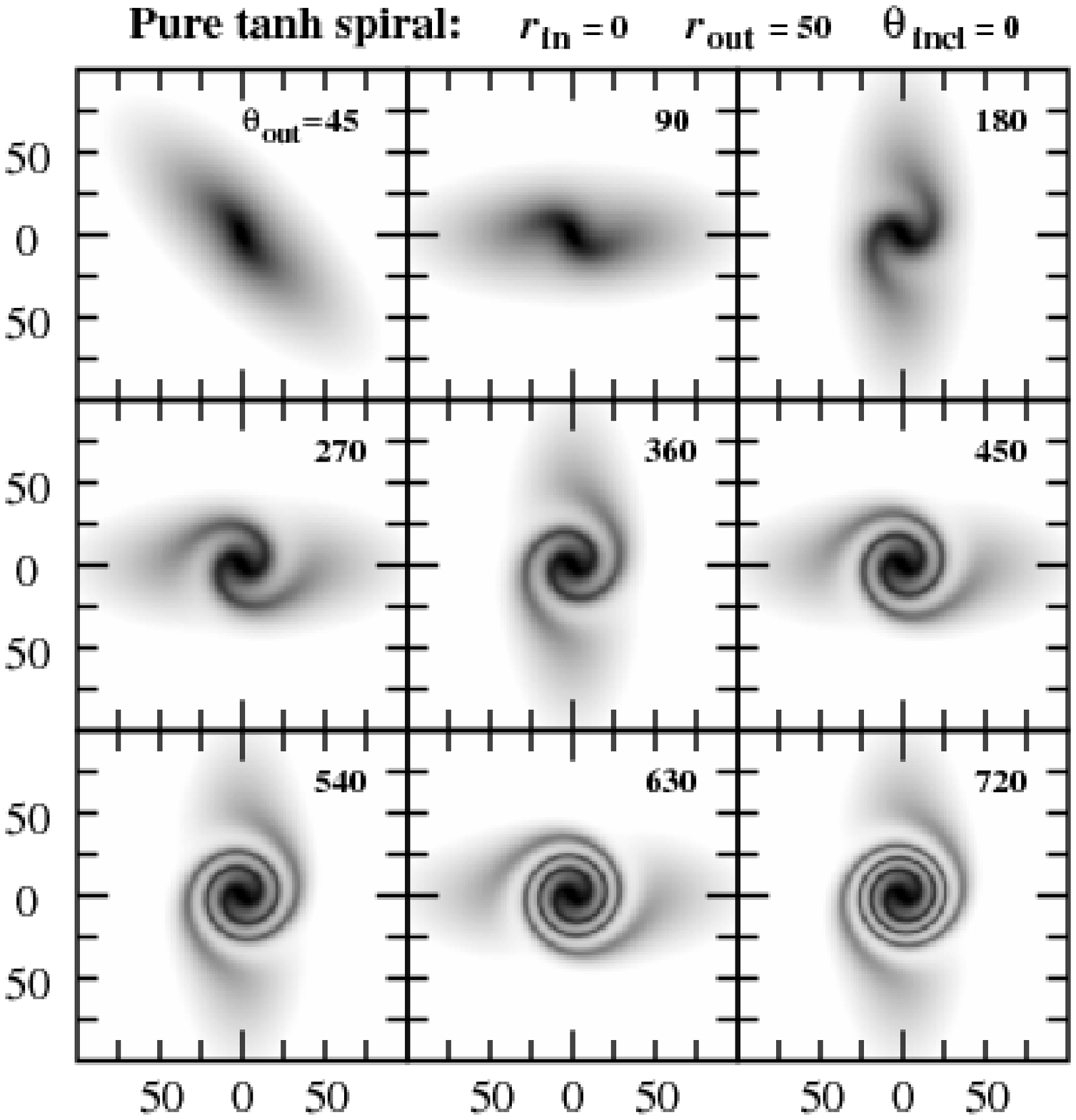}

    \figcaption{Examples of pure (i.e.  with power law $\alpha = 0$ or without
    logarithmic function) hyperbolic tangent coordinate rotation modifying an
    elliptical profile with axis ratio $q=0.4$.  Note that all panels share
    the same parameters as shown up top.  The spiral model has no bar.  The
    numbers within each panel show the amount of total winding (units in
    degrees) at the spiral rotation radius of 50.  Notice that outside $r=50$,
    the rotation angle becomes constant, due to the rotation function being a
    hyperbolic tangent, thereby creating the appearance of a flattened disk,
    even though there is not a separate disk component involved in the model.
    \label{fig:spiral1}}

\end{figure*}

\begin{figure*}
    \plotone{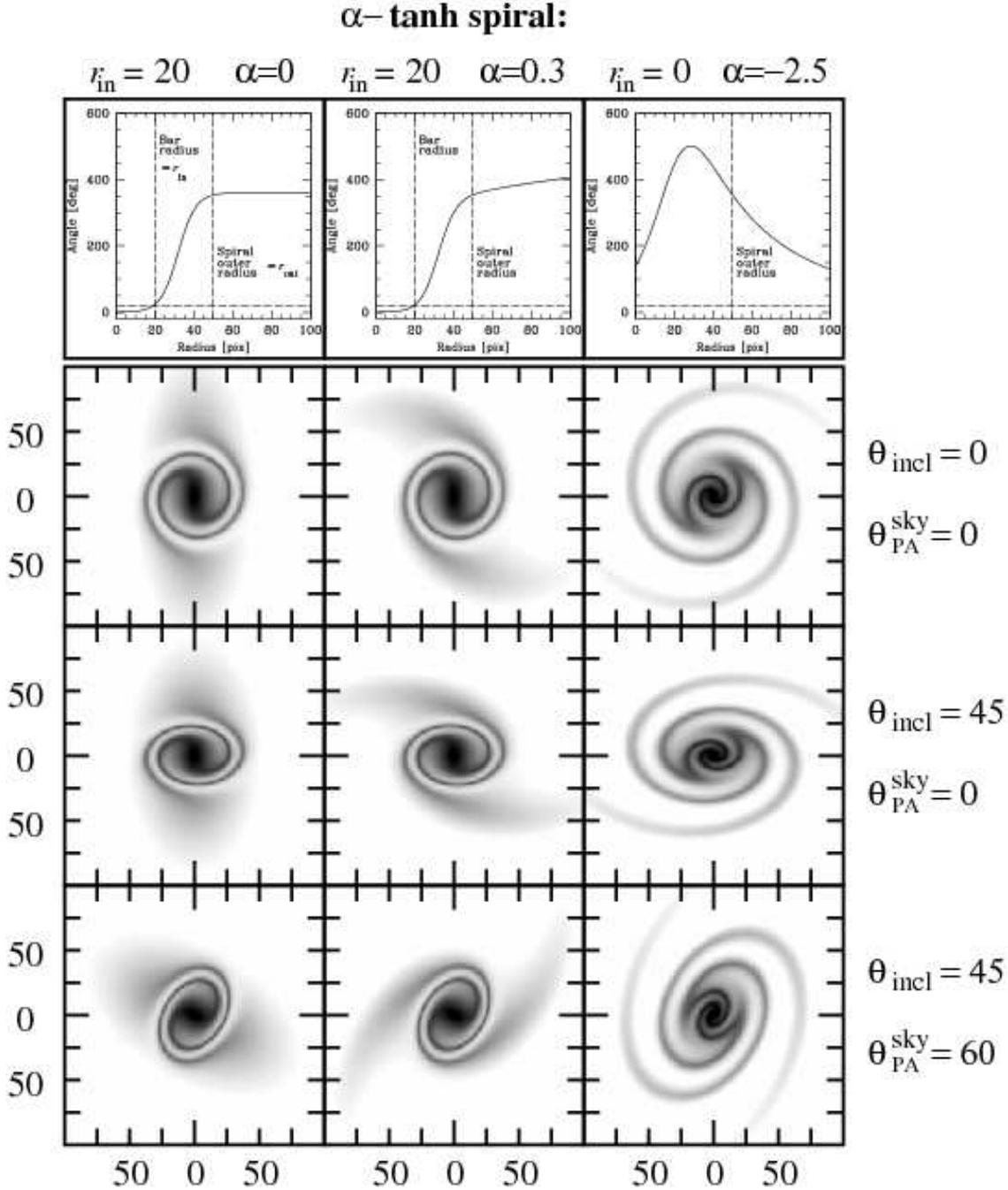}

    \figcaption{Examples of power law - hyperbolic tangent ($\alpha$-tanh)
    coordinate rotation modifying a face-on ($\theta_{\rm incl} = 0^\circ$)
    elliptical profile with axis ratio $q=0.4$.  The parameters of the
    rotation functions are shown on the top and right-hand side of the
    diagram.  The top panels show the spiral rotation angle as a function of
    radius for the panels in the same column. In the right-most column, the
    spiral arms reverse direction at $r=30$ because the spiral rotation
    function (top-right panel) decreases in rotation angle.
    \label{fig:spiral2}}

\end{figure*}

\begin{figure*}
    \centerline{\includegraphics[angle=0,height=8.cm,width=8.cm]{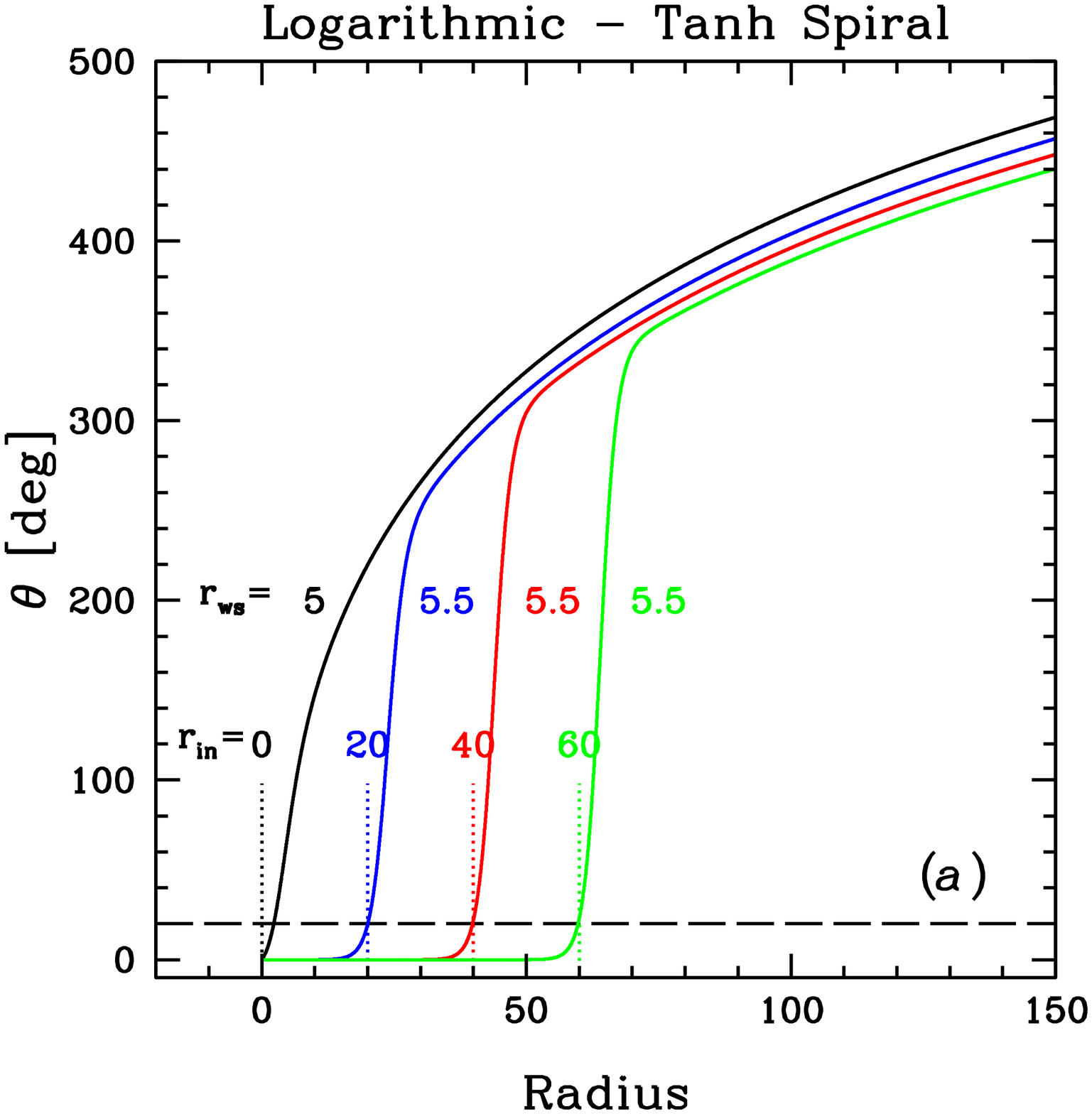}\hfill
                \includegraphics[angle=0,height=8.cm,width=8.cm]{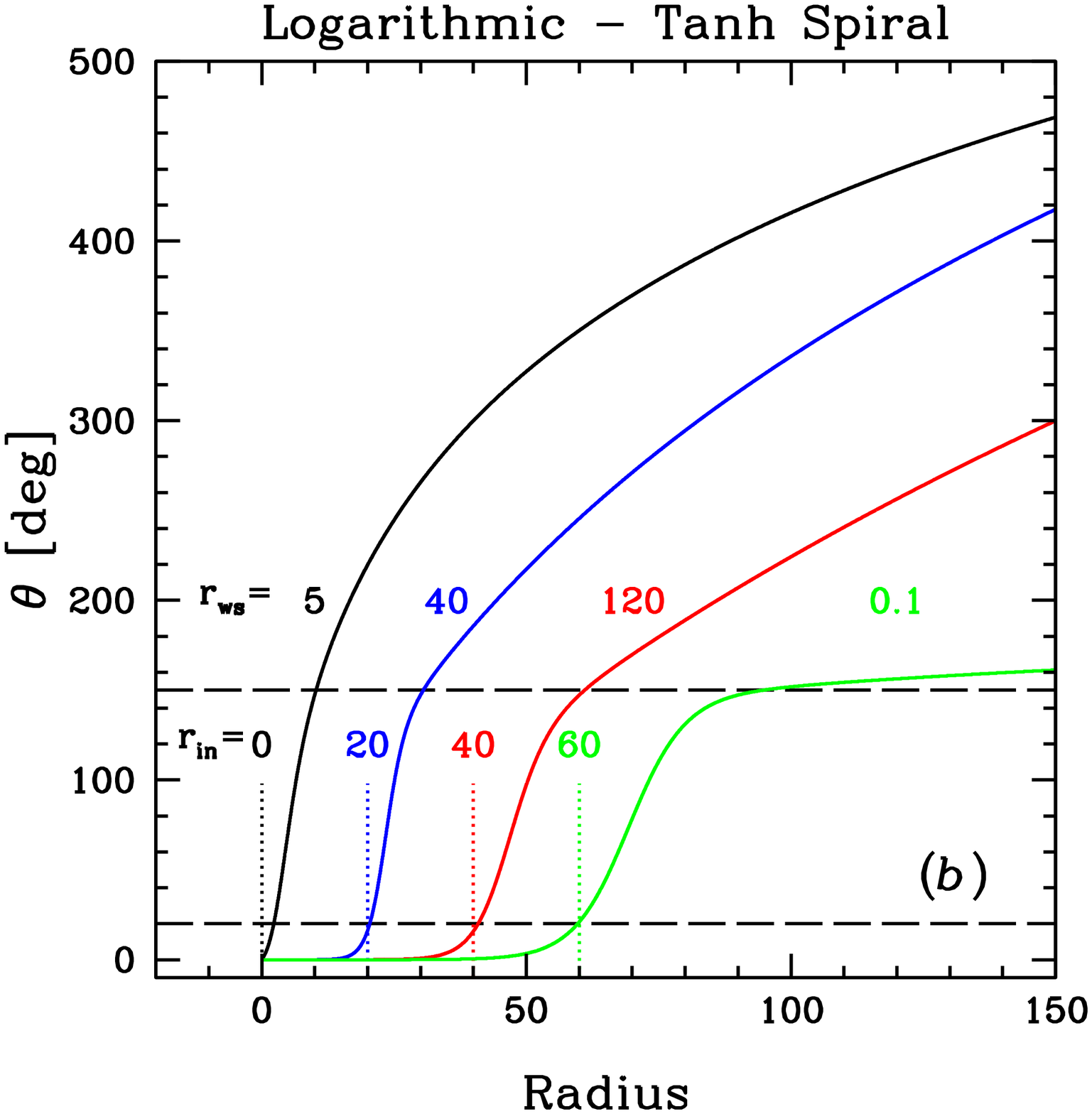}}

    \figcaption{Logarithmic - hyperbolic tangent spiral angular rotation
    functions.  ({\it a}) Examples of different bar radii, where the outer
    hyperbolic spiral radius is $r_{\rm out}=r_{\rm in} + 10$.  The lower
    horizontal dashed line shows the rotation angle at the ``bar'' radius
    ($r_{\rm in}$).  ({\it b}) Examples with different ``bar'' radii ($r_{\rm in}$) and winding-scale radii $r_{\rm ws}$, as indicated, illustrating the
    degree of flexibility of the spiral rotation rate.  The rotation angle at
    $r_{\rm out}$ is fixed to $150^\circ$, as shown by the upper horizontal
    dashed line.  The left-most, black curve is close to being a pure
    logarithmic function, recasted so that at $r=0$, the rotation angle
    $\theta=0^\circ$.  \label{fig:spirallog}}

\end{figure*}

\begin{figure*}
    \plotone {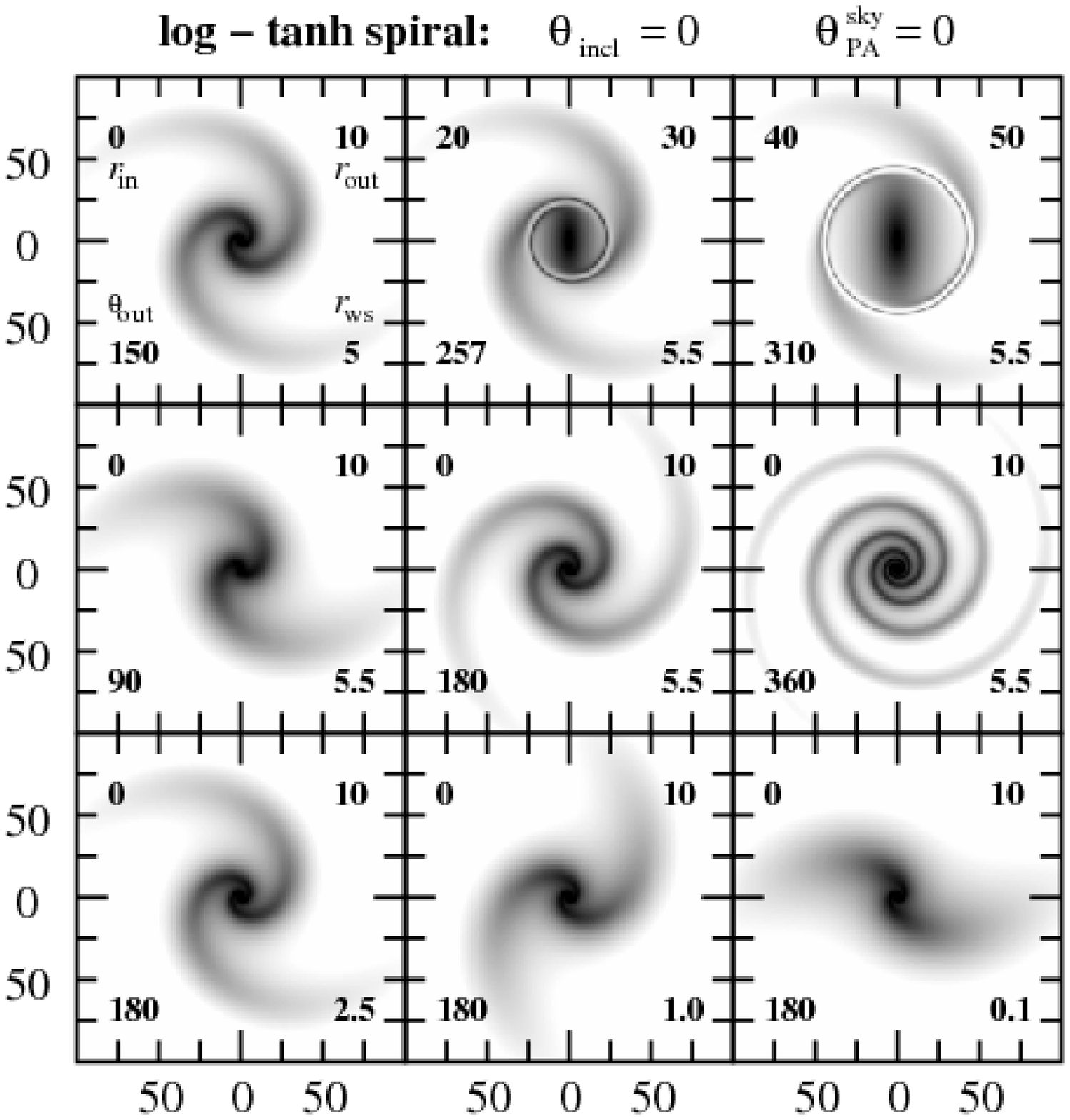}

    \figcaption{Logarithmic - hyperbolic tangent spiral (log-tanh) angular
    rotation examples, all face-on ($\theta_{\rm incl} = 0^\circ$) and $\theta^{\rm
    sky}_{\rm PA} = 0^\circ$.  The top-left panel shows the meaning of the
    rotation parameter values at the corners of each box.  As with the
    $\alpha$-tanh spirals, the log-tanh spiral can be tilted and rotated to
    any sky projection angle, or combined with Fourier modes to produce
    lopsided or multi-armed spiral structures (not shown), and with truncation
    function to produce an inner ring or an outer taper.  The top-left panel
    figure, for all practical purposes, is a pure logarithmic spiral with a
    winding scale radius $r_{\rm ws} = 5$. \label{fig:spiral3}}

\end{figure*}

\noindent {\bf Coordinate Rotation: The Concept} \ \ \ \ \ Sometimes the
isophotes of a galaxy can rotate as a function of radius, as in the case of
spiral galaxies.  To model spiral patterns, it is now possible to allow for
coordinate rotation in \galfit.  Coordinate rotation in \galfit\ means that
the flux within circular annuli overlayed on a model rotates as a function of
radius, i.e.  $\theta = f(r)$.  The functional form $f(r)$ can be fairly
arbitrary but the most familiar pattern in nature is that of a logarithmic
spiral, i.e.  $\theta \sim {\rm log}(r)$.  However, many spiral galaxies
deviate from logarithmic winding either in the inner region, for instance due to the
presence of a bar, or in the outer region, as might be due to tidal or non-relaxed
features.  These structures pose a problem when fitting galaxy images because
one cannot simply mask out regions of non-interest when the goal might be to
obtain the cleanest separation between a spiral and other embedded components.
Therefore a pure logarithmic spiral, while useful to {\it trace segments} of a
spiral, is often not ideal for {\it fitting} the whole galaxy, but ought to be
modified in some ways.  For this reason we introduce the concept of a
hyperbolic-tangent (tanh) modification to a logarithmic or a power-law spiral.

A pure tanh function looks like Figure~\ref{fig:spiralpow}{\it a}, showing
that $f(r)$ asymptotes to constant values at $r\rightarrow\pm
\infty$, which is a highly desirable feature.  As shown in
Figure~\ref{fig:spiralpow}{\it a}, the function can be scaled, stretched, and
shifted so that $\theta(r) \approx 0$ at $r<r_{\rm in}$: it is useful to model
a bar-like feature, which, by definition, has a constant PA as a function of
radius.  A tanh function is also useful in the upper asymptotic limit because
$f(r)$ at $r>r_{\rm out}$, when multiplied by another function $f_2(r)$, takes
on the form of $f_2(r)$, and the crosstalk within $r_{\rm in}$ is minimal, as
shown in Figure ~\ref{fig:spiralpow}{\it b}.  In short, a tanh function allows
for a {\it transition} between {\it two} functions:  a constant function at
$r<r_{\rm in}$ and another $r>r_{\rm out}$, for example a power-law or a
logarithmic function.  Moreover, the rate of that transition can be cleanly
{\it managed} and is easy to interpret.  For this reason a hyperbolic tangent
is also a function of choice later on in Section~\ref{sect:truncation} when we
present the idea of a truncation function.  \galfit\ allows for two types of
coordinate rotation functions, the power-law spiral ($\alpha$-tanh), and the
logarithmic spiral (log-tanh), both of which are motivated empirically.  We
note that even though the logarithmic spiral is favored more in the
literature, we find that the $\alpha$-tanh spiral is better able to capture
the range of spiral behaviors found in nature because of the one extra degree
of freedom in $\alpha$, which can simulate the behavior of the log-tanh spiral
over regimes of interest.  We therefore tend to prefer use of the
$\alpha$-tanh coordinate rotation by default.  We now give an overview of the
two types of coordinate rotation:

\medskip

\noindent {\bf Coordinate Rotation I:  Power-law - Hyperbolic Tangent ($\alpha$-tanh)} \ \ \ \
\ The term ``power law'' refers to the fact that the pure tanh function of
Figure~\ref{fig:spiralpow}{\it a} is multiplied by a function of the form
$\sim r^{\alpha}$.  The exact functional form of the rotation function is
lengthy (see Appendix A), but the schematic functional dependence of the
power-law spiral on the parameters is given by the following:

\begin {equation}
    \label{eqn:alphatanh}
    \theta (r) = \theta_{\rm out}\ \ {\rm tanh}\ \left(r_{\rm in}, r_{\rm
	   out}, \theta_{\rm incl}, \theta^{\rm sky}_{\rm PA}; r \right) \times\ 
	   \left[\frac{1}{2}\left(\frac{r}{r_{\rm out}} + 
	   1\right)\right]^{\alpha}.
\end {equation}

\noindent As defined, the power-law rotation starts to take hold beyond
$r=r_{\rm out}$, and below which the tanh transition dominates.
Figure~\ref{fig:spiralpow} shows a pure hyperbolic tangent rotation function
for several different values of the parameter $r_{\rm in}$ ({\it left}), and a
combination of ``bar'' ($r_{\rm in}$) parameter and the asymptotic power-law
slope $\alpha$ ({\it right}), where $r$ is the radial coordinate system and
$\theta_{\rm out}$ is the rotation angle roughly at $r_{\rm out}$.  The inner
radius, $r_{\rm in}$, is defined to be the radius where the rotation reaches
roughly $20^\circ$.  This angle corresponds fairly closely to our intuitive
notion of bar length based on examining images, but is not a rigorous,
physical definition.  The angle $\theta_{\rm incl}$ is the line-of-sight
inclination of the disk, where $\theta_{\rm incl}=0^\circ$ is face-on and
$\theta_{\rm incl}=90^\circ$ is perfectly edge-on.

To motivate intuition for the free parameters used in the coordinate rotation
definition, Figures~\ref{fig:spiral1} and \ref{fig:spiral2} show a progressive
series of images for the spiral rotation function with different combination
of parameter values.  For instance, Figure~\ref{fig:spiral1} shows a series of
images of pure hyperbolic tanh spiral with increasing maximum rotation angle
($\theta_{\rm out}$), all else being held constant at the values indicated at
the top.  The spiral arm winding increases with increasing $\theta_{\rm out}$,
and the winding gets tighter, but the body does not expand wider because
$r_{\rm out}$ is fixed.  It is also important to note that a face-on model
does not necessarily mean that the outer-most isophotes are round.  Rather,
the ellipticity of the outer-most isophotes is related to the asymptotic
behavior of the rotation function, which asymptotes to a constant PA beyond a
radius of $r_{\rm out}$ for a pure hyperbolic tangent ($\alpha=0$,
Figure~\ref{fig:spiralpow}{\it a}).  The isophotes only appear circular in the
main body of the spiral structure when it has a large number of windings.
Figure~\ref{fig:spiral2} shows several other examples of barred and unbarred
spirals, with progressively different $\alpha$ values, sky inclination angle,
and rotated to different sky position angles ($\theta^{\rm sky}_{\rm PA})$.
The parameters for each grey-scale figure are shown at the top and to the
right of the corresponding (column, row).  When the power-law index $\alpha$
is negative, the spiral pattern can reverse course after reaching a maximum
value (see right-most column of Figure~\ref{fig:spiral2}).

In summary, the hyperbolic tangent power-law function has 6 free parameters:
$\theta_{\rm out}, r_{\rm in}, r_{\rm out}, \alpha, \theta_{\rm incl}$, and
$\theta^{\rm sky}_{\rm PA}$.  The thickness of the spiral structure is
controlled by the axis ratio $q$ of the ellipsoid being modified by the
hyperbolic tangent, or by the Fourier modes that modify the ellipsoid.  To
create highly intricate and asymmetric spiral structures, Fourier modes can be
used in conjunction with coordinate rotation.

%\medskip

We note that the ``bar'' radius ($r_{\rm in}$) is a mathematical tool.  Even
though the $r_{\rm in}$ term in the coordinate rotation does look like a bar
when it is sufficiently positive, it should be regarded only as a {\it
mathematical construct} to grant the rotation function as much flexibility as
possible.  This construct {\it can} reflect reality, but it does not have to.
For instance, mathematically, a {\it negative} $r_{\rm in}$ radius
(Figure~\ref{fig:spiralpow}{\it b}) is perfectly sensible because of the way
Equations~\ref{eqn:alphatanh} and \ref{eqn:logtanh} (for logarithmic spirals,
below) are defined: a negative $r_{\rm in}$ value just means that the spiral
rotation function has a finite rotation angle at $r=0$ relative to the initial
ellipsoid out of which it is constructed.  When there is clearly no bar, the
$r_{\rm in}$ parameter can become quite negative; in this case, the fit is
often indistinguishable from one where the bar radius is 0.  Furthermore,
often times, one may not wish to create a bar and a spiral out of one smoothly
continuous function for various reasons, for instance because they may have
different widths, the spiral may not extend into the center, or the spiral may
start off in a ring.  In these situations, one can ``detach'' the bar from the
spiral by using a truncation function (see \S~\ref{sect:truncation}), by
instead creating a bar with a separate \sersic, Ferrer, or other function.
When this is done, a ``bar radius'' is still useful mathematically in the
coordinate rotation function, but it may bear no physical relation to the
physical bar.

%\smallskip 

Finally, we draw attention to some limitations of the spiral rotation
formulation.  While the $\alpha$-tanh rotation function works surprisingly
well for many spiral galaxies, the function is smooth, so ``kinks'' in the
spiral structure cannot yet be modeled, even though it is possible to do so by
allowing for ``kinks'' in the rotation function.  Lastly, the spiral structure
cannot wind back onto itself, because that would require the rotation function
to be multi-valued.

\bigskip

\noindent {\bf Coordinate Rotation II:  Logarithmic - Hyperbolic Tangent
(log-tanh) } \ \ \ \ \ The winding rate of spiral arms in late-type galaxies
is often thought to be logarithmic with radius rather than power law.  Thus,
\galfit\ also allows for a logarithmic-hyperbolic tangent coordinate rotation
function, which is defined as:

\begin{eqnarray}
    \label{eqn:logtanh}
    \theta (r) & = & \theta_{\rm out}\ {\rm tanh}\ 
        \left(r_{\rm in}, r_{\rm out}, \theta_{\rm incl}, 
	\theta^{\rm sky}_{\rm PA}; r \right) \times\  \notag \\
        & & \left[{\rm log} \left(\frac{r}{r_{\rm ws}} + 
        1\right) / {\rm log} \left(\frac{r_{\rm out}}{r_{\rm ws}} + 
        1\right) \right]. 
\end{eqnarray}

\medskip

\noindent Like the $\alpha$-tanh rotation function, the log-tanh function has
a hyperbolic tangent part that regulates the bar length and the speed of
rotation within $r_{\rm out}$.  Beyond $r_{\rm out}$ the asymptotic rotation
rate is that of the logarithm function, which has a winding scale radius of
$r_{\rm ws}$; the larger the winding scale radius, the {\it tighter} the
winding.  Thus, like the $\alpha$-tanh spiral, the log-tanh spiral rotation
function also has 6 free parameters:  $\theta_{\rm out}, r_{\rm in}, r_{\rm
out}, r_{\rm ws}, \theta_{\rm incl},$ and $\theta^{\rm sky}_{\rm PA}$.  Note
that in terms of capabilities, the $\alpha$-tanh function can often reproduce
the log-tanh function and more.  Therefore, the $\alpha$-tanh is probably a
more useful rotation function in practice.

Note that \galfit\ does not allow for a pure logarithmic spiral because such a
function has a negative-infinity rotation angle at $r=0$.  Therefore, in
\galfit, at $r=0$ the rotation function reaches $\theta=0$
(Figure~\ref{fig:spirallog}).  Lastly, it is also important to
keep in mind that the meaning of the ``bar radius,'' just as described in the
section for $\alpha$-tanh rotation function, is a mathematical construct.

\bigskip

\begin{figure*}
    \plottwo {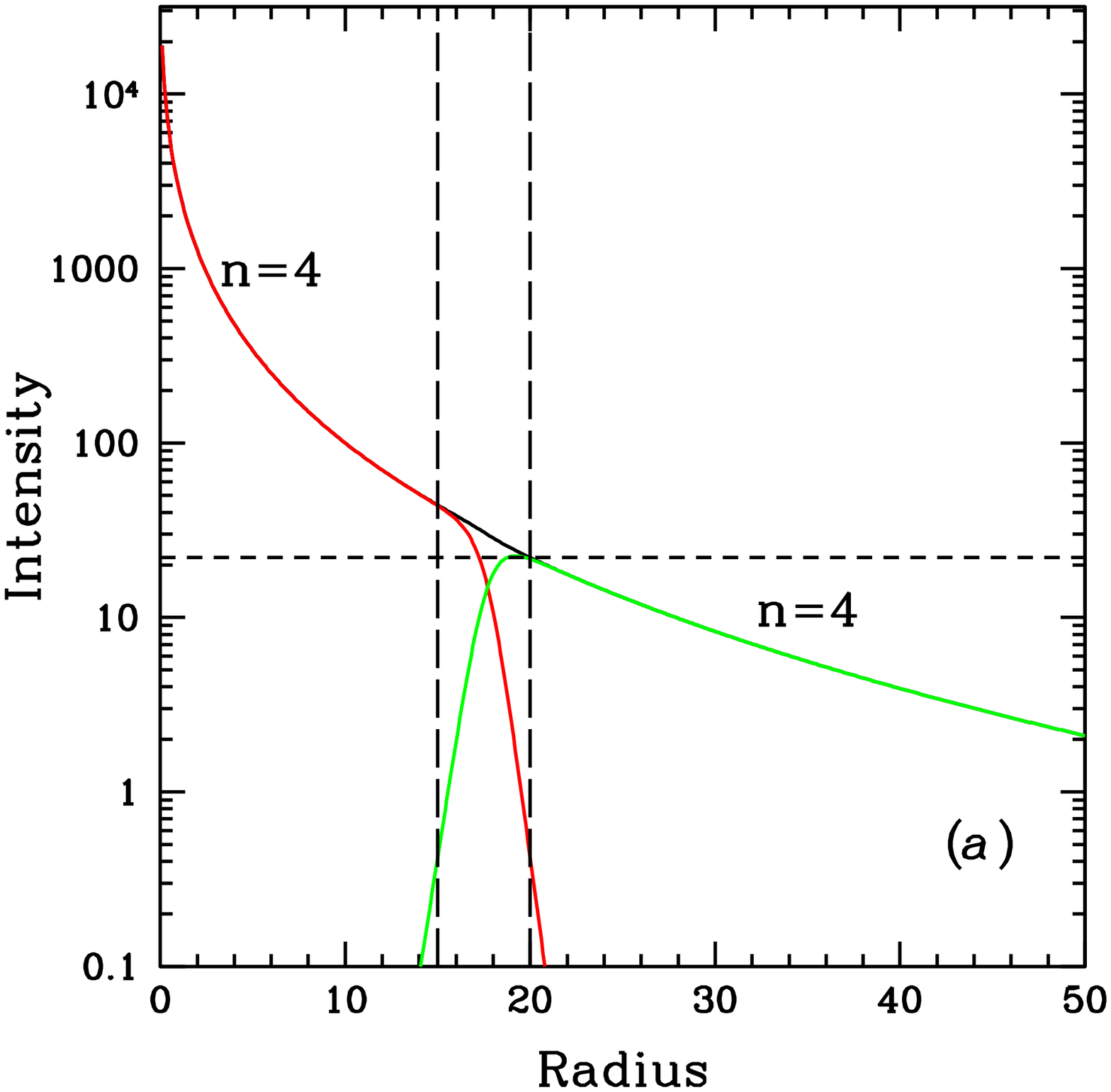} {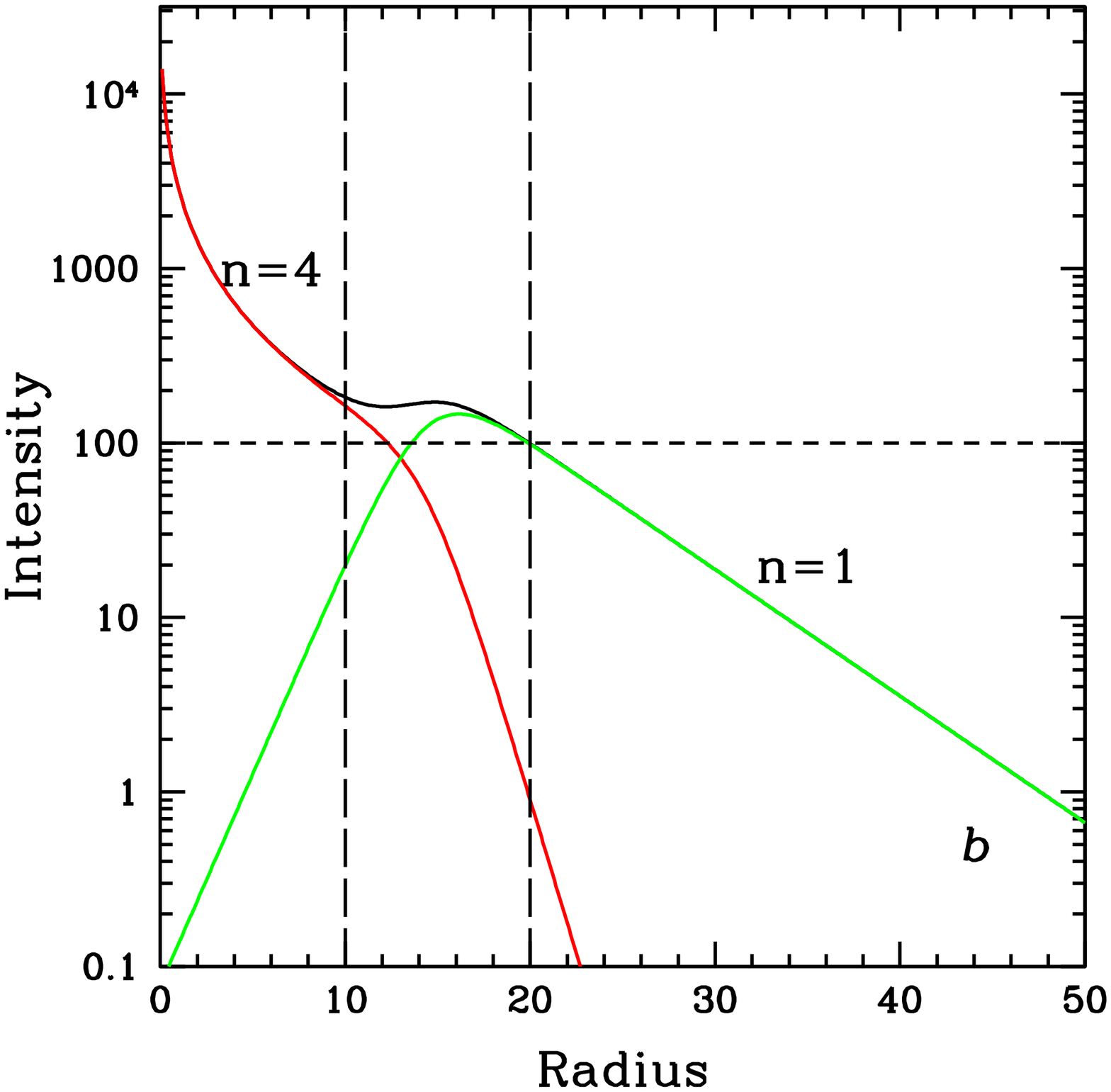}

    \figcaption{Examples of hyperbolic truncation functions on $n=4$ and $n=1$
    \sersic\ profiles.  ({\it a}) A continuous $n=4$ model represented as two
    truncated models of otherwise identical $r_e$, $n$, and central surface
    brightness, with truncation radii at $r=15$ and $r=20$, as marked by the
    vertical dashed lines.  The black curve is the sum of the inner and outer
    functions.  This shows that, outside the truncation region, there is very
    little ``crosstalk'' between the inner and outer components.  ({\it b}) A
    composite profile made up of an $n=4$ nucleus truncated in the wings and
    an $n=1$ truncated in the core, with truncation radii $r=10$ and $r=20$.
    Note that the hump in the summed model would give rise to a ring in a 2-D
    model. \label{fig:truncation1}}
\end{figure*}

\begin{figure*}
    \plotone {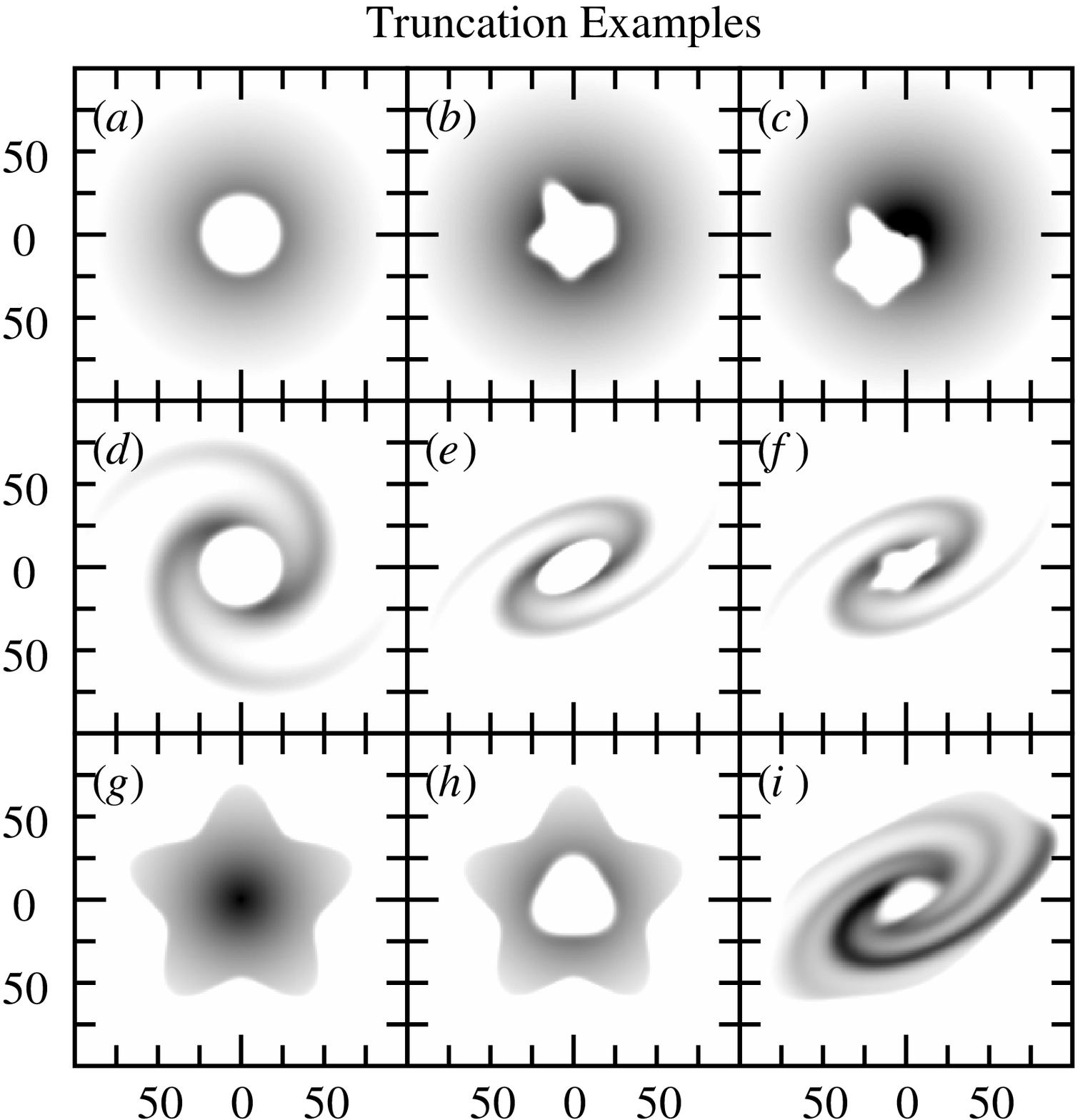}

    \figcaption{Examples of truncation functions acting on a {\it
    single-component} light profile of various shapes.  ({\it a}) Inner
    truncation of a round profile, creating a ring.  ({\it b}) The truncation
    function can be modified by Fourier modes, just like the light profile.
    ({\it c}) The truncation function can be offset in position relative to
    the light profile.  ({\it d}) The truncation function can act on a spiral
    model.  ({\it e}) The truncation can tilt in the same way as the spiral.
    ({\it f}) The truncation function can be modified by Fourier modes while
    acting on a spiral model.  ({\it g}) A round light profile is being
    truncated in the wing by a pentagonal (Fourier mode 5) truncation
    function.  ({\it h}) A round light profile is being truncated in the inner
    region by a triangular function (Fourier mode 3), and in the wing by a
    pentagonal function.  ({\it i}) A three-arm, lopsided, spiral light
    profile model is truncated in the wing by a pentagonal function, and in
    the inner region by a triangular function. \label{fig:truncation2}}

\end{figure*}

\section {THE TRUNCATION FUNCTION}

\label {sect:truncation}

Truncation functions allow for the possibility of creating rings, outer profile
cut-offs, dust lanes, or a composite profile in the sense that the inner region
behaves as one function and the outer behaves as another.  The truncation
function can modify both the radial profile and azimuthal shape.  A ring can
be created by truncating the inner region of a light profile.  Likewise, when
a galaxy has spiral arms that do not reach the center, it can be viewed as
being truncated in the inner region.

\subsection {General Principle}

In \galfit\ each truncation function can modify one or more light profile
models.  Also, any number of light profiles can share the {\it same}
truncation function.  The truncation function in \galfit\ is a hyperbolic
tangent function (see Equation~\ref{eqn:trunc} in Appendix~B).  Schematically,
a truncated component is created by multiplying a radial light profile
function, $f_{0,i}(x,y; ...)$, by one or more truncation functions, $P_m$ or
$1-P_n$ (depending on whether the type is an inner or an outer truncation), in
the following way:

\begin{eqnarray}
\label{eqn:trprod}
    f_i (x,y; ...) = f_{0,i} (x, y; x_{c,i}, y_{c,i} ...\ q_i, \theta_{{\rm PA}, i})\ \times \ \ \ \ \ \ \ \ \ \ \ \ \ \\
    \prod^m P_m \left(x, y; x_{c, m}, y_{c, m}, r_{{\rm break},m},
    \Delta r_{{\rm soft},m}, q_m, \theta_{{\rm PA},m}\right) \times \notag \\
    \prod^n \left[1-P_n \left(x, y; x_{c, n}, y_{c, n}, r_{{\rm
    break},n}, \Delta r_{{\rm soft},n}, q_n, \theta_{{\rm
    PA},n}\right)\right]. \notag
\end{eqnarray}

\noindent The break radius, $r_{\rm break}$, is defined to be the location
where the profile is 99\% of the original (i.e. untruncated) model flux at
that radius.  The parameter $\Delta r_{\rm soft}$ is the softening length, so
that $r=r_{\rm break} \pm \Delta r_{\rm soft}$ is where the flux drops to 1\%
of that of an untruncated model at the same radius (the $\pm$ sign depends on
whether the truncation is inner or outer).  The inner truncation function
($P_m$) tapers a light profile in the inner regions of a light profile,
whereas the outer truncation function ($1-P_n$) tapers a light profile in the
wings.

The behavior of the hyperbolic tangent function is ideal for truncation
because it asymptotes to 1 at the break radius $r\gtrsim r_{\rm break}$ and 0
at the softening radius $r <r_{\rm soft}$, and vice versa for the complement
function.  Thus, when multiplied to a light profile $f(r)$, the functional
behavior exterior to the break radius has intuitively obvious meanings.  For
example, as shown in Figure~\ref{fig:truncation1}{\it a}, if a
\sersic\ function with $n=4$ is truncated in the wings (shown in red), the
core has exactly an $n=4$ profile interior to $r_{\rm break}$ (marked with a
vertical dashed line), which is a free parameter to fit.  Likewise, an $n=4$
profile truncated in the core (green) has exactly an $n=4$ profile exterior to
the outer break radius.  Thus, when one sums two functions of different
\sersic\ indices $n$ (Figure~\ref{fig:truncation1}{\it b}) the asymptotic
profiles of the wing and core retain their original meaning, and there is very
little crosstalk outside of the truncation region (denoted by vertical dashed
lines in Figure~\ref{fig:truncation1}).

Use of the truncation functions is highly flexible.  There can be an
unrestricted number of inner and outer truncation functions for each light
profile model.  Furthermore, multiple light profile models can share in the
{\it same} truncation functions.  This is useful, for instance, when trying to
fit a dust lane (inner truncation) in a fairly edge-on galaxy that may affect
both the bulge and the disk components.  Just as with light profile models,
the truncation functions can be modified by Fourier modes, bending modes,
etc., independent of the higher order modes for the light profile they are
modifying.

\subsection {Different Variations of the Truncation Function}

Truncation models appear in many physical contexts, such as dust lanes, rings,
spirals that do not reach the center, joining a spiral with a bar, or cut-off
of the outer disk.  To allow the truncation parameters to be more intuitive to
understand given situations at hand, \galfit\ offers several variations.  In
addition to inner and outer truncations, truncation functions can share in the
same parameters as the parent light profile.  There are radial and
length/height truncations, softening radius vs.  softening length (default vs.
Type 2), inclined vs.  non-inclined (default vs. Type b) truncations, and,
lastly, four different ways to normalize the flux---the most sensible choice
depends on how a profile is truncated.  We now discuss each of these
variations in more detail.

\bigskip

\noindent {\bf Parameter Sharing.} \ \ \ \ \ In the most general form, each
truncation function has its own set of free parameters: $x_0, y_0, r_{\rm
break}, \Delta r_{\rm soft}, q$, and $\theta_{\rm PA}$.  However, by default,
the parameters $x_0, y_0, q$, and $\theta_{\rm PA}$ are tied to the light
profile model, and are activated only when the user explicitly specifies a
value for them.

\bigskip

\noindent {\bf Radial (``radial'') vs. Length (``length'') or Height
(``height'') Truncations.} \ \ \ \ \ The most useful type of truncation is one
that has radial symmetry to first order, i.e. where it has a center, an
ellipticity, and an axis ratio.  However, in the case of a perfectly edge-on disk
galaxy (``edgedisk'' model), an additional type is allowed that truncates
linearly in length or in height.  For instance, a dust lane running through the
length of the galaxy has an inner height truncation.  For the ``edgedisk''
profile, \galfit\ also allows for a radial truncation, as with all other
functions.  The one drawback to height and length truncations is that they
cannot be modified by Fourier and higher order modes like the radial
truncations.

\bigskip

\noindent {\bf Softening Length (``radial'') vs. Softening Radius
(``radial2'').} \ \ \ \ \ Sometimes, instead of softening {\it length} ($\Delta
r_{\rm soft}$), it is more useful for the fit parameter to be a softening {\it
radius} ($r_{\rm soft}$), especially when one desires to hold the parameter
fixed.  That is also allowed in \galfit\ as a Type 2 truncation function,
designated, for example, as ``radial2.''  The default option does not have a numerical
suffix.

\bigskip

\noindent {\bf Inclined (default, ``radial'') vs. Non-inclined (``radial-b'')
Truncations.} \ \ \ \ \ A spiral rotation function is an infinitesimally thin,
planar structure.  Nevertheless, it should be thought of as a 3-D structure
in the sense that the plane of the spiral can be rotated through three Euler
angles, not just in position angle on the sky.  When a truncation function is
modifying a spiral model, it is therefore sometimes useful to think about the
truncation in the plane of the spiral model.  When Fourier modes and radial
truncations are modifying a spiral structure, the default (``radial'') is for
the modification to take place in the plane of the spiral structure.  However,
there are some instances when that may not be ideal (e.g., a face-on spiral may
actually be ellipsoidal).  In those situations, one can choose ``radial-b'',
which would allow a truncation function to modify the spiral structure in the
plane of the sky, even though the spiral structure can tip and tilt as needed.

Lastly, the truncation function can be Type 2b (i.e. ``radial2-b'') as well.

\bigskip

\noindent {\bf Flux Normalization.} \ \ \ \ \ The most intuitive flux
normalization for a truncated profile is the total luminosity.  Unfortunately,
both the total luminosity and the derivative of the free parameters with
respect to the total luminosity are especially time-consuming to work out
computationally and slow down the iteration process. There are generally no
closed form analytic solutions to the problem.  Therefore, the alternative is
to allow for different ways to normalize a component flux.  The user may
choose whichever one is more sensible, given the situation and the science
task at hand.  The default depends on the truncation type:

\begin{itemize}

    \item Inner truncation:  the flux is normalized at the break radius.  This
	is most sensible for a ring model because this radius roughly
	corresponds to the peak flux of the ring.

    \item Outer truncation:  flux normalized at the center.

    \item Both inner and outer truncation:  same as the case for inner
        truncation.

\end{itemize}

However, there are many situations when the default is not desirable.
Instead, the user can choose the radius where the flux is normalized. To be
pedagogical, we explicitly show here the normalization for just the
\sersic\ function:

\begin{itemize}

    \item {\it function\ }: default (e.g., ``sersic,'' ``nuker,'' ``king,''
          etc.).  See the details for individual functions.

    \item {\it function1}: flux normalized at the center $r=0$ (i.e.
          $\Sigma_0$).  A function that is given originally by $f_{\rm
          orig}(r)$ is now defined as  $f_{\rm mod}(r) = \Sigma_0
          \frac{f_{\rm orig}(r)}{f_{\rm orig}(0)}$.  For the \sersic\ profile
          (i.e. called ``sersic1''), the profile function is redefined in
          the following way, written explicitly:

    \begin {equation}
	  \label{eqn:sersic1}
          f_{\rm mod}(r)=\Sigma_0\frac{\exp{\left[-\kappa
          \left(\left({\frac{r}{r_e}}\right)^{1/n} - 1\right)\right]}}
          {\exp{\left[\kappa\right]}}.
    \end {equation}

          For the Ferrer and King profiles, this normalization is the same as
	  the default normalization.

    \item {\it function2}: flux parameter is the surface brightness at a
	  model's native size parameter (parameter 4 of the light profile
	  model).  For a \sersic\ profile, called ``sersic2,'' this means the
	  effective radius $r_e$.  So, $f_{\rm mod}(r) = \Sigma_e\frac{f_{\rm
	  orig}(r)}{f_{\rm orig}(r_e)}$.  For example, a \sersic\ profile now
	  has the following explicit form:

    \begin {equation}
	\label{eqn:sersic2}
        f_{\rm mod}(r)=\Sigma_e
        \exp{\left[-\kappa \left(\left({\frac{r}{r_e}}\right)^{1/n} -
        1\right)\right]}.
    \end {equation}

          For the Nuker profile this normalization is the same as the default
	  normalization.

    \item {\it function3}:  flux parameter is the surface brightness
	  ($\Sigma_{\rm break}$) at the break radius ($r_{\rm break}$).  This
	  is the most useful situation when a truncation results in a
	  large-scale galaxy ring, so that the surface brightness parameter
	  corresponds closely to the peak of the light profile model.  When
	  the truncation is not concentric with the light profile model, this
	  kind of normalization is not very intuitive.  For ``radial''
	  truncation, $r_{\rm{break}}$ is parameter 4, whereas for
	  ``radial2,'' $r_{\rm{break}}$ is parameter 4 for outer truncation
	  and parameter 5 for inner truncation.  When the ``sersic3'' option
	  is chosen, the $r_{\rm{break}}$ parameter comes automatically from
	  the {\it first} truncation component with which a certain light
	  profile model is associated.

          In our example of the \sersic\ profile, $f_{\rm mod}(r) =
          \Sigma_{\rm break}\frac{f_{\rm orig}(r)}{f_{\rm orig}(r_{\rm
          break})}$.  For example, a \sersic\ profile now has the following
          explicit form:

    \begin {equation}
	\label{eqn:sersic3}
        f_{\rm mod}(r)=\Sigma_{\rm break} \frac{\exp{\left[-\kappa
        \left(\left({\frac{r}{r_e}}\right)^{1/n} - 1\right)\right]}}
        {\exp{\left[-\kappa \left(\left({\frac{r_{\rm
        break}}{r_e}}\right)^{1/n} - 1\right)\right]}}.
    \end {equation}

\end{itemize}

Figure~\ref{fig:truncation2} demonstrates just some of the possibilities
allowed when fitting truncations.  In addition to the regular ellipsoid shape,
the higher order modes like diskiness/boxiness parameters, bending modes, and
Fourier modes can also modify the shape of the truncation functions.  One can
also use the truncation function {\it on} a spiral model, on models with
Fourier and bending modes, and diskiness/boxiness models, some of which are
shown in Figures~\ref{fig:truncation2}{\it d}, \ref{fig:truncation2}{\it e },
16{\it f }, and \ref{fig:truncation2}{\it i }.  When a truncation function
acts on a spiral component, it can do so either in the plane of the disk
(``Type a'') or in the plane of the sky (``Type b,''; e.g., ``radial-b'').
While the default is in the plane of the disk, the parameters are more
intuitive in Type~b cases when the disk is tilted and rotated.

\smallskip

\subsection {Caveats about using the Truncation Function} 

The use of
truncation functions should be carefully supervised because unexpected things
can happen, such as the size or the concentration index of a component can
grow without bound.  This behavior is due to the fact that there are
degeneracies between the sharpness of truncation and the steepness/size of the
galaxy.  Therefore, truncation functions should only be used on objects that
clearly have truncations.

When two functions are joined by using a truncation function, the crosstalk
region is located in between the two truncation radii: it is worth bearing in
mind the definition that at the break and softening radii, the fluxes are 99\%
and 1\% that of the same model without truncation, respectively.  In other
words, the larger the truncation length, the larger the crosstalk region.
Therefore, when one (or more) of the parameters $r_{\rm break}$, $r_{\rm
break} + \Delta r_{\rm soft}$, or $r_{\rm soft}$ is either too small
($\lesssim $ few pixels) or larger than the image size, it probably indicates
that profile truncation parameters are not meaningful.  Rather, it more likely
reveals the fact that there is a mismatch between the light profile model and
the actual galaxy profile.

\section {INTERPRETATION, PARAMETER DEGENERACIES, UNIQUENESS, LOCAL MINIMA,
AND ERROR ANALYSIS}

\label{interpretation}

Now that we have introduced several ways to modify an ellipse into more exotic
shapes, a natural question to ask is how unique or robust are the
modifications.  A single-component ellipsoid fit can often be used to quantify
the global average profile of galaxies.  However, beyond that, decisions about
what procedure to use get to be more complicated.  On the one hand, the
science goal might call for fitting detailed structures inside a galaxy (e.g.,
a bulge, bar, nuclear star cluster, etc.).  On the other hand, doing so raises
concerns about parameter degeneracies, uniqueness, and local minima solutions
when the analysis becomes complex.  It is therefore useful to consider in some
depth what causes degeneracies and the different contexts in which they
appear.  Doing so allows for better understanding for how to deal with them
and how to properly interpret results from complex analysis.  For, not all
complex analyses are more suspect, nor are all simple analyses more robust.

The term ``degeneracy'' has a specific mathematical connotation, namely the
relation of $a+b=c$ is degenerate in $a$ and $b$ for a constant value of $c$.
In the galaxy fitting literature, ``degeneracy'' is often more loosely used to
also refer to ``non-unique'' or ``local minimum'' solutions (e.g., a fit
oriented at $90^\circ$ from the best orientation), or strong ``parameter
correlation'' (e.g., sky is anti-correlated with the \sersic\ index $n$).  We
will mostly not make the subtle distinctions here and instead will use the
term ``parameter degeneracy'' generically to refer to all such situations.

However, when fitting galaxies, it is more important to distinguish between
the aforementioned real degeneracies from ``pseudo'' ones.  Real degeneracies
refer to correlated parameters, local minima, and mathematically degenerate
solutions.  By contrast, ``pseudo'' degeneracies involve convergence issues
when an algorithm is used beyond its technical limits, or when users provide
bad input model priors to fit the data.  They may have nothing to do with real
degeneracies, yet the behavior of convergence may seem to suggest otherwise.
Whereas problems with real degeneracies are often resolvable by using full
spatial information of 2-D images, pseudo-degeneracy problems are solved
through experience and by using sound scientific or technical judgment, as we
elaborate further.

In this section, we discuss how most of the parameter degeneracy problems are
avoidable with proper input priors and proper fitting supervision, even when
large numbers of free parameters are involved.  We also discuss why, contrary
to popular notions, when it comes to avoiding model degeneracy and local
minima, it is not sufficient to only choose a model that is the simplest.
Rather, it is a judicious combination of {\it simplicity} and {\it realism}
that make for the most robust solutions.  Lastly, these discussions are
intimately connected to the issue of error analysis because error measurements
are nearly always dominated by systematic issues rather than photon noise in
galaxy fitting.  We therefore discuss why it is more important to quantify
model-dependent systematic errors rather than to rely on statistical
estimates.

We note that the discussions below are mostly based on experience, which we
present using practical examples rather than to show using rigorous proof.
Carrying out a rigorous proof is not only beyond the scope of this study, but
it is nearly impossible to do in a general manner because different scientific
applications have different sensitivities to different types of degeneracies.
We are also aware that presenting a full discussion of degeneracy issues lends
credence to the common notion that multi-component analysis is dangerously
complex.  However, the reality is not nearly so dire when one has a proper
understanding of the underlying issues and causes.

\subsection{True Numerical Degeneracies Caused by Correlated or Non-Unique
Parameters}

There are well-known situations when different parameters in one or more
functions are capable of modeling the same profile behavior.  This scenario is
the one most commonly referred to in generic discussions about model
degeneracies.  For instance, very large \sersic\ index values ($n\gtrsim4$)
have highly extended wings, the presence of which is non-unique with the sky
parameter.  A high $n$, caused by profile mismatch or poor model prior, can
often suppress the sky estimate.  It is therefore advisable to estimate the
sky independent of the fit, and to hold it fixed to the best estimate.  As a
second example, in the Nuker profile (Equation~\ref{eqn:nuker}), there are
three parameters ($\alpha$, $\beta$, and $\gamma$) that control the
inner/outer slopes and sharpness of the bending (Figure~\ref{fig:nuker}).
When the break radius $r_b$ of a Nuker profile is sufficiently small and
profile mismatch sufficiently large, model discrimination relies entirely on
the power law $\frac{\gamma-\beta}{\alpha}$.  Because there are numerous ways
to yield a specific value for $\frac{\gamma-\beta}{\alpha}$ in the model, it
leads to a degenerate situation involving three parameters.

As another example, a low-amplitude second Fourier mode and the first bending
mode (shear) can both be degenerate with the axis ratio $q$ parameter of an
ellipse, therefore they should not be used together except in obvious
situations where doing so is useful.  Lastly, in the spiral rotation function,
the periodicity of the rotation function can sometimes be a source of
``degeneracy.'' Multiple windings can approximate a smooth continuous model,
whether or not there is a spiral structure present.  For instance, a classical
\sersic\ ellipsoid can be simulated by a spiral model with a very large
$\theta_{\rm out}$.  While the fit is not good and easy to diagnose by an end
user, it is nevertheless a numerically allowed solution.

The above situations are not meant to be a complete laundry list, but they are
the most common situations.  In complex analysis, one always needs to be
circumspect about the potential hazards of mixing and matching different
functions whose parameters produce similar profile behaviors.  Even though
\galfit\ allows for a great deal of flexibility in the analysis, it is
ultimately up to the user to decide on what to allow, based on the goals of
the science, and to understand when potentially degenerate parameters may be
used effectively.

The above discussion may also seem to imply degeneracies or non-uniqueness are
too numerous for complex analysis to be practical or reliable.  That notion is
only true when it is not possible to verify the results of a fit and to try
out other solutions.  Such a scenario is more common for large scale galaxy
surveys, in which automated, detailed, analysis is admittedly quite difficult
to conduct sensibly.  However, even in those scenarios, there are many
situations where mutually coupled parameters do not affect the other main
parameters of scientific interest: degeneracies in the Fourier modes often do
not have any bearing on the total luminosity or size of a component.
Moreover, when an analysis is done manually, it is reassuring that the
problems are almost always easy to recognize and remedy when they do happen,
even by simple inspection of the model and residual images.

\subsection{Pseudo-Degeneracies Caused by Technical Conditions (e.g.,
Model Profile Resolution, Parameter Boundaries)}

Occasionally, what appears to be numerical degeneracy problems may be caused
by someone using a code outside the algorithm's physical capabilities.  As
such, it is a pseudo-degeneracy.  Different algorithms have different
limitations that affect convergence, whether the code is gradient descent,
Metropolis, or otherwise.  This situation may appear like parameter degeneracy
because restarting the fit does indeed yield a different solution, but in fact
the code may be hamstrung in its convergence.  For example, gradient descent
algorithms require the calculation of a gradient, and thus can run into
problems when the gradient cannot be calculated properly.  In simulated
annealing algorithms \citep[e.g.,][]{press92}, parameter boundaries and
annealing speed control the algorithmic behavior:  anneal too quickly, the
solution may settle into a local minimum.  To search larger parameter spaces
requires longer annealing times.

While all algorithms have conditions under which they perform poorly,
pseudo-degeneracies can always be recognized and mitigated.
\galfit\ is based on a Levenberg-Marquardt subroutine that performs the
least-squares minimization.  In part a gradient descent algorithm, the
convergence behavior is affected by the calculation of gradient images that
determines the direction of steepest $\chi^2$ descent.  When the gradient
images are problematic, they affect the convergence to a proper solution.
There are three main problematic situations.  The first, and most common,
occurs when a model becomes extremely compact (FWHM $\lesssim 0.5$ pixel), so
that the profile gradient cannot be resolved: all the gradient information in
the model fits into a single pixel.  This situation mostly arises when working
with high-contrast imaging data, such as quasar host galaxy decomposition,
when one of the subcomponents may be used to reduce the strong residuals
caused by a PSF mismatch.  A similar situation arises when a model is very
thin (axis ratio $q \lesssim 0.05$) and the object is compact; here, the
gradient does not exist along one spatial direction because of a lack of pixel
resolution.  Another rare example occurs when the inclination angle of a
spiral rotation component is close to perfectly face-on ($\theta_{\rm incl}
\rightarrow0$ in Equations~\ref{eqn:alphatanh} and \ref{eqn:logtanh}), when
the derivative image for the inclination parameter approaches zero.

Another abnormal numerical behavior may occur when one places parameter
constraints on a model to prevent some parameters from wandering too far from
their initial values.  Doing so may cause poor convergence by forcing the
solution into a tight ``corner,'' when the best solution is somewhere beyond
it.  A typical situation is where there are other sources in the image that
are not masked or fitted by models, but that are sufficiently luminous to
influence the fit of the target of interest.  In this situation, no amount of
effort will produce a sensible solution, because the best solution is outside
of the parameter boundaries, even though the desired solution may be within.
Pseudo-degeneracies occur in this situation both because there is an abnormal
condition imposed and because the input prior for the model is poor.

While technical issues with code operation add a layer of complexity to image
analysis, in practice the majority of situations one encounters are
straightforward to recognize by observing when the parameters take on
extremely large or small values.  However, clearly recognizing the problem as
being pseudo degeneracies is the key.  Once diagnosed, these situations are
easy to guard against, by holding those parameters fixed when they go below
certain values.  In practice, technical issues are not problematic even when
\galfit\ is used for automated analysis \citep{haeussler07} \footnote{While
these conditions can always be anticipated in advance, implementing a solution
in the code is more tricky, because the act of doing so may also induce other
convergence difficulties.  This leads to a false sense of security about the
robustness of a solution.}.

\subsection {Pseudo-Degeneracies Caused by Bad Input Model Priors}

One of the most common causes of degeneracy problems in galaxy fitting
analysis comes from using input priors that are not well suited to the data.
The most common ``input priors'' involve the choice of the type or the number
of components in a model\footnote{An input prior does not refer to the
accuracy of the initial parameter guesses.}.  Input priors are ideal when the
number of components of a model used in a fit matches the number of luminous
components in a galaxy.  However, often times one may choose to use either
fewer or more components than needed by the data.

A common example where the input prior is bad is when one uses fewer model
components than called for by the data.  Two of the main reasons for doing so
are to reduce the number of components/free parameters, or to allow automated
analysis, where it is not yet possible to tailor fits to individual galaxies.
This approach is often an intentional course of action taken by many studies,
especially when it comes to automating two component analysis; the goal,
ostensibly, is to decompose a galaxy into bulge and disk components.
Seemingly reasonable and justifiable on the notion of reducing the potential
for degeneracy, the approach is generally regarded by most people to be a
positive attribute, rather than a source of problem itself.  Yet, that
intuitive notion conflicts with the basic principle of how least squares
algorithms work, and leads to perhaps the most common causes of
(pseudo-)degeneracy problems cautioned by literature.

To understand why using fewer components than necessary is bad, it is
important to appreciate that galaxy fitting analysis is fundamentally flux
weighted.  Thus, when a luminous structure is not accounted for, other
subcomponents try to compensate, however imperfectly, for its presence.  For
instance, one may use a two-component model fit to a galaxy that has a bulge,
disk, and bar.  Doing so may have several different outcomes.  One solution is
where one component is a sum of (disk+bar) while the other is the bulge.
Another can be (bulge+bar) and disk, or perhaps a compromise (e.g., bulge +
0.7 bar; disk + 0.3 bar).  Which scenario occurs depends on the relative
contrast (i.e. flux weighting) of the bar to the bulge and disk, and
potentially on the initial parameters of the three components; small
perturbations may ``bump'' the solution out from one minimum into another.  It
is quite possible for there to be a ``global minimum'' solution to this
problem.  However, when the most meaningful solution, physically, is simply
dis-allowed by the input prior, a globally minimum $\chi^2$ cannot lend much
credence to the reality of the model components.

An input model prior might also be bad if the model involves using {\it more}
subcomponents than inherently present in a galaxy.  In this situation, the
results depend strongly on the degree of profile mismatch between the model
function and the data.  If there is significant mismatch, all the components
cooperate to reduce the residuals.  For instance, it is always possible to fit
multiple exponential models to a single-component de Vaucouleurs profile.  If
the goal is to obtain the total flux, the sum would do a better job than using
a single exponential.  However, individually, the structural parameters may
not have much physical meaning.  

Another example involving model prior is in the area of high-contrast imaging,
where the goal is to deblend a central, unresolved, point source from a
diffuse underlying object (e.g., quasar and host galaxy).  To do so reliably
requires an accurate PSF model for the unresolved source, or else the
residuals may overwhelm the extended object, causing unreliable fits.  Here,
the prior is the PSF model.  Quantifying how the prior affects the fitting
results involves trying out different PSFs, or to include extra components to
account for the PSF residuals, depending on the science goal.

These examples illustrate some of the most common situations where the
reliability of a fit depends less on the number of free parameters, and more
on having a proper model to describe the data.  Beyond a single-component
analysis, the need to make such a decision means that it will be difficult to
automate highly detailed decompositions of galaxies.  However, while
multi-subcomponent fitting is difficult to automate, it is reassuring that
making a wise decision, interactively, is often not particularly difficult
when a science goal is clearly defined.  Moreover, for galaxy surveys, where
the goal is to fit single-component profiles to galaxies, multi-{\it object}
decomposition is quite feasible to automate \citep[e.g., M.  Barden et al.
2009, in preparation;][]{haeussler07}.

\medskip

In summary, pseudo-degeneracy conditions exist because least-squares fitting
fundamentally involves flux weighting:  when luminous flux distributions are
present in an image, the models are attracted toward them to reduce the
residuals.  Therefore, when all components are not properly modeled, the
result may be tricky to interpret not because of potential for model
degeneracies, but that the solution may have no physical meaning even if there
is a global minimum.  The solution is to increase the complexity of the
analysis progressively until all luminous components are properly accounted.
This process does not imply, however, that it is necessary to account for
every component inside a galaxy for the solution to have any meaning, only
that components of similar flux ratios ought to be simultaneously accounted in
detailed analysis; with a few exceptions (e.g.  locally dominant features like
nuclear star cluster, nuclear ring), components with low fluxes generally do
not significantly disturb the parameters of the much more luminous
sub-components.

\subsection {Parameter Degeneracies Can be Broken by Spatial Information in
2-D}

One of the most common notions regarding fitting degeneracy is that the more
free parameters there are the greater is the potential for degeneracy
problems.  However, the sheer number of parameters is often not itself an
indication of a potential for parameter crosstalk.  Consider, for instance,
that it is equally robust to fit thousands of well-separated stars as it is to
fit an isolated one.  The same is true for galaxies, even though they are
considerably more extended and may overlap:  in large-scale image simulations,
\citet{haeussler07} studied automated batch analysis of galaxies using one
\sersic\ profile per galaxy. They find that simultaneously fitting overlapping
or neighboring objects using multiple components (often 3--10 \sersic\ models
at a time) recovers the input simulated parameters more accurately than
fitting a galaxy singly while masking out the neighbors.  

Indeed, spatially well-localized sources, like a bar, ring, or off-nuclear
star clusters, are virtually free from degeneracies caused by crosstalk with
other components.  A galaxy bar is well determined because it is more
elongated, has a flatter radial profile, and is more sharply defined than the
surrounding bulge and disk components, despite being embedded within.  Compact
objects that are off-centered may also be well determined if the rest of the
galaxy can be modeled accurately.  Contrary to notions that more model
components lead to greater degeneracy, it is important to consider the
qualitative aspects of those components:  not accounting for strong features
explicitly can yield a less reliable and less physically meaningful fit
because the solution is a compromise between the different subcomponents.

\subsection {Measurement Uncertainties, Parameter Correlation, and Parameter
Degeneracies}

\label{subsect:degeneracies}

The issue of parameter degeneracies closely ties into the topic of measurement
uncertainties, especially when the result of the analysis may depend on the
input model in fitting galaxies.  When the model fits the data perfectly (i.e.
the residuals are only due to Poisson noise) it is possible to infer parameter
uncertainties from the covariance matrix of free parameters, which is produced
during least-squares minimization by the Levenberg-Marquardt algorithm.  In
galaxy fitting, ideal situations are often not realized because the
differences between the data and the model profile involve not only random
(e.g., Poisson) sources, but also systematics from non-stochastic (e.g.,
profile function or shape mismatch, neighboring galaxies, etc.), and
stochastic factors (overall smoothness, for instance due to star clusters).
The one exception is under low signal-to-noise (S/N) situations, when Poisson
noise exceeds model imperfection.  In most other situations, non-random
factors dominate the residuals, causing uncertainties inferred from covariance
matrices to be underestimated.  Therefore, it is frequently not very
meaningful in galaxy fitting to cite measurement uncertainties for the fitting
parameters in the traditional sense.

One way to quantify uncertainties, possible in large galaxy surveys, is to
allow the {\it scatter} of the data points in physical relations (e.g., radius
vs.  luminosity, luminosity vs.  metallicity, etc.) to articulate the overall
uncertainty of the measurements, even if individual errors could not be easily
obtained.  Such a scatter inherently involves a convolution of several error
sources: the intrinsic scatter present in a physical relation, Poisson
measurement error, stochastic and non-stochastic systematic errors due to
model imperfections.  Intrinsic scatter, being a fact of nature, remains
present in physical relationships even should the data have infinite S/N, and
even if the models are perfect fits to the data.  Intrinsic scatter is often a
scientifically interesting quantity, but it is difficult to differentiate from
scatter caused by systematic and stochastic errors, which do not vanish given
infinite S/N.

In the absence of large galaxy surveys, it is then important to quantify
stochastic and non-stochastic systematic errors for individual objects.  Some
example situations include the black hole mass vs.  galaxy relation studies
\citep{kormendy95, gebhardt00a, ferrarese00} and the fundamental plane
\citep{djorgovski87}.

In general it is very difficult to pin-point all the causes of non-stochastic
systematic errors in an analysis, and to quantify their magnitude.  However,
one common cause is profile model mismatch:  to the extent that one does not
know the intrinsic model of a galaxy a priori, the uncertainty in measuring
the parameters is wedded to one's assumptions about the model.  Therefore, the
process of quantifying systematic, model-dependent errors involves exploring
the degree of parameter coupling, by trying out different models and seeing
how the parameters of key scientific interest change.  Another source of
systematic error is due to comparing results from different algorithms.  In
this scenario, the most sensible practice is therefore to only compare
parameters that are derived using the {\it same fitting technique} (rather
than 1-D vs.  2-D), and using the {\it same pixel and flux weighting scheme}
(instead of Poisson vs.  non-Poisson) during analysis.

In contrast, stochastic errors arising from general non-smoothness of a galaxy
profile are caused by, for example, star forming patches, dust lanes, etc..
Existing on small scales and widely dispersed, non-smoothness cannot be easily
identified and modeled in a practical manner using multiple components.  Even
if it is possible to do so, whether they ought to be fitted explicitly,
masked, or not treated at all, falls under the purview of the science goal.
Stochastic fluctuations often influence the analysis in a manner analogous to
having large correlated noise in the data.  If the fluctuations can be
quantified, one possible solution is to include them in the fit as a variance
term of $\chi^2$ (Equation~\ref{chi2}).  To estimate the fluctuations requires
first obtaining a smooth underlying model, which is not always easy to do if
galaxies have steep and/or irregular profiles.

While it is generally difficult to disentangle stochastic from non-stochastic
sources of systematic errors, there also do not seem to be obvious benefits
for doing so from a scientific standpoint.  For most applications, one should
only be interested in the overall magnitude of the systematic errors in a
collective sense.  One way forward is therefore to understand which parameters
are most strongly coupled, then compare the results of different solutions
judging by which ones are physically plausible.  For instance, one common
interest in bulge-to-disk decomposition is to quantify the uncertainty of the
\sersic\ index $n$.  We know that the \sersic\ index $n$ takes on a large
value when a profile has both a steep core {\it and} an extended wing (see
Figure~\ref{fig:sersic}).  Therefore, quantifying systematic errors in
measuring the \sersic\ $n$ might involve masking or fitting nuclear
sources/neighboring contamination, trying out different PSFs, or fitting the
disk by allowing for different disk \sersic\ index values.  Properly judging
the causes of systematic errors and accounting for them often would lead to
more natural fits and more sensible parameter values, without the need to hold
certain parameters fixed to preconceived answers.

In exploring the parameter space as described, there is often a concern that
parameter degeneracies are too numerous or problematic to understand, which
brings the discussion back full circle.  As discussed in previous sections,
when the cause of parameter degeneracy is properly identified, and when the
model priors are well conceived, our experience has been that spatial
information in 2-D can often effectively break many potential degeneracies
between the components.  Even when the size, luminosity, and central
concentration, of the different components correlate they often interact in
fairly superficial ways, and do not dramatically change what the model
components represent physically.  However, in situations where cross-talk is
significant and there is no reason to prefer one solution over another (when
the input prior is befitting), then differences in the answer speak to the
degree of the parameter uncertainty that is of key interest to quantify,
rather than to avoid, because ultimately the models are empirically motivated.

\medskip

In summary, to the extent that the results may depend on model assumptions,
parameter exploration is the only viable way to quantify true measurement
errors in the fit parameters.  Thus, when used properly, parameter
coupling/degeneracy, rather than complicating the interpretation, offers a
deeper insight into the reliability of the overall analysis.  We illustrate
the above ideas more explicitly in the following examples.

\section {EXAMPLES OF DETAILED DECOMPOSITION}

\label{sect:examples}

To demonstrate how to use the new features to extract complex structures, we
analyze five galaxies that are well resolved:  IC 4710, an edge-on disk
galaxy, Arp 147, M51, and NGC 289.  These galaxies are chosen because they
represent examples where traditional analysis using perfectly ellipsoid models
tend to leave some question as to what is physically being measured and to the
robustness of the photometry and decomposition.  The primary purpose here is
to illustrate the basic building blocks of galaxy morphology, not to address
what are the most ``scientifically interesting'' or useful applications---the
scope of which is far too broad to address.  As such, each individual example
is not intended to necessarily be ``interesting'' in its own right.  For
instance, while parameterizing a ring galaxy like Arp 147 may not itself be
too worthwhile scientifically, the concept has other relevance to deblending
Einstein rings from lensing galaxies in the image plane of strong
gravitational lenses, or separating a ring from a bulge, disk, and bar in
spiral galaxies.  Indeed, these are heuristic examples meant to generate new
ideas for potentially interesting applications, and to illustrate the dynamic
range of capabilities in our new approach.

Another goal of this section is to illustrate two seemingly contradictory
notions when it comes to galaxy morphology analysis:

\begin{itemize} 

    \item {\it Sometimes it is not necessary to perform ``full-blown''
	  analysis, including spiral structures, Fourier modes, rings, etc..}
	  The detailed analysis below will show when it is {\it not} necessary
	  to utilize the full machinery in order to meet the science
	  requirements, such as when the interest is to only quantify global
	  properties.  However...

    \item {\it Sometimes it is necessary to perform full-blown analysis.} In
	  situations where detailed decomposition matters (e.g., quantifying
	  bulge-disk-bar fractions) the most reliable analysis is to make full
	  use of the machinery available.

\end{itemize}

\noindent Indeed, the availability of new tools does not in any way invalidate
or weaken the conclusions of hundreds of studies that came before this one---quite the contrary.  Rather, the main message is that given the new
capabilities, it is more important now than ever to weigh the relative
benefits of sophistication against the drawback of increased difficulty and
time, whereas no such options existed before.

\subsection {IC 4710}

\begin {figure*} 
    \centerline {\includegraphics[angle=90,width=6.5truein]{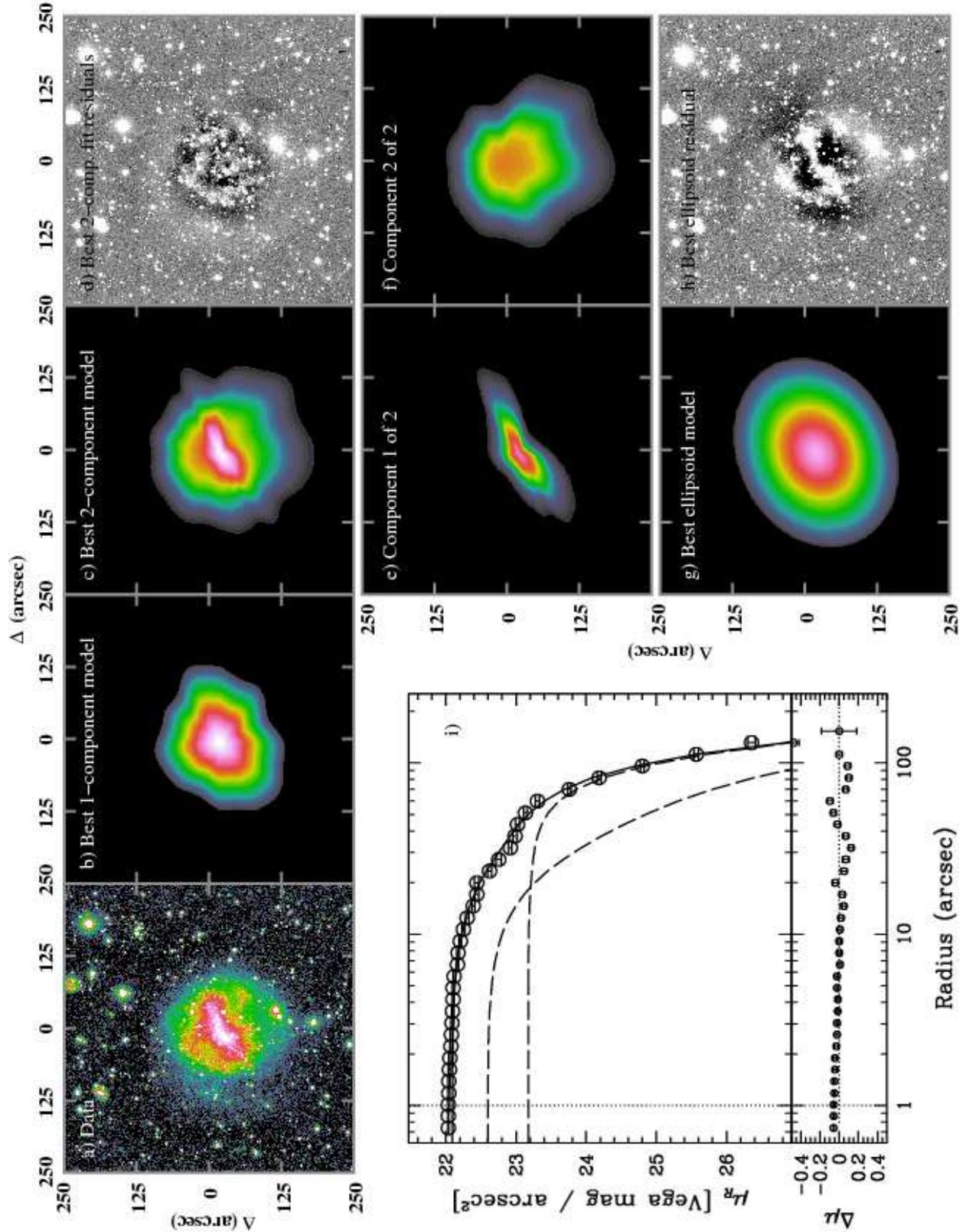}}

    \caption {Detailed analysis of IC 4710.  ({\it a}) Original data.  ({\it
    b}) Best single-component \sersic\ profile fit with Fourier modes $m=1$ to
    $m=10$. ({\it c}) Best two-component \sersic\ profile fit each with
    Fourier modes, corresponding to the parameters shown in
    Table~\ref{ic4710-table}. ({\it d}) Best-fit residuals.  ({\it e})
    Component 1 of 2 in the best-fit model of Panel ({\it c}).  ({\it f})
    Component 2 of 2 in the best-fit model.  ({\it g}) A traditional
    single-component ellipsoid fit. ({\it h}) Residuals from the model in
    Panel ({\it g}). ({\it i}) 1-D surface brightness profile.  The individual
    components are shown as dashed lines, and the solid line coursing through
    the data is the sum of the two components.  The lower panel shows the
    residuals of data $-$ model.}

    \vskip 0.15truein
    \label{ic4710}
\end {figure*}

\begin{deluxetable*}{|c|c|r|rrrrrrr|l|}
\tabletypesize{\scriptsize}
\tablewidth{0pt}
\tablecaption {IC 4710 Fitting Results}
\tablehead{  
  & \#  & --- sersic --- & $\Delta x$ [\arcsec] & $\Delta y$ [\arcsec] &  mag    &   $r_e$ [\arcsec]  &   $n$     &  $q$      &  $\theta_{\rm{PA}}$ [deg]   & Comments \\ 
  &     & fourier &    ---      & \multicolumn{2}{r}{mode: ampl. \& phase [deg]} & \multicolumn{2}{r}{mode: ampl. \& phase [deg]}  & \multicolumn{2}{r|}{mode: ampl. \& phase [deg]} & }
\startdata
Best fit  &  1 &    --- sersic   ---  &  $      0.00$  &  $      0.00$  &  $     13.71$  &  $     48.92$  &  $      0.55$  &  $      0.32$  &  $    -63.36$  & \\ 
          &    &                      &  $      0.16$  &  $      0.08$  &  $      0.00$  &  $      0.17$  &  $      0.00$  &  $      0.00$  &  $      0.07$  & \\ 
Inner     &    &             fourier  &       ---      & 1: $    0.16$  & 1: $  -97.40$  & 3: $    0.17$  & 3: $  -17.95$  & 4: $    0.06$  & 4: $   17.67$  & \\ 
component &    &                      &       ---      & 1: $    0.00$  & 1: $    1.18$  & 3: $    0.00$  & 3: $    0.21$  & 4: $    0.00$  & 4: $    0.32$  & \\ 
          &    &             fourier  &       ---      & 5: $    0.05$  & 5: $   18.37$  & 6: $   -0.06$  & 6: $   13.22$  & 7: $    0.03$  & 7: $   -1.20$  & \\ 
          &    &                      &       ---      & 5: $    0.00$  & 5: $    0.34$  & 6: $    0.00$  & 6: $    0.23$  & 7: $    0.00$  & 7: $    0.63$  & \\ 
          &    &             fourier  &       ---      & 8: $    0.05$  & 8: $   10.92$  & 9: $    0.01$  & 9: $    4.55$  & 10: $    0.03$  & 10: $   -6.68$  & \\ 
          &    &                      &       ---      & 8: $    0.00$  & 8: $    0.17$  & 9: $    0.00$  & 9: $    1.63$  & 10: $    0.00$  & 10: $    0.29$  & \\ 
Outer     &  2 &    --- sersic   ---  &  $      1.97$  &  $     26.28$  &  $     12.49$  &  $     57.24$  &  $      0.37$  &  $      0.90$  &  $     41.38$  & \\ 
Component &    &                      &  $      0.24$  &  $      0.19$  &  $      0.00$  &  $      0.09$  &  $      0.00$  &  $      0.00$  &  $      0.85$  & \\ 
          &    &             fourier  &       ---      & 1: $   -0.31$  & 1: $  -39.25$  & 3: $    0.03$  & 3: $   55.46$  & 4: $    0.03$  & 4: $  -27.65$  & \\ 
          &    &                      &       ---      & 1: $    0.00$  & 1: $    0.86$  & 3: $    0.00$  & 3: $    1.31$  & 4: $    0.00$  & 4: $    0.89$  & \\ 
          &    &             fourier  &       ---      & 5: $    0.04$  & 5: $  -15.73$  & 6: $    0.02$  & 6: $  -12.47$  & 7: $    0.01$  & 7: $   16.56$  & \\ 
          &    &                      &       ---      & 5: $    0.00$  & 5: $    0.87$  & 6: $    0.00$  & 6: $    1.15$  & 7: $    0.00$  & 7: $    1.74$  & \\ 
          &    &             fourier  &       ---      & 8: $   -0.03$  & 8: $  -13.49$  & 9: $    0.01$  & 9: $  -19.63$  & 10: $    0.02$  & 10: $  -15.64$  & \\ 
          &    &                      &       ---      & 8: $    0.00$  & 8: $    0.86$  & 9: $    0.00$  & 9: $    0.91$  & 10: $    0.00$  & 10: $    0.92$  & \\ 
          &    &     merit  &  \multicolumn{2}{c}{$\chi^2$ = 167438.77} &  \multicolumn{2}{c}{$N_{\rm{dof}}$ = 127966} &  $N_{\rm{free}}$ = 53 &  \multicolumn{2}{c|}{$\chi^2_\nu$ = 1.31} & \\ 
\colrule
Single    &  1 &    --- sersic   ---  &  $      0.00$  &  $      0.00$  &  $     12.15$  &  $     60.37$  &  $      0.69$  &  $      0.82$  &  $    -63.51$  & \\ 
component &    &                      &  $      0.06$  &  $      0.05$  &  $      0.00$  &  $      0.12$  &  $      0.00$  &  $      0.00$  &  $      0.31$  & \\ 
          &    &     merit  &  \multicolumn{2}{c}{$\chi^2$ = 247304.81} &  \multicolumn{2}{c}{$N_{\rm{dof}}$ = 128009} &  $N_{\rm{free}}$ = 10 &  \multicolumn{2}{c|}{$\chi^2_\nu$ = 1.93} & \\ 
\colrule
Single    &  1 &    --- sersic   ---  &  $      0.00$  &  $      0.00$  &  $     12.15$  &  $     59.00$  &  $      0.69$  &  $      0.83$  &  $    -64.43$  & \\ 
component &    &                      &  $      0.12$  &  $      0.09$  &  $      0.00$  &  $      0.09$  &  $      0.00$  &  $      0.00$  &  $      0.24$  & \\ 
with      &    &             fourier  &       ---      & 1: $   -0.05$  & 1: $   72.02$  & 3: $   -0.07$  & 3: $   27.55$  & 4: $   -0.02$  & 4: $   -7.37$  & \\ 
Fourier   &    &                      &       ---      & 1: $    0.00$  & 1: $    2.39$  & 3: $    0.00$  & 3: $    0.31$  & 4: $    0.00$  & 4: $    0.48$  & \\ 
modes     &    &             fourier  &       ---      & 5: $    0.02$  & 5: $    5.54$  & 6: $   -0.01$  & 6: $   -7.72$  & 7: $   -0.02$  & 7: $   15.40$  & \\ 
          &    &                      &       ---      & 5: $    0.00$  & 5: $    0.54$  & 6: $    0.00$  & 6: $    0.80$  & 7: $    0.00$  & 7: $    0.33$  & \\ 
          &    &             fourier  &       ---      & 8: $    0.01$  & 8: $    0.28$  & 9: $    0.01$  & 9: $    3.16$  & 10: $   -0.02$  & 10: $    1.12$  & \\ 
          &    &                      &       ---      & 8: $    0.00$  & 8: $    0.50$  & 9: $    0.00$  & 9: $    0.63$  & 10: $    0.00$  & 10: $    0.25$  & \\ 
          &    &     merit  &  \multicolumn{2}{c}{$\chi^2$ = 235140.44} &  \multicolumn{2}{c}{$N_{\rm{dof}}$ = 127991} &  $N_{\rm{free}}$ = 28 &  \multicolumn{2}{c|}{$\chi^2_\nu$ = 1.84} & 
\enddata

\tablecomments{Best-fitting parameters for IC 4710.  The meaning of the object
parameters is shown at the top for each model component.  The statistical
uncertainties for each model component, based on the covariance matrix of the
fit, are shown in the row underneath the best-fitting model parameters.  
Systematic uncertainties due to imperfect model-data match are typically
1\%--10\% for the fluxes, 10\%--20\% for the sizes, and 20\%--30\% for the \sersic\
index.  For the Fourier modes, the phase angle is relative to the major axis
of the light profile component.  Note that the sky parameters are not shown.
The ``{\it Best fit}'' parameters (top section) correspond to Panel ({\it c}) in
Figure~\ref{ic4710}, ``{\it Single component}'' parameters (middle section)
correspond to Panel ({\it g}), and ``{\it Single component with Fourier modes}''
parameters (bottom section) correspond to Panel ({\it b}).}

\label{ic4710-table}

\end{deluxetable*}

IC 4710 is an SB(s)m galaxy, which has a bar-like feature in the middle of a
roundish outer structure, as shown in the $R$-band image of
Figure~\ref{ic4710}, which comes from the CINGS (Carnegie-Irvine Nearby Galaxy
Survey) project\footnote{\tt
http://users.obs.carnegiescience.edu/lho/projects/CINGS/CINGS.html}.  Prior to
the analysis, we masked out the stars using the SExtractor software
\citep{bertin96}.  We analyze this galaxy using, for comparison, both one- and
two-component regular and higher order models with Fourier modes, shown in
Figures~\ref{ic4710}{\it b-i}.  The best-fit parameters are given in
Table~\ref{ic4710-table}, which illustrates three different sets of analysis
parameters:  best fit using two components (Figure~\ref{ic4710}{\it c}), a
model using just the traditional ellipsoid component (Figure~\ref{ic4710}{\it
g}), and the same single-component model with Fourier modes added
(Figure~\ref{ic4710}{\it b}).  Figure~\ref{ic4710}{\it i} shows the radial
surface brightness profile of the data and the individual subcomponents of the
best model.

There are several points to understand from comparing detailed and simple
analyses.  The best-fitting ellipsoid model (Figure~\ref{ic4710}{\it g}) is
oriented more parallel to the bar-like, higher surface brightness component
than the lower surface brightness body; however, the ellipsoid model is much
broader than the bar (Figure~\ref{ic4710}{\it e}).  This happens because a
single-component fit is a compromise between the various subcomponents of a
galaxy, and, as such, it reflects neither one perfectly.  Allowing the
azimuthal shape to change by adding 9 Fourier modes results in a shape shown
in Figure~\ref{ic4710}{\it b}.  Note that because the profile is restricted to
having a \sersic\ functional form in every direction radially from the peak,
the shape does not have complete freedom to take on any shape, as opposed to a
shapelet or wavelet-type Fourier inversion: it is merely a higher order
perturbation of the best-fitting ellipse.  Indeed, in comparing
single-component fit parameters in Table ~\ref{ic4710-table} for the two
models, the main \sersic\ structural parameters hardly budged, despite the
Fourier model having 18 more free parameters.  Therefore, the marginal returns
in using more free parameters is negligible in this situation when it comes to
the main \sersic\ structural parameters.  However, if the scientific interest
is to quantify the global symmetry, then the higher order modes are of
interest.

Another point of interest is how higher order models affect the accuracy of
the global photometry.  It is natural to expect when a model is unrealistic
for a galaxy that the photometry is also unreliable.  In Figure
~\ref{ic4710}{\it c}, it is evident that a two-component model is more
appropriate than the single-component fits of Figure ~\ref{ic4710}{\it b} and
\ref{ic4710}{\it g}.  However, when the flux of the two-component model is
summed, one finds that the difference with the single-component
fits is only 0.03 mag.  This and subsequent examples illustrate empirically
that the process of least-squares minimization using even na{\"i}ve ellipsoids is
often capable of providing accurate photometry to within 0.1 to 0.2 mag, even
if the galaxy shape departs from ellipsoid models quite drastically.

Lastly, Figures~\ref{ic4710}{\it e} and \ref{ic4710}{\it f}\ demonstrate that
it is quite feasible to unambiguously disentangle embedded components that
have different shapes, using higher order Fourier modes.  Despite there being
a large number of parameters, it is visually clear based on
Figures~\ref{ic4710}{\it e} and \ref{ic4710}{\it f} that parameter degeneracy
is not an issue, because the shapes of the components are quite different.  In
part, this is possible because of how Fourier modes are implemented in
\galfit:  the profile function is preserved in every direction radially from
the peak, even in situations where the shape is irregular, as in
Figure~\ref{ic4710}{\it e}.

\subsection {GEMS Edge-on Galaxy}

\begin {figure*} 
    \centerline {\includegraphics[angle=90,width=6.5truein]{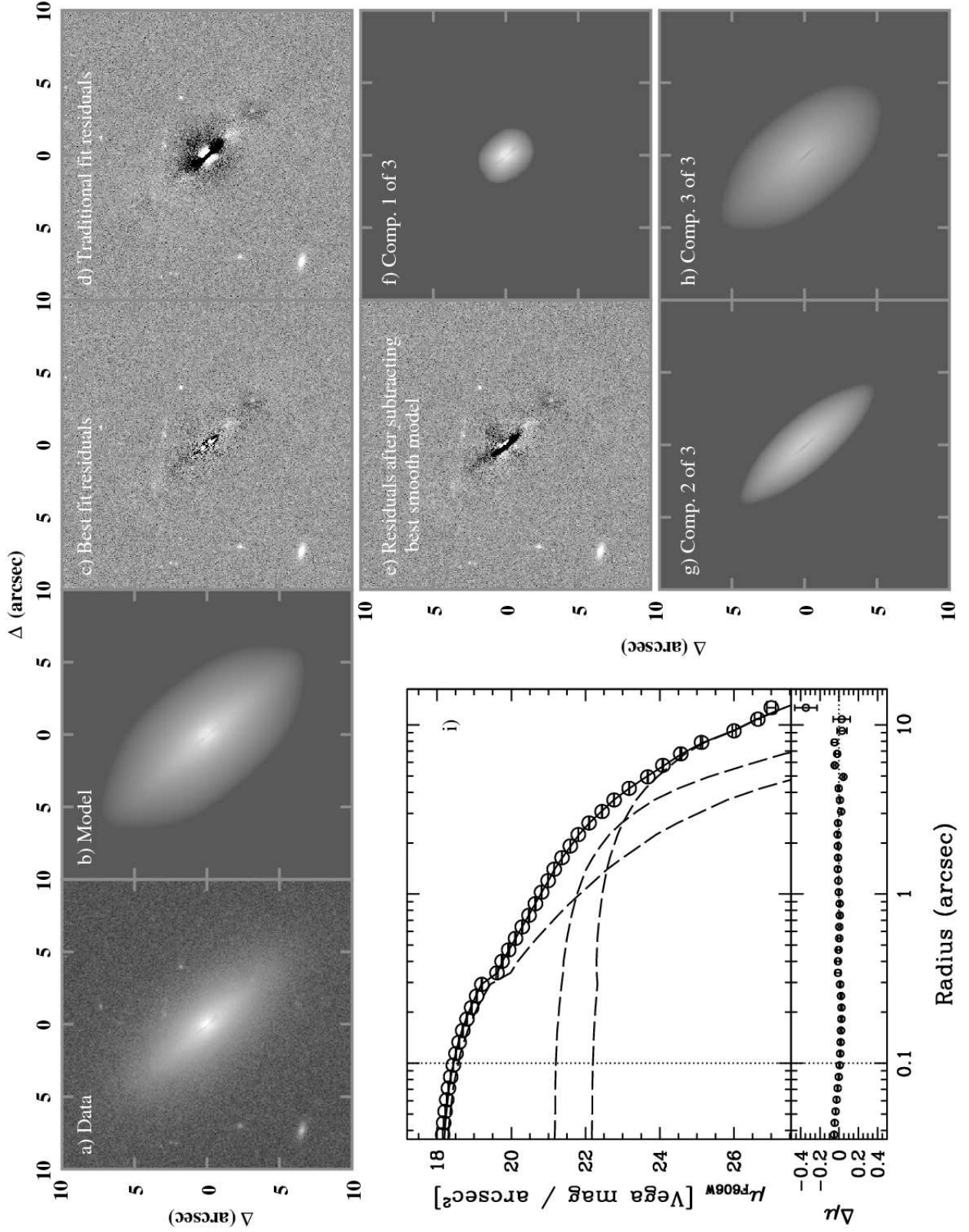}}

    \caption {Detailed analysis of an edge-on disk galaxy from GEMS.  ({\it
    a}) Original data.  ({\it b}) Best two-component \sersic\ profile fit each
    with Fourier modes, corresponding to the parameters shown in
    Table~\ref{gal07-table}. ({\it c}) Best-fit residuals.  ({\it d}) The fit
    residuals using traditional (i.e. purely ellipsoid) models without masking
    the dust lane.  ({\it e}) Residuals after subtracting the best traditional
    models, masking out the dust lane.  ({\it f}) The bulge component of the
    best-fit model. ({\it g}) The edge-on disk component of the best-fit
    model.  ({\it h}) The extended halo component of the best-fit model.
    ({\it i}) 1-D surface brightness profile.  The individual components are
    shown as dashed lines, and the solid line coursing through the data is the
    sum of the different components.  The lower panel shows the residuals of
    data $-$ model.}

    \vskip 0.15truein
    \label{gems}
\end {figure*}

\begin{deluxetable*}{|c|c|r|rrrrrrr|l|}
\tabletypesize{\scriptsize}
\tablewidth{0pt}
\tablecaption {GEMS Disk Galaxy Fitting Results}
\tablehead{  
  & \#  & --- sersic2 --- & $\Delta x$ [\arcsec] & $\Delta y$ [\arcsec] &  mag/arcsec$^2$  &   $r_e$ [\arcsec]  &   $n$     &  $q$      &  $\theta_{\rm{PA}}$ [deg]   &  Comments \\ 
  &     & fourier &    ---      & \multicolumn{2}{r}{mode: ampl. \& phase [deg]} & \multicolumn{2}{r}{mode: ampl. \& phase [deg]}  & \multicolumn{2}{r|}{mode: ampl. \& phase [deg]} & \\
  & \#  & --- radial ---  & $\Delta x$ [\arcsec] & $\Delta y$ [\arcsec] & --- & $r_{\rm break}$ [\arcsec] & $\Delta r_{\rm soft}$  & $q$  & $\theta_{\rm{PA}}$ [deg] & }
\startdata
Best      &  1 &   --- sersic2 / ---  &  $      0.00$  &  $      0.00$  &  $     19.80$  &  $      0.40$  &  $      1.60$  &  $      0.72$  &  $     44.00$  & Trunc. by comp.\\ 
fit       &    &                      &  $      0.00$  &  $      0.00$  &  $      0.01$  &  $      0.00$  &  $      0.01$  &  $      0.00$  &  $      0.16$  & inner: 4       \\ 
          &    &             fourier  &       ---      & 1: $   -0.04$  & 1: $  -92.16$  & 3: $   -0.00$  & 3: $  -51.36$  & 4: $    0.03$  & 4: $   -6.61$  & \\ 
          &    &                      &       ---      & 1: $    0.00$  & 1: $    3.23$  & 3: $    0.00$  & 3: $   30.65$  & 4: $    0.00$  & 4: $    0.53$  & \\ 
          &    &             fourier  &       ---      & 5: $   -0.01$  & 5: $   -5.74$  & 6: $    0.01$  & 6: $   -7.31$  &       ---      &       ---      & \\ 
          &    &                      &       ---      & 5: $    0.00$  & 5: $    2.35$  & 6: $    0.00$  & 6: $    1.74$  &       ---      &       ---      & \\ 
          &  2 &   --- sersic2 / ---  &  $\{  0.00\}$  &  $\{  0.00\}$  &  $     22.19$  &  $      2.29$  &  $      0.85$  &  $      0.31$  &  $     41.30$  & Trunc. by comp.\\ 
          &    &                      &  $\{  0.00\}$  &  $\{  0.00\}$  &  $      0.02$  &  $      0.01$  &  $      0.01$  &  $      0.00$  &  $      0.03$  & inner: 4       \\ 
          &    &             fourier  &       ---      & 1: $   -0.04$  & 1: $  -24.97$  & 3: $    0.02$  & 3: $   26.37$  & 4: $   -0.02$  & 4: $   -0.04$  & \\ 
          &    &                      &       ---      & 1: $    0.00$  & 1: $    1.82$  & 3: $    0.00$  & 3: $    1.03$  & 4: $    0.00$  & 4: $    0.79$  & \\ 
          &    &             fourier  &       ---      & 5: $    0.00$  & 5: $    6.19$  & 6: $   -0.01$  & 6: $    6.62$  &       ---      &       ---      & \\ 
          &    &                      &       ---      & 5: $    0.00$  & 5: $    5.63$  & 6: $    0.00$  & 6: $    1.35$  &       ---      &       ---      & \\ 
          &  3 &   --- sersic2 / ---  &  $\{  0.00\}$  &  $\{  0.00\}$  &  $     23.92$  &  $      4.45$  &  $      1.08$  &  $      0.49$  &  $     41.06$  & Trunc. by comp.\\ 
          &    &                      &  $\{  0.00\}$  &  $\{  0.00\}$  &  $      0.03$  &  $      0.05$  &  $      0.02$  &  $      0.00$  &  $      0.10$  & inner: 4       \\ 
          &    &             fourier  &       ---      & 1: $    0.04$  & 1: $    4.59$  & 3: $    0.01$  & 3: $   -3.37$  & 4: $   -0.01$  & 4: $   40.55$  & \\ 
          &    &                      &       ---      & 1: $    0.00$  & 1: $    1.16$  & 3: $    0.00$  & 3: $    1.99$  & 4: $    0.00$  & 4: $    3.94$  & \\ 
          &    &             fourier  &       ---      & 5: $    0.01$  & 5: $   -7.19$  & 6: $   -0.01$  & 6: $   -3.25$  &       ---      &       ---      & \\ 
          &    &                      &       ---      & 5: $    0.00$  & 5: $    1.40$  & 6: $    0.00$  & 6: $    0.73$  &       ---      &       ---      & \\ 
          &  4 &    --- radial   ---  &  $      0.02$  &  $     -0.14$  &       ---      &  $      1.48$  &  $      1.48$  &  $      0.09$  &  $     41.38$  & Truncates comp.\\ 
          &    &                      &  $      0.00$  &  $      0.00$  &       ---      &  $      0.01$  &  $      0.02$  &  $      0.00$  &  $      0.05$  & inner: 1 2 3   \\ 
          &    &             fourier  &       ---      & 1: $   -0.12$  & 1: $ -156.83$  & 3: $    0.10$  & 3: $  -20.62$  & 4: $    0.19$  & 4: $    9.20$  & \\ 
          &    &                      &       ---      & 1: $    0.00$  & 1: $    1.23$  & 3: $    0.00$  & 3: $    0.52$  & 4: $    0.00$  & 4: $    0.15$  & \\ 
          &    &             fourier  &       ---      & 5: $    0.10$  & 5: $    2.17$  & 6: $    0.14$  & 6: $    6.62$  &       ---      &       ---      & \\ 
          &    &                      &       ---      & 5: $    0.00$  & 5: $    0.19$  & 6: $    0.00$  & 6: $    0.10$  &       ---      &       ---      & \\ 
          &    &     merit  &  \multicolumn{2}{c}{$\chi^2$ = 1474348.38} &  \multicolumn{2}{c}{$N_{\rm{dof}}$ = 1435846} &  $N_{\rm{free}}$ = 64 &  \multicolumn{2}{c|}{$\chi^2_\nu$ = 1.03} & \\
\colrule
Tradit.   &  1 &    --- sersic2  ---  &  $      0.00$  &  $      0.00$  &  $     19.98$  &  $      0.38$  &  $      1.32$  &  $      0.74$  &  $     42.75$  & \\ 
ellipsoid &    &                      &  $      0.00$  &  $      0.00$  &  $      0.00$  &  $      0.00$  &  $      0.01$  &  $      0.00$  &  $      0.31$  & \\ 
model     &  2 &    --- sersic2  ---  &  $\{  0.00\}$  &  $\{  0.00\}$  &  $     22.78$  &  $      2.57$  &  $      0.85$  &  $      0.25$  &  $     41.21$  & \\ 
with      &    &                      &  $\{  0.00\}$  &  $\{  0.00\}$  &  $      0.03$  &  $      0.02$  &  $      0.01$  &  $      0.00$  &  $      0.06$  & \\ 
dust      &  3 &    --- sersic2  ---  &  $\{  0.00\}$  &  $\{  0.00\}$  &  $     23.12$  &  $      3.34$  &  $      1.77$  &  $      0.49$  &  $     41.30$  & \\ 
masking   &    &                      &  $\{  0.00\}$  &  $\{  0.00\}$  &  $      0.04$  &  $      0.04$  &  $      0.03$  &  $      0.00$  &  $      0.08$  & \\ 
          &    &     merit  &  \multicolumn{2}{c}{$\chi^2$ = 1478767.62} &  \multicolumn{2}{c}{$N_{\rm{dof}}$ = 1434511} &  $N_{\rm{free}}$ = 18 &  \multicolumn{2}{c|}{$\chi^2_\nu$ = 1.03} & 
\enddata

\tablecomments{Best-fitting parameters for an edge-on disk galaxy in GEMS.
See Table~\ref{ic4710-table} for details.  The curly braces (\{...\}) around
parameters indicate that they are coupled relative to the first component.
Note that the flux amplitude of {\it sersic2} is normalized to the surface
brigtness at $r_{\rm e}$, as defined in Equation~\ref{eqn:sersic2}.  The ``{\it
Best fit}'' parameters (top section) correspond to Panel ({\it b}) in
Figure~\ref{gems}, ``{\it Traditional ellipsoid model}'' parameters (bottom
section) produce residuals shown in Panel ({\it c}), and the model is not
shown.}

\label{gal07-table}

\end{deluxetable*}

This edge-on galaxy (Figure~\ref{gems}, Table~\ref{gal07-table}) comes from
the GEMS (Galaxy Evolution from Morphology and SED, \citep{rix04}) project,
which is an {\it HST}\ imaging survey of the {\it Chandra}\ Deep Field-South.
Belying a benign morphological appearance is a dust lane
(Figure~\ref{gems}{\it e}) that courses through the center, complicating the
traditional ellipsoid fitting technique.

The analysis of even this simple object can be quite involved.  The
best-fitting model involves three components: a fairly compact bulge, an
edge-on disk component, and an puffy stellar halo enveloping both.  Since the
halo component is more luminous than the bulge component, a two-component
model fit would naturally ascribe the halo component to the bulge, despite
there being a distinctly rounder component at the center.  Like the previous
example, each of the three components (Figure~\ref{gems}{\it f-h}) is modified
by Fourier modes.  Furthermore, the best fit includes an actual model for the
dust lane (component 4, Table~\ref{gal07-table}).  The dust lane is modeled by
an inner truncation function as discussed in Section~\ref{sect:truncation}.

A truncation model is shown as a model ``component'' in the fit; it is unique
because it is not a light profile model, and one cannot generate an image to
see what it looks like.  Instead, its influence is to be seen on all the light
profile models (i.e. components 1--3; Figure~\ref{gems}{\it f-h}), where it
reduces the light by the same fraction for all components.  In every other
way, the truncation function behaves exactly like a light profile model:  it
can have its own centroid (or not), and it can be modified by Fourier modes,
as shown in Table~\ref{gal07-table}.  The benefit of using a single truncation
model for all three light profile models is not only to reduce the degrees of
freedom, but it is also physically motivated because foreground dust
attenuates all background light sources by an equal fractional amount.
Nevertheless, if desired, it is also possible to allow each component to be
attenuated differently.

This example also demonstrates how the result of the analysis depends on the
input prior of the model.  In the fit using traditional ellipsoid parameters,
a mask is used to minimize the effect of the dust on the analysis.  Yet, the
effects cannot be completely removed.  As shown in Table~\ref{gal07-table},
the inclusion of the truncation model can significantly affect the structural
parameters:  the surface brightnesses can differ by 0.8 mag arcsec$^{-2}$, and
the sizes by 10\%--20\%, even in this seemingly uncomplicated situation.
Moreover, the differences far surpass the statistical uncertainties shown in
Table~\ref{gal07-table}.  To the extent that it is not possible to judge which
model is more physically correct, both measurements ought to be treated as
equally valid.  In that situation, the uncertainties, due entirely to model
assumptions, are roughly $\sim 0.4$ mag in surface brightness and $\sim10\%$
in size.

\subsection {Arp 147}

\begin {figure*} 
    \centerline {\includegraphics[angle=90,width=6.5truein]{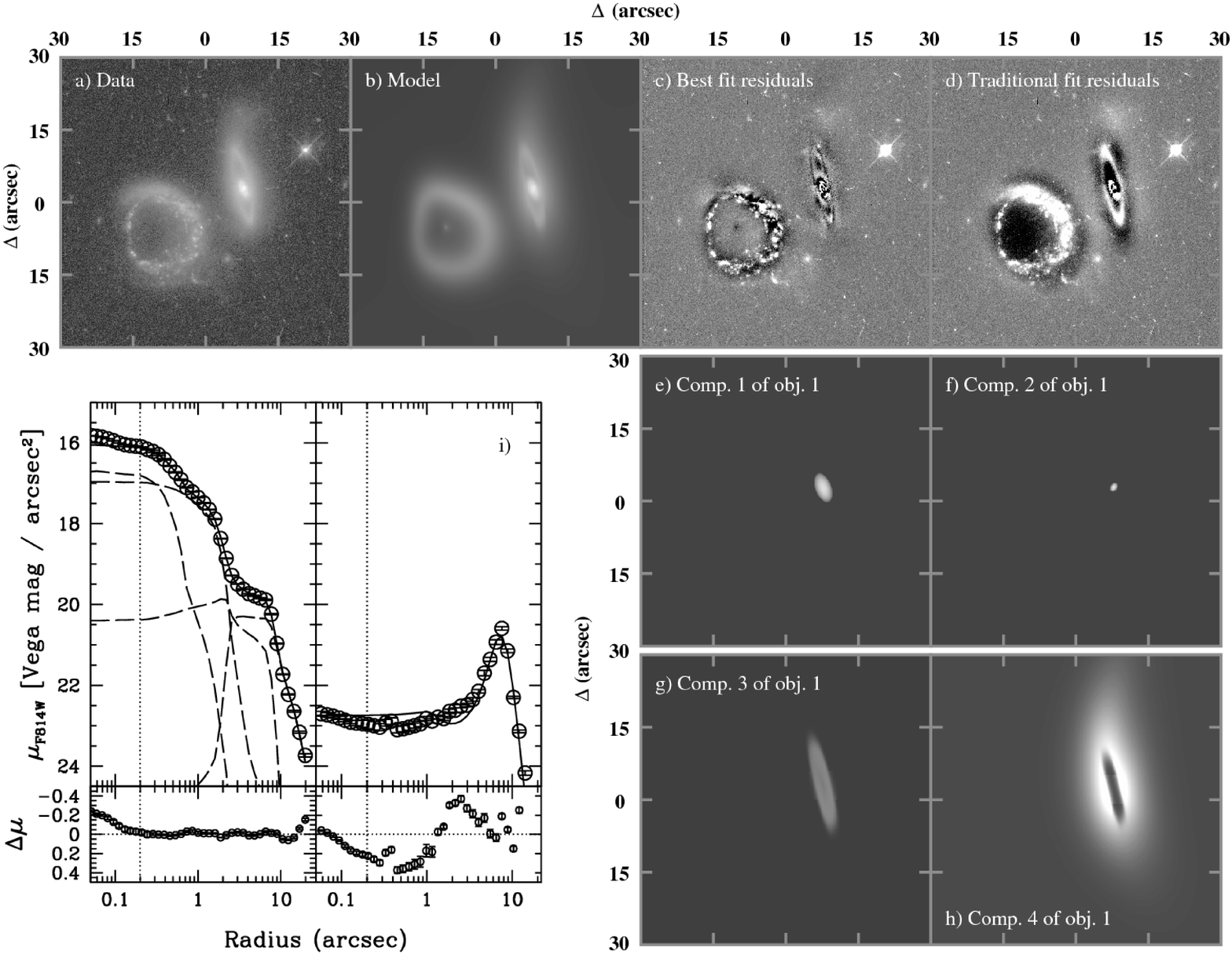}}

    \caption {Detailed analysis of Arp 147. ({\it a}) Original data.  ({\it b})
    Best \sersic\ profile fits of the two galaxies, all with Fourier modes,
    corresponding to the parameters shown in Table~\ref{arp147-table}. ({\it
    c}) Best-fit residuals.  ({\it d}) The fit residuals using traditional,
    i.e.  axisymmetric ellipsoidal model components.  ({\it e}) The bulge
    component of the right-hand galaxy in Panel~({\it b}).  ({\it f}) The inner
    fine-structure component of the best-fit model.  ({\it g}) The ring
    component of the best-fit model. ({\it h}) The extended tidal-feature-like
    component of the best-fit model.  ({\it i}) 1-D surface brightness profile
    of the two galaxies.  The individual components are shown as dashed lines,
    and the solid line coursing through the data is the sum of the different
    components.  The lower panel shows the residuals of data $-$ model.}

    \vskip 0.15truein
    \label{arp147}
\end {figure*}

\begin{deluxetable*}{|c|c|r|rrrrrrr|l|}
\tabletypesize{\scriptsize}
\tablewidth{0pt}
\tablecaption {Arp 147 Fitting Results}
\tablehead{  
  & \#  & --- sersic --- & $\Delta x$ [\arcsec] & $\Delta y$ [\arcsec] &  mag    &   $r_e$ [\arcsec]  &   $n$     &  $q$      &  $\theta_{\rm{PA}}$ [deg]   & Comments \\ 
  & \#  & --- sersic3 --- & $\Delta x$ [\arcsec] & $\Delta y$ [\arcsec] &  mag/arcsec$^2$  &   $r_e$ [\arcsec]  &   $n$     &  $q$      &  $\theta_{\rm{PA}}$ [deg]   &  \\ 
  &     & fourier &    ---      & \multicolumn{2}{r}{mode: ampl. \& phase [deg]} & \multicolumn{2}{r}{mode: ampl. \& phase [deg]}  & \multicolumn{2}{r|}{mode: ampl. \& phase [deg]} & \\
  &     & bending &    ---      & \multicolumn{2}{r}{mode: amplitude [\arcsec]} & \multicolumn{2}{r}{mode: amplitude [\arcsec]}  & \multicolumn{2}{r|}{mode: amplitude [\arcsec]} & \\
  & \#  & --- radial --- & --- & --- & --- & $r_{\rm break}$ [\arcsec] & $\Delta r_{\rm soft}$  & $q$  & $\theta_{\rm{PA}}$ [deg] & }
\startdata
Galaxy 1  &  1 &    --- sersic   ---  &  $      0.00$  &  $      0.00$  &  $     15.49$  &  $      1.26$  &  $      0.47$  &  $      0.50$  &  $    194.70$  & Bulge. \\ 
          &    &                      &  $      0.00$  &  $      0.00$  &  $      0.00$  &  $      0.00$  &  $      0.00$  &  $      0.00$  &  $      0.05$  & Inner fine\\ 
          &  2 &    --- sersic   ---  &  $      0.07$  &  $      0.13$  &  $     17.39$  &  $      0.35$  &  $      0.43$  &  $      0.60$  &  $    150.75$  & structure. \\ 
          &    &                      &  $      0.00$  &  $      0.00$  &  $      0.01$  &  $      0.00$  &  $      0.01$  &  $      0.00$  &  $      0.39$  & \\ 
          &  3 &   --- sersic3 / ---  &  $     -0.08$  &  $      0.33$  &  $     23.84$  &  $      1.21$  &  $      0.96$  &  $      0.18$  &  $    187.42$  & Trunc. by comp.\\ 
          &    &                      &  $      0.01$  &  $      0.03$  &  $      0.02$  &  $      0.02$  &  $      0.01$  &  $      0.00$  &  $      0.03$  & inner: 5 (ring).  \\ 
          &    &             fourier  &       ---      & 1: $    0.03$  & 1: $   49.11$  & 3: $    0.02$  & 3: $   15.69$  & 4: $    0.04$  & 4: $    4.24$  & \\ 
          &    &                      &       ---      & 1: $    0.00$  & 1: $    9.08$  & 3: $    0.00$  & 3: $    1.04$  & 4: $    0.00$  & 4: $    0.45$  & \\ 
          &  4 &   --- sersic3 / ---  &  $      0.09$  &  $     -0.34$  &  $     22.04$  &  $      3.81$  &  $      1.94$  &  $      0.42$  &  $    184.08$  & Trunc. by comp.\\ 
          &    &                      &  $      0.01$  &  $      0.02$  &  $      0.01$  &  $      0.04$  &  $      0.03$  &  $      0.00$  &  $      0.04$  & inner: 5  (Tidal     \\ 
          &    &             fourier  &       ---      & 1: $    0.16$  & 1: $   21.49$  & 3: $    0.09$  & 3: $   18.36$  & 4: $   -0.01$  & 4: $   16.13$  & feature). \\ 
          &    &                      &       ---      & 1: $    0.00$  & 1: $    0.82$  & 3: $    0.00$  & 3: $    0.17$  & 4: $    0.00$  & 4: $    1.55$  & \\ 
          &    &             bending  &       ---      & 2: $   -0.14$  &       ---      &       ---      &       ---      &       ---      &       ---      & \\ 
          &    &                      &       ---      & 2: $    0.00$  &       ---      &       ---      &       ---      &       ---      &       ---      & \\ 
          &  5 &    --- radial   ---  &       ---      &       ---      &       ---      &  $     10.94$  &  $      6.00$  &  $      0.18$  &  $    187.61$  & Truncates comp.\\ 
          &    &                      &       ---      &       ---      &       ---      &  $      0.02$  &  $      0.06$  &  $      0.00$  &  $      0.03$  & inner: 3 4     \\ 
          &    &             fourier  &       ---      & 1: $    0.05$  & 1: $   39.00$  & 3: $    0.02$  & 3: $   19.88$  & 4: $    0.04$  & 4: $    3.88$  & \\ 
          &    &                      &       ---      & 1: $    0.00$  & 1: $    4.97$  & 3: $    0.00$  & 3: $    1.18$  & 4: $    0.00$  & 4: $    0.39$  & \\ 
Galaxy 2  &  6 &   --- sersic3 / ---  &  $    -18.29$  &  $     -7.93$  &  $     22.14$  &  $      0.78$  &  $      1.85$  &  $      0.79$  &  $    187.91$  & Trunc. by comp.\\ 
          &    &                      &  $      0.01$  &  $      0.01$  &  $      0.00$  &  $      0.01$  &  $      0.01$  &  $      0.00$  &  $      0.12$  & inner: 7  (Ring) \\ 
          &    &             fourier  &       ---      & 1: $    0.23$  & 1: $ -113.97$  & 3: $    0.07$  & 3: $   15.27$  & 4: $   -0.02$  & 4: $   23.33$  & mag$_{\rm tot}=14.90$\\ 
          &    &                      &       ---      & 1: $    0.00$  & 1: $    0.53$  & 3: $    0.00$  & 3: $    0.12$  & 4: $    0.00$  & 4: $    0.39$  & \\ 
          &  7 &    --- radial   ---  &       ---      &       ---      &       ---      &  $     10.77$  &  $      6.08$  &  $      0.82$  &  $    195.16$  & Truncates comp.\\ 
          &    &                      &       ---      &       ---      &       ---      &  $      0.01$  &  $      0.01$  &  $      0.00$  &  $      0.19$  & inner: 6       \\ 
          &    &             fourier  &       ---      & 1: $    0.17$  & 1: $ -149.22$  & 3: $    0.07$  & 3: $    4.98$  & 4: $    0.02$  & 4: $  -31.52$  & \\ 
          &    &                      &       ---      & 1: $    0.00$  & 1: $    0.99$  & 3: $    0.00$  & 3: $    0.17$  & 4: $    0.00$  & 4: $    0.38$  & \\ 
          &    &     merit  &  \multicolumn{2}{c}{$\chi^2$ = 714735.38} &  \multicolumn{2}{c}{$N_{\rm{dof}}$ = 357760} &  $N_{\rm{free}}$ = 77 &  \multicolumn{2}{c|}{$\chi^2_\nu$ = 2.00} & \\ 
\colrule
Tradit.   &  1 &    --- sersic   ---  &  $      0.00$  &  $      0.00$  &  $     15.33$  &  $      1.07$  &  $      0.90$  &  $      0.62$  &  $    193.49$  & Bulge. \\ 
ellipsoid &    &                      &  $      0.00$  &  $      0.00$  &  $      0.00$  &  $      0.00$  &  $      0.00$  &  $      0.00$  &  $      0.11$  & \\ 
model     &  2 &    --- sersic   ---  &  $     -0.12$  &  $      0.55$  &  $     14.86$  &  $      6.63$  &  $      0.43$  &  $      0.34$  &  $    184.20$  & Disk. \\ 
          &    &                      &  $      0.00$  &  $      0.01$  &  $      0.00$  &  $      0.01$  &  $      0.00$  &  $      0.00$  &  $      0.04$  & \\ 
Galaxy 2  &  3 &    --- sersic   ---  &  $    -15.61$  &  $     -8.66$  &  $     15.09$  &  $      8.28$  &  $      0.12$  &  $      0.80$  &  $    201.65$  & Ring \\ 
          &    &                      &  $      0.01$  &  $      0.01$  &  $      0.00$  &  $      0.01$  &  $      0.00$  &  $      0.00$  &  $      0.29$  & galaxy. \\ 
          &    &     merit  &  \multicolumn{2}{c}{$\chi^2$ = 1435193.25} &  \multicolumn{2}{c}{$N_{\rm{dof}}$ = 357813} &  $N_{\rm{free}}$ = 24 &  \multicolumn{2}{c|}{$\chi^2_\nu$ = 4.01} & 
\enddata

\tablecomments{Best-fitting parameters for Arp 147.  See
Table~\ref{ic4710-table} for details.  Note that the flux amplitude of {\it
sersic3} is normalized to the surface brigtness at $r_{\rm break}$, as defined
in Equation~\ref{eqn:sersic3}, whereas {\it sersic} magnitude means the total flux.
The ``{\it Best fit}'' parameters (top section) correspond to Panel ({\it b}) in
Figure~\ref{arp147}, ``{\it Traditional ellipsoid model}'' parameters (bottom
section) produce residuals shown in Panel ({\it d}), and the model is not
shown.  The free parameters for the sky are not listed.}

\label{arp147-table}

\end{deluxetable*}

The {\it HST}/F814W image of the field Arp 147 contains two ring galaxies
(Figure~\ref{arp147}, Table~\ref{arp147-table}), one of which has a bulge-like
component with a tidally disturbed outer region (Galaxy 1), and the other is a
pure ring (Galaxy 2).  The best-fitting model for Galaxy 2 is a
single-component ring, modified by Fourier modes, as seen in
Figure~\ref{arp147}{\it b}, whereas Galaxy 1 requires two ring components, a
bulge, and an inner fine-structure component (Figures~\ref{arp147}{\it e-h}).
The fine-structure component of Galaxy 1 can be easily seen in the surface
brightness profile as an upturn within $r=0\farcs2$ of Figure~\ref{arp147}{\it
i} (left).  In addition, the tidal component is slightly bent, which is
modeled elegantly using the bending modes of Equation~\ref{eqn:bending}.  As
in the case of a dust lane, the ring model comes about by truncating the inner
region of a pure \sersic\ profile (see Section~\ref{sect:truncation}.  The
only difference here is that the truncation radii are a larger fraction of the
galaxy size.  Whereas for the edge-on galaxy, it makes more sense to normalize
the flux at the effective radius (Equation~\ref{eqn:sersic2}), for ring
galaxies, normalizing the flux at the break radius
(Equation~\ref{eqn:sersic3}) is more intuitive, because it is closer to the
peak of the profile model.  In fact, the peak of the ring is about half-way
between $r_{\rm break}$ and $r_{\rm break} + \Delta r_{\rm soft}$, but the
exact location depends on the profile type.

It is again instructive to compare a traditional fit using simple
\sersic\ ellipsoid models (Table~\ref{arp147-table}, bottom) with more
sophisticated analysis (Table~\ref{arp147-table}, top).  In terms of the total
flux for Galaxy 1, the magnitude of the most sophisticated model is $m=14.18$,
compared to $m=14.32$ for a model based on classical ellipsoids.
Interestingly, a single-component fit (not shown) to Galaxy 1 yields a
magnitude of $m=14.15$.  For Galaxy 2, we know from the outset that classical
ellipsoid models are entirely inappropriate to use.  Yet, despite every reason
to believe that the photometry would be inaccurate, we find that the total
flux of the traditional ellipsoid fit is only 0.2 magnitude different from the
most realistic ring model.  These two examples show once again that a
single-component \sersic\ ellipsoid fit to complicated galaxies can produce
quite accurate measurement of the total flux.

It is sometimes desirable to conduct bulge-to-disk (B/D) decompositions, and
Galaxy 1 is an ideal candidate to conduct a comparison.  In the traditional
ellipsoid model (Table~\ref{arp147-table}, bottom), the B/D ratio is 0.65.
The more sophisticated model (Table~\ref{arp147-table}, top) requires summing
the ring+tidal feature components to obtain the disk component, which yields
14.65 mag, thus a B/D ratio of 0.54.  In this situation, most of the
differences arise from measuring the disk component, which differs by 0.2
mag, whereas the bulge component is quite robust, with a difference of only
0.01 mag.

It is also of interest to understand how the structural parameters are
affected by different model choices, in particular for the ring Galaxy 2.
Whereas the effective radius for the ring model is only 0\farcs78, for the
ellipsoid model it is 8\farcs28.  This is understandable, bearing in mind that
the ring has a radius of nearly 8\arcsec.  To a classical \sersic\ profile,
the galaxy appears to have a very flat (in fact, a deficit) core, which leads
to a low \sersic\ index of $n=0.12$.  As most of the flux is at 8\arcsec,
beyond which the ring flux quickly fades, the ring radius is closely related
to the effective radius for a classical \sersic\ model.  For the
inner-truncated ring model (component 7), however, the physical size of the
ring is captured by the break radius $r_{\rm break}$ parameter, whereas the
$r_e$ term no longer has the classical meaning of the effective radius (i.e.
half the light is within $r_e$).  Instead, $r_e$ for component 7 is
essentially an exponentially declining scale length parameter, given by
Equation~\ref{eqn:sersic3}.  As the flux dies away quickly beyond the peak, as
shown in Figure~\ref{arp147}{\it i}, the scale length $r_e$ for the ring model
must therefore be quite small.  The differences in the $r_e$ parameter between
the traditional model and the truncated model are therefore only due to
definitions, and not due to systematic or random measurement uncertainties.

\subsection{M51}

\begin {figure*} \centerline
    {\includegraphics[angle=90,width=6.5truein]{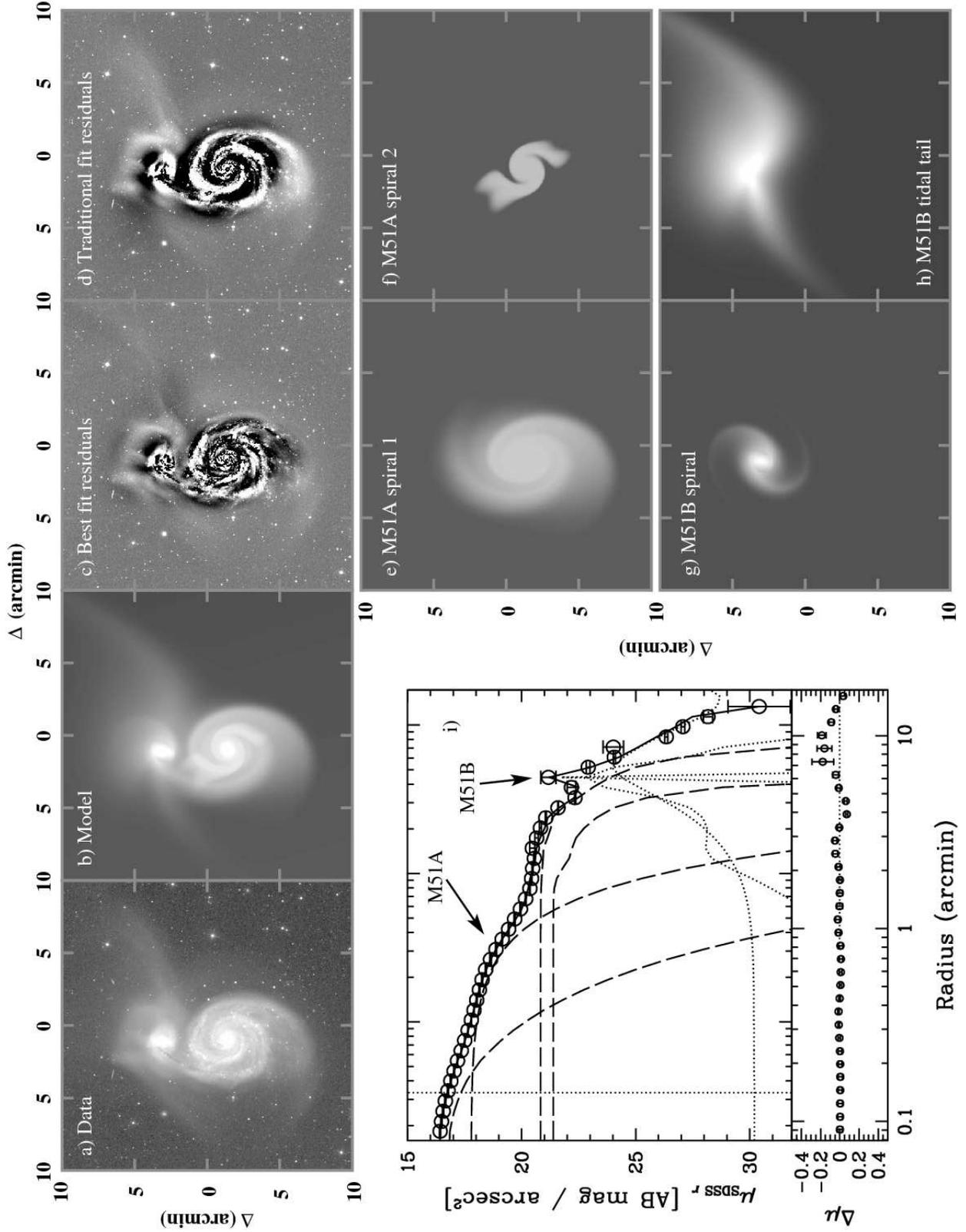}} 
    
    \caption {Detailed analysis of M51.  ({\it a}) Original data.  ({\it b})
    Best \sersic\ profile fits of the M51A and B, all with Fourier modes,
    corresponding to the parameters shown in Table~\ref{m51-table}. ({\it c})
    Best-fit residuals.  ({\it d}) The fit residuals using traditional,
    axisymmetric, ellipsoidal model components.  ({\it e}) The extended
    grand-design spiral component of M51A model in Panel~({\it b}).  ({\it f})
    The inner fine-structure spiral component of the best-fit model.  ({\it g})
    The spiral component of M51B. ({\it h}) The extended tidal feature-like
    component of M51B, using simultaneous bending and Fourier modes.  A bulge
    component is present but not shown in the figures of M51A and B.  {\it
    i}) 1-D surface brightness profile of the two galaxies.  The individual
    components are shown as dashed lines, and the solid line coursing through
    the data is the sum of the different components.  The lower panel shows the
    residuals of data $-$ model.}

    \vskip 0.15truein \label{m51}
\end {figure*}

\begin{deluxetable*}{|c|c|r|rrrrrrr|l|}
\tabletypesize{\scriptsize}
\tablewidth{0pt}
\tablecaption {M51 Fitting Results}
\tablehead{  
  & \#  & --- sersic --- & $\Delta x$ [\arcmin] & $\Delta y$ [\arcmin] &  mag    &   $r_e$ [\arcmin]  &   $n$     &  $q$      &  $\theta_{\rm{PA}}$ [deg]   & Comments \\ 
  &     & power &    ---      &    $r_{\rm{in}}$ [\arcmin]&  $r_{\rm{out}}$ [\arcmin] &  $\theta_{\rm{rot}}$ [deg] & $\alpha$   &  $\theta_{\rm{incl}}$ [deg] &  $\theta_{\rm{sky}}$ [deg]& \\ 
  &     & fourier &    ---      & \multicolumn{2}{r}{mode: ampl. \& phase [deg]} & \multicolumn{2}{r}{mode: ampl. \& phase [deg]}  & \multicolumn{2}{r|}{mode: ampl. \& phase [deg]} & \\
  &     & bending &    ---      & \multicolumn{2}{r}{mode: amplitude [\arcmin]} & \multicolumn{2}{r}{mode: amplitude [\arcmin]}  & \multicolumn{2}{r|}{mode: amplitude [\arcmin]} & }
\startdata
Best      &  1 &    --- sersic   ---  &  $      0.00$  &  $      0.00$  &  $     13.09$  &  $      0.04$  &  $      1.18$  &  $      0.91$  &  $    -15.25$  & Compound \\ 
fit       &    &                      &  $      0.00$  &  $      0.00$  &  $      0.04$  &  $      0.00$  &  $      0.04$  &  $      0.01$  &  $      4.74$  & bulge.\\ 
          &  2 &    --- sersic   ---  &  $\{  0.00\}$  &  $\{  0.00\}$  &  $     10.49$  &  $      0.26$  &  $      0.67$  &  $      0.88$  &  $    -65.31$  & Compound\\ 
M51A      &    &                      &  $\{  0.00\}$  &  $\{  0.00\}$  &  $      0.00$  &  $      0.00$  &  $      0.01$  &  $      0.00$  &  $      0.66$  & bulge.\\ 
          &  3 &    --- sersic   ---  &  $\{  0.00\}$  &  $\{  0.00\}$  &  $      8.50$  &  $      2.78$  &  $      0.35$  &  $      0.75$  &  $    -87.22$  & Compound\\ 
          &    &                      &  $\{  0.00\}$  &  $\{  0.00\}$  &  $      0.00$  &  $      0.00$  &  $      0.00$  &  $      0.00$  &  $     39.53$  & spiral.\\ 
          &    &               power  &       ---      &  $     -1.29$  &  $      4.28$  &  $   -718.11$  &  $      0.29$  &  $     40.42$  &  $    -82.20$  & \\ 
          &    &                      &       ---      &  $      0.20$  &  $      0.03$  &  $     41.39$  &  $      0.02$  &  $      0.05$  &  $      0.10$  & \\ 
          &    &             fourier  &       ---      & 1: $   -0.07$  & 1: $  109.10$  & 3: $    0.03$  & 3: $    4.07$  & 4: $    0.02$  & 4: $  -36.57$  & \\ 
          &    &                      &       ---      & 1: $    0.00$  & 1: $    0.44$  & 3: $    0.00$  & 3: $    0.43$  & 4: $    0.00$  & 4: $    0.35$  & \\ 
          &    &             fourier  &       ---      & 5: $    0.02$  & 5: $   24.34$  &       ---      &       ---      &       ---      &       ---      & \\ 
          &    &                      &       ---      & 5: $    0.00$  & 5: $    0.32$  &       ---      &       ---      &       ---      &       ---      & \\ 
          &  4 &    --- sersic   ---  &  $\{  0.00\}$  &  $\{  0.00\}$  &  $     10.06$  &  $      1.88$  &  $      0.14$  &  $      0.39$  &  $      5.45$  & Compound\\ 
          &    &                      &  $\{  0.00\}$  &  $\{  0.00\}$  &  $      0.00$  &  $      0.00$  &  $      0.00$  &  $      0.00$  &  $   4190.51$  & spiral.\\ 
          &    &               power  &       ---      &  $      0.66$  &  $      2.34$  &  $   -172.67$  &  $     -0.11$  &  $     -0.01$  &  $     15.58$  & \\ 
          &    &                      &       ---      &  $      0.02$  &  $      0.01$  &  $      3.51$  &  $      0.01$  &  $   1100.08$  &  $   4190.50$  & \\ 
          &    &             fourier  &       ---      & 1: $   -0.15$  & 1: $   25.39$  & 3: $    0.02$  & 3: $  -32.12$  & 4: $    0.15$  & 4: $    8.38$  & \\ 
          &    &                      &       ---      & 1: $    0.00$  & 1: $    0.62$  & 3: $    0.00$  & 3: $    1.50$  & 4: $    0.00$  & 4: $    0.19$  & \\ 
          &    &             fourier  &       ---      & 5: $    0.02$  & 5: $   -4.02$  &       ---      &       ---      &       ---      &       ---      & \\ 
          &    &                      &       ---      & 5: $    0.00$  & 5: $    0.82$  &       ---      &       ---      &       ---      &       ---      & \\ 
M51B      &  5 &    --- sersic   ---  &  $     -0.19$  &  $      4.44$  &  $     12.06$  &  $      0.05$  &  $      0.89$  &  $      0.62$  &  $    -72.22$  & Compound \\ 
          &    &                      &  $      0.00$  &  $      0.00$  &  $      0.01$  &  $      0.00$  &  $      0.01$  &  $      0.00$  &  $      0.38$  & bulge. \\ 
          &  6 &    --- sersic   ---  &  $     -0.16$  &  $      4.43$  &  $     11.93$  &  $      0.18$  &  $      1.06$  &  $      0.81$  &  $     -2.79$  & Compound \\
          &    &                      &  $      0.00$  &  $      0.00$  &  $      0.02$  &  $      0.00$  &  $      0.03$  &  $      0.01$  &  $      1.45$  & bulge. \\
          &  7 &    --- sersic   ---  &  $     -0.45$  &  $      5.19$  &  $      9.93$  &  $      2.51$  &  $[    1.00]$  &  $      0.62$  &  $    -96.19$  & Tidal\\ 
          &    &                      &  $      0.01$  &  $      0.01$  &  $      0.00$  &  $      0.01$  &       ---      &  $      0.00$  &  $      0.36$  & structure. \\ 
          &    &             bending  &       ---      & 2: $    0.03$  & 3: $   -0.15$  &       ---      &       ---      &       ---      &       ---      & \\ 
          &    &                      &       ---      & 2: $    0.02$  & 3: $    0.00$  &       ---      &       ---      &       ---      &       ---      & \\ 
          &    &             fourier  &       ---      & 1: $    0.34$  & 1: $   17.20$  & 3: $   -0.25$  & 3: $   32.55$  & 4: $    0.14$  & 4: $   -3.73$  & \\ 
          &    &                      &       ---      & 1: $    0.00$  & 1: $    0.81$  & 3: $    0.00$  & 3: $    0.40$  & 4: $    0.00$  & 4: $    0.43$  & \\ 
          &    &             fourier  &       ---      & 5: $    0.03$  & 5: $    7.32$  &       ---      &       ---      &       ---      &       ---      & \\ 
          &    &                      &       ---      & 5: $    0.00$  & 5: $    1.16$  &       ---      &       ---      &       ---      &       ---      & \\ 
          &  8 &    --- sersic   ---  &  $     -0.10$  &  $      4.52$  &  $     10.20$  &  $      0.90$  &  $      0.72$  &  $      0.58$  &  $    -46.52$  & Bar and \\ 
          &    &                      &  $      0.00$  &  $      0.00$  &  $      0.00$  &  $      0.00$  &  $      0.00$  &  $      0.00$  &  $      0.56$  & spiral. \\ 
          &    &               power  &       ---      &  $      0.88$  &  $      1.08$  &  $     46.34$  &  $      1.60$  &  $     42.29$  &  $     52.50$  & \\ 
          &    &                      &       ---      &  $      0.00$  &  $      0.00$  &  $      0.66$  &  $      0.01$  &  $      0.16$  &  $      0.25$  & \\ 
          &    &             fourier  &       ---      & 1: $    0.07$  & 1: $  103.72$  & 3: $    0.05$  & 3: $   28.79$  & 4: $    0.01$  & 4: $   -0.11$  & \\ 
          &    &                      &       ---      & 1: $    0.00$  & 1: $    1.59$  & 3: $    0.00$  & 3: $    0.52$  & 4: $    0.00$  & 4: $    4.25$  & \\ 
          &    &             fourier  &       ---      & 5: $    0.01$  & 5: $   23.12$  &       ---      &       ---      &       ---      &       ---      & \\
          &    &                      &       ---      & 5: $    0.00$  & 5: $    1.09$  &       ---      &       ---      &       ---      &       ---      & \\
	  &    &     merit  &  \multicolumn{2}{c}{$\chi^2$ = 34279512.00} &  \multicolumn{2}{c}{$N_{\rm{dof}}$ = 632434} &  $N_{\rm{free}}$ = 104 &  \multicolumn{2}{c|}{$\chi^2_\nu$ = 54.20} & \\
\colrule
Trad.     &  1 &    --- sersic   ---  &  $      0.00$  &  $      0.00$  &  $     10.05$  &  $      0.33$  &  $      1.75$  &  $      0.85$  &  $    -62.09$  & \\ 
ellipsoid &    &                      &  $      0.00$  &  $      0.00$  &  $      0.00$  &  $      0.00$  &  $      0.01$  &  $      0.00$  &  $      0.55$  & \\ 
model     &  2 &    --- sersic   ---  &  $     -0.06$  &  $     -0.13$  &  $      8.47$  &  $      2.21$  &  $      0.33$  &  $      0.75$  &  $     26.69$  & \\ 
M51A      &    &                      &  $      0.00$  &  $      0.00$  &  $      0.00$  &  $      0.00$  &  $      0.00$  &  $      0.00$  &  $      0.13$  & \\ 
M51B      &  3 &    --- sersic   ---  &  $     -0.18$  &  $      4.40$  &  $      8.93$  &  $      2.54$  &  $      8.02$  &  $      0.92$  &  $    -44.73$  & \\ 
          &    &                      &  $      0.00$  &  $      0.00$  &  $      0.03$  &  $      0.11$  &  $      0.09$  &  $      0.00$  &  $      1.24$  & \\ 
          &  4 &    --- sersic   ---  &  $     -0.04$  &  $      4.90$  &  $     10.66$  &  $      1.45$  &  $      1.66$  &  $      0.57$  &  $     71.33$  & \\ 
          &    &                      &  $      0.00$  &  $      0.00$  &  $      0.02$  &  $      0.02$  &  $      0.02$  &  $      0.00$  &  $      0.42$  & \\ 
          &    &     merit  &  \multicolumn{2}{c}{$\chi^2$ = 42126720.00} &  \multicolumn{2}{c}{$N_{\rm{dof}}$ = 632507} &  $N_{\rm{free}}$ = 31 &  \multicolumn{2}{c|}{$\chi^2_\nu$ = 66.60} & 
\enddata

\tablecomments{Best-fitting parameters for M51.  See Table~\ref{arp147-table}
for details. The ``{\it Best fit}'' parameters (top section) correspond to
Panel ({\it b}) in Figure~\ref{m51}, ``{\it Traditional ellipsoid model}''
parameters (bottom section) produce residuals shown in Panel ({\it d}), and the
model is not shown.  The free parameters for the sky are not listed.  The
parameter in square brackets, [...], is held constant in the fit.  The curly
braces (\{...\}) around parameters indicate that they are coupled relative
to the first component.}

\label{m51-table}
\end{deluxetable*}

The classical Whirlpool galaxy is a beautiful system where a grand-design
spiral, M51A, is interacting with another spiral, M51B (Figure~\ref{m51},
Table~\ref{m51-table}).  In addition to there being obvious spiral structures
for both galaxies, there are large tidal disturbances that emerge from M51B,
as seen in the SDSS $r$-band image provided by D. Finkbeiner.  Because they
are closely overlapping, a desirable goal is to deblend M51A and B, as well
as to model the spiral and tidal structures, simultaneously.

As with previous examples, we fit this galaxy using both the most
sophisticated analysis (Table~\ref{m51-table}, top) in our toolbox, and
comparing the results to the traditional axisymmetric ellipsoids
(Table~\ref{m51-table}, bottom) analysis.  The traditional analysis requires
two components each, in order to decompose a galaxy ostensibly into a bulge
and a disk.  The reduction in $\chi^2_\nu$ between the two methods is modest,
because most of the residuals come from high-frequency starforming regions
that are not removed by models which are fundamentally smooth, despite being
modified by radial Fourier modes and spiral rotations.

In the most detailed analysis of M51A, we use two spiral arm components and
two components for the bulge.  There is actually not a strong need to use two
components for the bulge except to better capture the detailed profile shape,
which has an inflection at $r \approx 0\farcm4$, as seen in
Figure~\ref{m51}{\it i}.  On the other hand, the use of two spiral components
is necessary because the spiral arm has a ``kink'' in the rotation that cannot
be created by using a single smooth rotation function.  The spiral structures
are modified by Fourier modes to create both a slight lopsidedness and other
subtle features.  Because there are more degrees of freedom in a two-arm
spiral, the higher order Fourier modes also can ``see'' detailed structures,
like the reverse flaring of the spiral structure in Figure~\ref{m51}{\it f}.

For M51B, we employ three components in the fit, a bulge (component 5 in
Table~\ref{m51-table}, top), a tidal feature component (component 6), and a
spiral function (component 7), which model the three most visually striking
components.  The tidal feature is mostly obtained by using the second and third
bending modes of Equation~\ref{eqn:bending}, as illustrated in
Figure~\ref{fig:bending}.  However, bending modes 2 and 3 are symmetric
functions, so the high degree of asymmetry comes about because of combined
action with the Fourier modes, which is shown to have a high amplitude of 0.23
for the $m=1$ mode, as well as moderate values for other modes.  Incidentally,
despite the complexity of the higher order structures, all the parameter
values are determined automatically by \galfit\ without the need for an user
to provide initial guesses (i.e. initially all 0 values) and without tweaking
at any point in the analysis (which is hardly feasible anyhow).

For even those who are experienced with detailed parametric fitting, one of
the alarming facts about this analysis is that it employs 103 free parameters
in the best-fit model.  So there are natural concerns about parameter
degeneracies.  However, as we have discussed in
Section~\ref{subsect:degeneracies}, parameter degeneracies do not arise purely
based on the number of free parameters, but rather on the types of parameters
involved.  The availability of spatial information in 2-D provides one of the
most important ways to break parameter degeneracies.  We see this explicitly
in Figures~\ref{m51}{\it e-h}, where there is little evidence that the
subcomponents for M51A are strongly influenced by M51B, and vice versa.
Furthermore, within each galaxy, the subcomponents are so different in shape,
both qualitatively and quantitatively, that the amount of crosstalk between
them is also not significant.  Therefore, despite the extreme complexity of
this system, and the use of 103 free parameters, we find that degeneracies
between the parameters are not an issue.  Or, if they exist, they do so at a
low enough level that they do not significantly affect the main parameters of
interest, like the luminosity of the subcomponents, or the profile shapes and
sizes.

There {\it are}, however, seemingly degenerate conditions that have little to
do with parameter coupling.  Instead, these are attributed to the fact that
M51 has many of non-smooth structural features, caused by dust lanes,
star-forming regions, tidal disturbances, and so forth.  Such spatially
localized features, if strong enough, can influence \galfit\ to ``lock'' on to
them if the initial conditions happened to be sufficiently close.  The
consequences appear as degeneracies when, in fact, there are many small local
minima solutions.  This graininess in the $\chi^2$ terrain introduces slight
perturbations to the models, and may even cause fairly large shape differences
in the final solutions.  However, to a large extent, it rarely affects the
main parameters of interest, such as the luminosity of a particular component
or its size, which are determined by much more global features than the
nuisances of local fluctuations to which higher order parameters are more
sensitive.

To gain some intuitive insight into the effects of complex analysis, it is
instructive to compare simple and complex methods with regard to global and
subcomponent properties.  In terms of the total luminosity, here we find
excellent agreement between sophisticated and traditional analysis,
respectively, of $m_{r}=8.24$ vs.  $m_{r}=8.25$ for M51A, and $m_{r} = 8.80$
vs.  $m_{r}=8.73$ for M51B.  While this level of agreement may at first seem
surprising, it is expected given the basic premise of least-squares
minimization.  In fact, even a single-component fit to M51A yields
$m_{r}=8.0$, and for M51B $m_{r}=9.0$, which are both quite close to the
overall best-fit models, despite the complications in the image.  The main
reason for the discrepancy here is the uncertainty in the sky, due to there
being a large gradient.  This fundamentally sets the limit on the accuracy of
the photometry to perhaps no better than 0.1 to 0.2 magnitude, independent of
the analysis method.

The most sensitive benchmark for understanding differences in the analysis is
in detailed decompositions.  Here we compare the bulge-to-disk decomposition
results.  In the traditional ellipsoid analysis, we find a B/D ratio of 0.23
for M51A and 4.9 for M51B.  The large B/D ratio for M51B is clearly
unphysical, and is driven by the large \sersic\ index ($n=8.0$) of the bulge
component, which is increased to accommodate the flux in the outskirts due to
tidal features.  In the most detailed analysis, the B/D ratio for M51A is
0.16, whereas for M51B it is merely 0.17.  Examining the bulge of M51A more
closely, we find that the detailed analysis yields a total flux of 10.38 mag,
whereas the traditional analysis extracts a brighter bulge of 10.05 mag.  The
differences come from the fact that the light of the inner spiral is in part
driving up the \sersic\ index of the bulge when it is not properly accounted.
It is probably safe to conclude that a magnitude of 10.05 is a firm upper
limit to the bulge luminosity.

Finally, it is worthwhile to compare how the disk parameters differ between
the analyses to gain an understanding for how coordinate rotation affects the
interpretation of the parameters for the spiral models.  From
Table~\ref{m51-table}, we find that the \sersic\ index of the simple and
complex models are essentially identical for M51A, at $n\approx0.33$.  The
interpretation for M51B is more complicated, because the ``disk'' in an
ellipsoidal model is not qualitatively the same structure as the spiral
analysis.  In fact, it is necessary to hold the \sersic\ index of the tidal
component 6 fixed in the analysis.  Nevertheless there are clearly
quantitative differences in that the simple analysis is larger by 55\% in $n$.
With regard to the effective radius, the traditional analysis of M51A finds
the disk size to be about 2\farcm2, which compares favorably with the spiral
model size of 2\farcm8, or a 25\% difference.  Furthermore, the disk
magnitudes for M51A differ only by 0.03 mag between simple and complex.

These comparisons therefore demonstrate that despite the complex analysis
being much more realistic looking, fundamentally the meaning of the structural
parameters (size, luminosity, concentration index) are unchanged from the
original definition, even in the situation of spiral components.  This is an
useful fact because our prior intuitions, honed on fitting ellipsoidal models,
continue to be applicable.  We note that the generally good agreement between
detailed and simplistic analysis witnessed here and in previous examples is
not entirely coincidental.  It so happens because all shapes are fundamentally
perturbations of the best-fitting ellipsoidal model, even if the result bears
no resemblance to the original ellipse.

\subsection {NGC 289}

\begin {figure*} 
    \centerline {\includegraphics[angle=90,width=6.5truein]{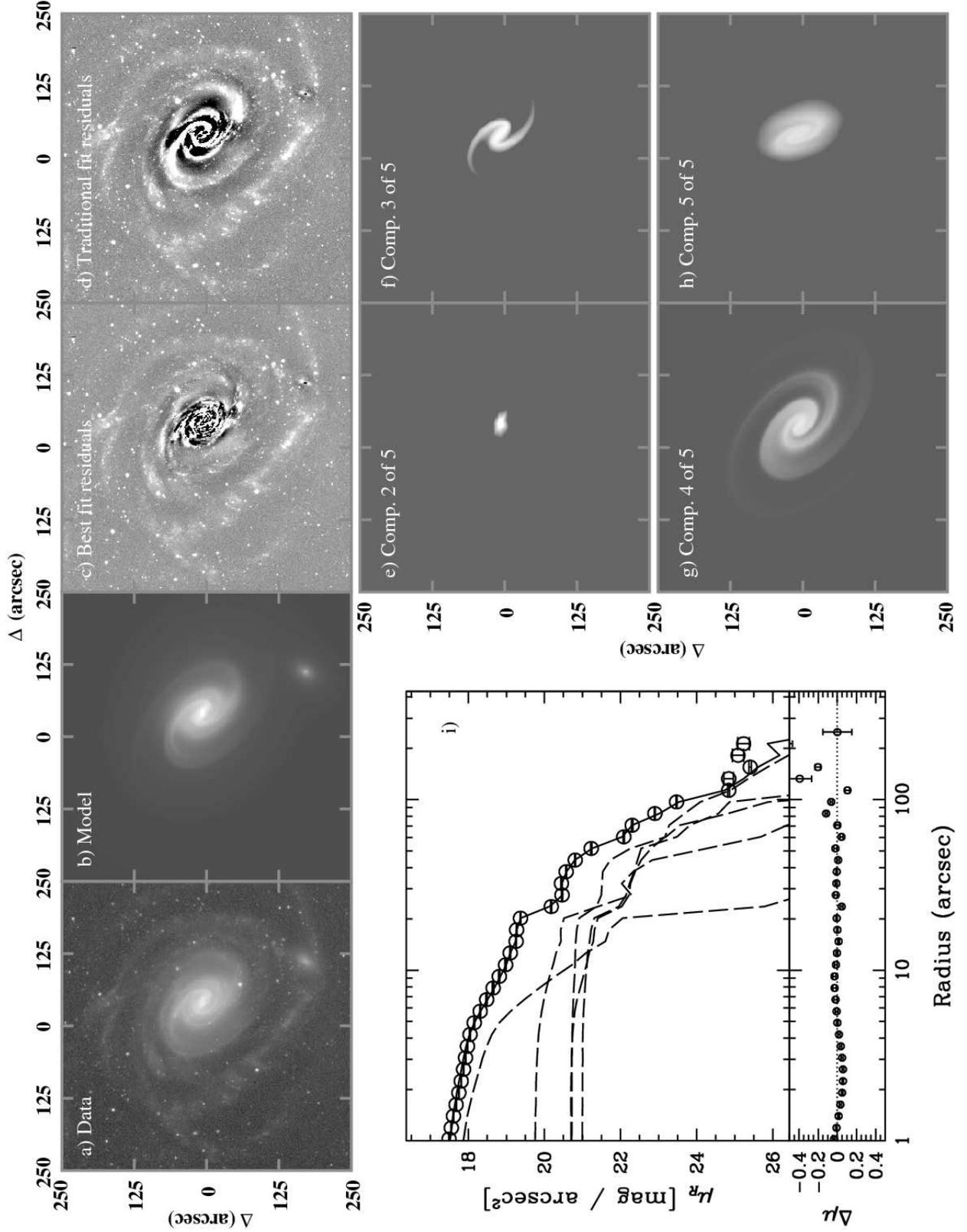}}

    \caption {Detailed analysis of NGC 289 from CINGS. ({\it a}) Original data.  ({\it b})
    Best \sersic\ profile fits with spiral rotation functions and Fourier
    modes, corresponding to the parameters shown in Table~\ref{n289-table}.
    ({\it c}) Best-fit residuals.  ({\it d}) The fit residuals using
    traditional, axisymmetric, ellipsoidal model components.  ({\it e})
    The fine details of the inner bar structure of Panel~({\it b}).  ({\it f}) Spiral
    component 1 of 3 of the best-fit model.  ({\it g}) The spiral component 2
    of 3. ({\it h}) The spiral component 3 of 3.  A bulge component is present
    but not shown in the figures.  ({\it i}) 1-D surface brightness profile of
    the galaxy.  The individual components are shown as dashed lines,
    and the solid line coursing through the data is the sum of the different
    components.  The lower panel shows the residuals of data $-$ model.}

    \vskip 0.15truein
    \label{n289}
\end {figure*}

\begin{deluxetable*}{|c|c|r|rrrrrrr|l|}
\tabletypesize{\scriptsize}
\tablewidth{0pt}
\tablecaption {NGC~289 Fitting Results}
\tablehead{  
  & \#  & --- sersic --- & $\Delta x$ [\arcsec] & $\Delta y$ [\arcsec] &  mag    &   $r_e$ [\arcsec]  &   $n$     &  $q$      &  $\theta_{\rm{PA}}$ [deg]   & Comments \\ 
  &     & power &    ---      &    $r_{\rm{in}}$ [\arcsec]&  $r_{\rm{out}}$ [\arcsec] &  $\theta_{\rm{rot}}$ [deg] & $\alpha$   &  $\theta_{\rm{incl}}$ [deg] &  $\theta_{\rm{sky}}$ [deg]& \\ 
  &     & fourier &    ---      & \multicolumn{2}{r}{mode: ampl. \& phase [deg]} & \multicolumn{2}{r}{mode: ampl. \& phase [deg]}  & \multicolumn{2}{r|}{mode: ampl. \& phase [deg]} & }
\startdata
Best      &  1 &    --- sersic   ---  &  $      0.00$  &  $      0.00$  &  $     11.69$  &  $     64.01$  &  $      1.72$  &  $      0.78$  &  $     61.27$  & Bulge. \\ 
fit       &    &                      &  $      0.09$  &  $      0.10$  &  $      0.01$  &  $      0.75$  &  $      0.03$  &  $      0.00$  &  $      0.43$  & \\ 
          &  2 &    --- sersic   ---  &  $     -2.63$  &  $     -1.92$  &  $     13.27$  &  $      6.05$  &  $      1.02$  &  $      0.51$  &  $     77.83$  & Inner \\ 
NGC 289   &    &                      &  $      0.02$  &  $      0.01$  &  $      0.01$  &  $      0.04$  &  $      0.01$  &  $      0.00$  &  $      0.29$  & bar. \\ 
          &    &             fourier  &       ---      & 1: $    0.10$  & 1: $   63.87$  & 3: $   -0.05$  & 3: $   -4.23$  & 4: $   -0.05$  & 4: $  -20.77$  & \\ 
          &    &                      &       ---      & 1: $    0.01$  & 1: $    4.87$  & 3: $    0.00$  & 3: $    0.92$  & 4: $    0.00$  & 4: $    0.95$  & \\ 
          &    &             fourier  &       ---      & 5: $    0.03$  & 5: $    4.62$  & 6: $    0.06$  & 6: $   -2.09$  &       ---      &       ---      & \\ 
          &    &                      &       ---      & 5: $    0.00$  & 5: $    1.32$  & 6: $    0.00$  & 6: $    0.61$  &       ---      &       ---      & \\ 
          &  3 &    --- sersic   ---  &  $     -2.25$  &  $     -2.77$  &  $     12.30$  &  $     32.86$  &  $      0.54$  &  $      0.32$  &  $    -85.94$  & Spiral \\
          &    &                      &  $      0.02$  &  $      0.02$  &  $      0.01$  &  $      0.15$  &  $      0.00$  &  $      0.00$  &  $      0.53$  & comp. 1. \\ 
          &    &               power  &       ---      &  $     19.23$  &  $     34.40$  &  $     85.51$  &  $      1.48$  &  $     52.11$  &  $    136.18$  & \\ 
          &    &                      &       ---      &  $      0.14$  &  $      0.14$  &  $      0.94$  &  $      0.02$  &  $      0.08$  &  $      0.10$  & \\ 
          &    &             fourier  &       ---      & 1: $    0.14$  & 1: $  -97.28$  & 3: $   -0.05$  & 3: $   -8.55$  & 4: $    0.02$  & 4: $    3.41$  & \\ 
          &    &                      &       ---      & 1: $    0.00$  & 1: $    0.74$  & 3: $    0.00$  & 3: $    1.13$  & 4: $    0.00$  & 4: $    2.09$  & \\ 
          &    &             fourier  &       ---      & 5: $    0.02$  & 5: $   -3.77$  & 6: $    0.01$  & 6: $   10.09$  &       ---      &       ---      & \\ 
          &    &                      &       ---      & 5: $    0.00$  & 5: $    0.99$  & 6: $    0.00$  & 6: $    3.36$  &       ---      &       ---      & \\ 
          &  4 &    --- sersic   ---  &  $     -4.54$  &  $     -4.10$  &  $     12.13$  &  $     52.28$  &  $      0.74$  &  $      0.56$  &  $    -32.85$  & Spiral \\
          &    &                      &  $      0.04$  &  $      0.03$  &  $      0.01$  &  $      0.22$  &  $      0.01$  &  $      0.00$  &  $     28.48$  & comp. 2. \\ 
          &    &               power  &       ---      &  $    -26.32$  &  $     71.31$  &  $    450.01$  &  $      0.77$  &  $     53.30$  &  $    140.24$  & \\ 
          &    &                      &       ---      &  $      4.45$  &  $      0.59$  &  $     25.64$  &  $      0.06$  &  $      0.05$  &  $      0.07$  & \\ 
          &    &             fourier  &       ---      & 1: $   -0.12$  & 1: $   84.16$  & 3: $    0.06$  & 3: $   32.07$  & 4: $   -0.06$  & 4: $  -30.37$  & \\ 
          &    &                      &       ---      & 1: $    0.00$  & 1: $    0.64$  & 3: $    0.00$  & 3: $    0.45$  & 4: $    0.00$  & 4: $    0.38$  & \\ 
          &    &             fourier  &       ---      & 5: $    0.04$  & 5: $   -6.47$  & 6: $    0.02$  & 6: $  -20.35$  &       ---      &       ---      & \\ 
          &    &                      &       ---      & 5: $    0.00$  & 5: $    0.32$  & 6: $    0.00$  & 6: $    0.81$  &       ---      &       ---      & \\ 
          &  5 &    --- sersic   ---  &  $     -2.40$  &  $     -1.90$  &  $     11.82$  &  $     50.67$  &  $      0.46$  &  $      0.68$  &  $    -45.45$  & Spiral \\ 
          &    &                      &  $      0.04$  &  $      0.04$  &  $      0.00$  &  $      0.30$  &  $      0.00$  &  $      0.00$  &  $     35.81$  & comp. 3. \\ 
          &    &               power  &       ---      &  $     -9.09$  &  $     75.87$  &  $    411.91$  &  $     -0.04$  &  $     64.81$  &  $    112.75$  & \\ 
          &    &                      &       ---      &  $      4.42$  &  $      0.69$  &  $     36.56$  &  $      0.01$  &  $      0.13$  &  $      0.09$  & \\ 
          &    &             fourier  &       ---      & 1: $   -0.11$  & 1: $  -33.85$  & 3: $    0.01$  & 3: $    1.93$  & 4: $   -0.00$  & 4: $    9.85$  & \\ 
          &    &                      &       ---      & 1: $    0.00$  & 1: $    1.08$  & 3: $    0.00$  & 3: $    3.10$  & 4: $    0.00$  & 4: $    6.71$  & \\ 
          &    &             fourier  &       ---      & 5: $    0.00$  & 5: $  -16.26$  & 6: $    0.02$  & 6: $   -6.02$  &       ---      &       ---      & \\ 
          &    &                      &       ---      & 5: $    0.00$  & 5: $   10.08$  & 6: $    0.00$  & 6: $    0.58$  &       ---      &       ---      & \\ 
Neighbor  &  6 &    --- sersic   ---  &  $     67.90$  &  $   -181.06$  &  $     14.69$  &  $     20.03$  &  $      1.93$  &  $      0.72$  &  $    -34.31$  & \\ 
galaxy    &    &                      &  $      0.04$  &  $      0.05$  &  $      0.01$  &  $      0.34$  &  $      0.03$  &  $      0.01$  &  $      0.99$  & \\ 
          &    &     merit  &  \multicolumn{2}{c}{$\chi^2$ = 158680.39} &  \multicolumn{2}{c}{$N_{\rm{dof}}$ = 150419} &  $N_{\rm{free}}$ = 103 &  \multicolumn{2}{c|}{$\chi^2_\nu$ = 1.05} & \\ 
\colrule
Tradit.   &  1 &    --- sersic   ---  &  $      0.00$  &  $      0.00$  &  $     11.03$  &  $     41.90$  &  $      1.62$  &  $      0.74$  &  $     54.96$  & ``bulge''?\\ 
ellipsoid &    &                      &  $      0.04$  &  $      0.03$  &  $      0.01$  &  $      0.33$  &  $      0.01$  &  $      0.00$  &  $      0.21$  & \\ 
model     &  2 &    --- sersic   ---  &  $      2.01$  &  $     -3.52$  &  $     11.69$  &  $     36.98$  &  $      0.29$  &  $      0.55$  &  $     23.85$  & disk \\ 
          &    &                      &  $      0.04$  &  $      0.05$  &  $      0.01$  &  $      0.06$  &  $      0.00$  &  $      0.00$  &  $      0.12$  & \\ 
          &  3 &    --- sersic   ---  &  $      0.02$  &  $     -1.10$  &  $     12.70$  &  $     10.83$  &  $      1.24$  &  $      0.43$  &  $     67.37$  & bar \\ 
          &    &                      &  $      0.02$  &  $      0.01$  &  $      0.03$  &  $      0.08$  &  $      0.01$  &  $      0.00$  &  $      0.19$  & \\ 
Neighbor  &  4 &    --- sersic   ---  &  $     70.60$  &  $   -179.97$  &  $     14.71$  &  $     19.62$  &  $      1.90$  &  $      0.72$  &  $    -34.01$  & \\ 
galaxy    &    &                      &  $      0.04$  &  $      0.05$  &  $      0.01$  &  $      0.34$  &  $      0.03$  &  $      0.01$  &  $      1.11$  & \\ 
          &    &     merit  &  \multicolumn{2}{c}{$\chi^2$ = 200361.19} &  \multicolumn{2}{c}{$N_{\rm{dof}}$ = 150491} &  $N_{\rm{free}}$ = 31 &  \multicolumn{2}{c|}{$\chi^2_\nu$ = 1.33} &
\enddata

\tablecomments{Best-fitting parameters for NGC 289.  See
Table~\ref{ic4710-table} for details.  The ``{\it Best fit}'' parameters (top
section) correspond to Panel ({\it b}) in Figure~\ref{n289}, ``{\it Traditional
ellipsoid model}'' parameters (bottom section) produce residuals shown in
Panel ({\it d}), and the model is not shown.  The free parameters for the
sky are not listed.}

\label{n289-table}

\end{deluxetable*}

NGC 289 is an SAB(rs)bc galaxy, with a weak bar and a complex inner spiral
system (Figure~\ref{n289}, Table~\ref{n289-table}) that resembles a ring.
Upon closer examination, the ring appearance comes about because there exists
a bifurcation in the spiral structure that connects up with the opposing
spiral arm.  Furthermore, the bar is also multi-component, with the inner
component oriented at an angle nearly $45^\circ$ from the strong outer bar.

The best-fit analysis involves three spiral components, an inner and an outer
bar component, and a bulge (Table~\ref{n289-table}, top).  All except for the
bulge component are modified by five Fourier modes, and are shown in
Figures~\ref{n289}{\it e-h}.  The requirement of components 3 and 4
(Figures~\ref{n289}{\it f} and {\it g}) is clear, because they are what form
the most striking and intricate patterns in the center, while the requirement
of component 5 (Figure \ref{n289}{\it h}) is only evident in the residuals,
and makes up some of the diffuse light within the inner 60\arcsec\ region.
Although it does not seem like an essential component, the inner bar structure
(Figure~\ref{n289}{\it e}) qualitatively affects the detailed residual pattern
at the center, and is therefore included.  When all the detailed inner
structures are properly accounted for, it is straightforward to infer the
bulge component, and assess the uncertainties by varying different parameters
of the bulge.  Doing so does not affect the inner fine-structures because they
are sharp and well localized.

Conducting the same decomposition using traditional ellipsoid models
(Table~\ref{n289-table}, bottom), we opted to fit three components, ostensibly to
model a bulge, disk, and a bar.  The result produces residuals seen in
Figure~\ref{n289}{\it d}, revealing the intricate details of the inner spiral
system.  From the fit, even though the disk and the bar component are
sensible, the bulge component is actually fitting a diffuse disk component,
which, in retrospect, is that shown in Figure~\ref{n289}{\it h}.  Because that
inner spiral component is quite luminous, and because there exists a bulge
component superposed on top of it, this quasi-bulge model is almost 0.7 mag brighter than
that inferred through the detailed modeling above.  Adding a fourth ellipsoid
model is not possible, because the central spiral residuals are so great that
they completely suppress the addition of another component, causing the flux
to go to zero.

Once again, comparing the total luminosity between the best-fit model with the
ellipsoid fit, we find an excellent agreement of $m=10.37$ mag vs. $m=10.42$,
respectively.  For a single-component ellipsoid fit, there is also an
excellent agreement of $m=10.46$, despite the main structural details not
being unaccounted.

\section {DISCUSSION and CONCLUSIONS}

\label{sect:conclusion}

This study is a proof of concept for how to conduct more realistic
image-fitting analysis using purely parametric functions, by breaking free
from traditional assumptions about axisymmetry.  We introduced several new
ideas, including the use of Fourier and bending modes, spiral rotation
functions, and truncation functions.  These features can be used individually,
or combined in arbitrary ways.  While these features are individually quite
simplistic, used collectively they proliferate a dizzying array of
possibilities.  Even so, the interpretation of each component remains
intuitive, down to the meaning of each fitting parameter.  Indeed, the
interpretation of the traditional ellipsoidal profile parameters, such as
those for the \sersic\ function, remains essentially unchanged under
modification.  We then applied these techniques to five case studies,
illustrating that highly complex and intricate structures can be modeled using
fully parametric techniques.

There are many practical applications for these techniques.  For instance, the
Fourier modes are useful for quantifying the average global symmetry of a
galaxy, and can easily be automated for galaxy surveys.  It is also
possible to disentangle bright from faint asymmetries, and to conduct more robust 
bulge-to-disk decompositions in some galaxies.  It would be useful to
quantify how much of the total flux is in a bulge versus the tidally distorted
component, which has implications for issues such as late- vs.  early-stage 
mergers, or major vs. minor mergers.

More than just a presentation of new techniques, one of the main purposes of
this study is to highlight a method to more realistically quantify measurement
uncertainties in high-S/N images.  In galaxy fitting, the most desirable goal
will always be to obtain a fit with the lowest $\chi^2$, using the simplest
model.  In the past, this idea was closely tied to the practice of using one-
or two-component ellipsoid models, out of necessity.  Simplicity is not
necessarily congruent with propriety or reality.  This study promotes the
notion that simple models can be realistic.  It also opens up new
possibilities for more detailed analysis depending on the image complexity.
However, this possibility is both a blessing and a curse.  For, the fact there
is not one generic solution for any galaxy leads to the following conundrum in
image analysis, but one that illustrates the merit of our approach:

{\it ``What model should one adopt, how much detail is enough, and what about
degeneracies?''} \ \ \ \ We have shown that detailed decomposition analysis
can be arbitrarily sophisticated.  It is for that same reason there is often
not a single, unique answer.  However, the essential fact, as seen through our
examples and other detailed analysis outside of this work, is that the
marginal return of adding complexity quickly diminishes.  Therefore, the above
conundrum is in practice easy to address by conducting analyses of varying
sophistication without prejudice, then judging the outcome by taking a clear
view of what goal is to be achieved.  If different solutions yield the same
result for a desired science goal (e.g., bulge luminosity, bulge-to-disk
ratio, average size, total luminosity, etc.), then it does not matter which
model one adopts.  If they yield different outcomes, then the most realistic
analysis ought to be the more true.  However, if it is not possible to decide
on the correct model, the different results by definition give an estimate of
the model-dependent {\it measurement uncertainty.} This last attribute, rather
than being a perceived weakness, is fundamentally that which makes the
analysis quantitatively rigorous.

Despite the flexibility allowed by the models, this paper is merely an initial
demonstration of concept and leaves many issues unsolved.  Currently, the
formulation of the spiral rotation function is fairly rigid, and cannot
produce arms that wind back onto itself (although that can be approximated
using the ring feature in \galfit).  The amount of curvature in the bending
modes can only fit arcs and not fuller semi-circles (which can partly be
modeled using a lopsided ring).  There remains substantial room for future
growth in profile types, shape definitions, and toward spatial-spectral
decompositions for integral-field data.

\bigskip

\acknowledgements

CYP gratefully acknowledges discussions with and suggestions from many people
over the course of this work, including Lauren. A.  MacArthur who greatly
improved this manuscript, J.  Greene, C.  Brasseur, D. McIntosh, J.  Hesser,
T. Puzia, K.  Jahnke, S.  Zibetti, E.  Bell, A.  Barth, E.  Laurikainen, M.
Barden, B.  H{\"a}ussler, M.  Gray, and \galfit\ users over the years.  We
thank the anonymous referee for an expeditious review that improved the
discussions.  A number of individuals not only very kindly provided us high
quality images, but also granted us permission to analyze and use them for
this study, including M.  Seigar and A.  Barth for IC 4710 and NGC 289 from
the Carnegie Irvine Nearby Galaxy Survey, D.  Finkbeiner for the SDSS image of
M51, J.  Caldwell and the GEMS collaboration for the edge-on galaxy from their
survey.  Several others, including J. Lee, M.  Kim, and H.  Hernandez,
contributed galaxy images which motivated the development of certain \galfit\
features.  The various contributions and input from all made the development
of this software a greatly enjoyable endeavor, and to whom CYP is greatly
indebted.  This work was made possible by the generous financial support of
the Herzberg Institute of Astrophysics, National Research Council of Canada,
through the Plaskett Fellowship program, by STScI through the
Institute/Giacconi Fellowship, and by visitor programs of the
Max-Planck-Institut F{\"u}r Astronomie, operated by the Max-Planck Society of
Germany.

This research has made use of the NASA/IPAC Extragalactic Database (NED) which
is operated by the Jet Propulsion Laboratory, California Institute of
Technology, under contract with the National Aeronautics and Space
Administration.

Based on observations made with the NASA/ESA Hubble Space Telescope, and
obtained from the Hubble Legacy Archive, which is a collaboration between the
Space Telescope Science Institute (STScI/NASA), the Space Telescope European
Coordinating Facility (ST-ECF/ESA) and the Canadian Astronomy Data Centre
(CADC/NRC/CSA).

Funding for the SDSS and SDSS-II has been provided by the Alfred P. Sloan
Foundation, the Participating Institutions, the National Science
Foundation, the U.S. Department of Energy, the National Aeronautics and
Space Administration, the Japanese Monbukagakusho, the Max Planck Society,
and the Higher Education Funding Council for England. The SDSS Web Site is
http://www.sdss.org/.

The SDSS is managed by the Astrophysical Research Consortium for the
Participating Institutions. The Participating Institutions are the American
Museum of Natural History, Astrophysical Institute Potsdam, University of
Basel, University of Cambridge, Case Western Reserve University, University of
Chicago, Drexel University, Fermilab, the Institute for Advanced Study, the
Japan Participation Group, Johns Hopkins University, the Joint Institute for
Nuclear Astrophysics, the Kavli Institute for Particle Astrophysics and
Cosmology, the Korean Scientist Group, the Chinese Academy of Sciences
(LAMOST), Los Alamos National Laboratory, the Max-Planck-Institute for
Astronomy (MPIA), the Max-Planck-Institute for Astrophysics (MPA), New Mexico
State University, Ohio State University, University of Pittsburgh, University
of Portsmouth, Princeton University, the United States Naval Observatory, and
the University of Washington.

\begin{appendix}

{\centerline {A --- HYPERBOLIC TANGENT ROTATION FUNCTION}}

\bigskip

The hyperbolic tangent (${\rm tanh} (r_{\rm in}, r_{\rm out}, \theta_{\rm
incl}, \theta^{\rm sky}_{\rm PA};r)$) portion of the $\alpha$-tanh
(Equation~\ref{eqn:alphatanh}) and log-tanh (Equation~\ref{eqn:logtanh})
rotation functions is given by Equation~\ref{eqn:tanh} below.  The constant
CDEF is defined such that at the mathematical ``bar radius'' $r_{\rm in}$, the
rotation angle $\theta$ reaches $20^\circ$.  This definition is entirely
empirical.  The above Figure shows a pure tanh rotation function, where the
rotation angle reaches a maximum $\theta_{\rm out}$ near $r=r_{\rm out}$.
Beyond $r_{\rm out}$, the rotation angle levels off at $\theta_{\rm out}$.
This function is multiplied with either a logarithmic or a power-law function
to produce the desired asymptotic rotation behavior seen in more realistic
galaxies (see Section~\ref{sect:azimuthal}).

\begin{figure}
\epsscale {0.5}
\plotone{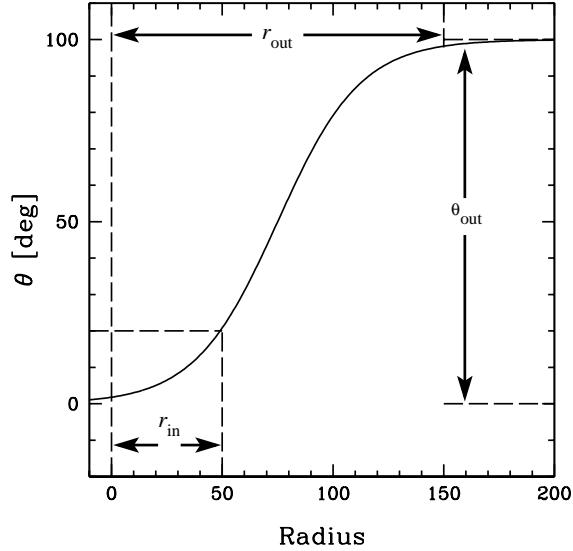}

\caption {Appendix Figure A: Schematics of a hyperbolic tangent rotation
    function.  $r_{\rm in}$ is the inner radius where the rotation angle
    reaches 20 degrees relative to the PA of the best fitting ellipse of a
    component.  Below $r_{\rm in}$ the function flattens out to zero degrees.
    $r_{\rm out}$ is the outer radius, beyond which the function flattens out
    to a constant rotation angle, and $\theta_{\rm out}$ is the total amount
    of rotation out to $r_{\rm out}$.}

\end{figure}

\begin{equation}
    {\rm CDEF} = 0.23\ \ \ \ \ {\rm (constant\ for\ ``bar" \ definition)}
\end{equation}

\begin{equation}
    A = \frac{2 \times {\rm CDEF}}{ |\theta_{\rm out}| + {\rm CDEF}} - 1.00001
\end{equation}

\begin{equation}
    B = \left(2 - {\rm tanh}^{-1} (A)\right)
		 \left(\frac{r_{\rm out}}{r_{\rm out}-r_{\rm in}}\right)
\end{equation}

\begin{equation}
    r = \sqrt {\Delta x^2 + \Delta y^2}\ \ \ \ \ \ {\rm (circular-centric\ distance)}
\end{equation}

\begin{equation}
\label{eqn:tanh}
    {\rm tanh} (r_{\rm in}, r_{\rm out}, \theta_{\rm incl}, \theta^{\rm
		sky}_{\rm PA}; r) \equiv 0.5 \times \left({\rm tanh}\
		\left[B\left(\frac{r}{r_{\rm out}} - 1\right) + 2\right] +1
		\right)
\end{equation}

\bigskip
\bigskip

{\centerline {B --- HYPERBOLIC TANGENT TRUNCATION FUNCTION}}

\bigskip

The hyperbolic tangent truncation function (${\rm tanh} (x_0, y_0; r_{\rm
break}, r_{\rm soft}, q, \theta_{\rm PA})$) (see
Section~\ref{sect:truncation}) is very similar to the coordinate rotation
formulation in Appendix~A, except for different constants that define the flux
ratio at the truncation radii:  at $r=r_{\rm break}$ the flux is 99\% of the
untruncated model profile, whereas at $r=r_{\rm soft}$ the flux is 1\%.  With
this definition, Equation~\ref{eqn:trunc} is the truncation function:

\begin{equation}
    B = 2.65 - 4.98 \left(\frac{r_{\rm break}}{r_{\rm break}-
	r_{\rm soft}}\right)
\end{equation}

\begin{equation}
\label{eqn:trunc}
    {\rm tanh} (x_0, y_0; r_{\rm break}, r_{\rm soft}, q, \theta_{\rm PA})
		\equiv 0.5 \times \left({\rm tanh}\ \left[(2-B)\frac{r}{r_{\rm
		break}} + B\right] +1\right)
\end{equation}

\noindent Note that the radius $r$ is a generalized radius (as opposed to a
circular-centric distance), i.e. one that is perturbed by $C_0$, bending
modes, or Fourier modes, of the truncation function.  When the softening length
($\Delta r_{\rm soft}$) is used as a free parameter, it is defined as $\Delta
r_{\rm soft} = r_{\rm break} - r_{\rm soft}$.

\end {appendix}

\bibliography{references}

\end {document}